\documentclass[twocolumn,
amsmath,amssymb,prb,
aps,longbibliography, groupedaddress,superscriptaddress
]{revtex4-2}

\usepackage{graphicx}
\usepackage{dcolumn}
\usepackage{bm}
\usepackage{braket}
\usepackage{mathtools}
\usepackage{bbm}
\usepackage{comment}
\usepackage{makecell}
\usepackage{afterpage}
\usepackage{siunitx}
\usepackage{nicefrac}
\usepackage{amsthm}
\usepackage[export]{adjustbox}
\usepackage{tikz}
\usetikzlibrary{positioning, shapes, arrows.meta}

\usepackage{xcolor}

\newcommand{\T}{\mathcal T}
\newcommand{\Z}{\mathcal Z}
\newcommand{\U}{\mathcal U}
\newcommand{\id}{\mathrm{id}}

\newcommand{\dagg}{^\dagger}
\newcommand{\proj}[1]{\ket{#1}\!\!\bra{#1}}

\usepackage{hyperref}
\hypersetup{
	colorlinks=true,
	linkcolor=blue,
	citecolor=blue,
	filecolor=magenta,
	urlcolor=cyan
}
\usepackage[capitalize]{cleveref}

\DeclareMathOperator{\tr}{Tr}
\DeclareMathOperator{\Tr}{Tr}

\def\dif{{\rm d}}

\newtheorem{theorem}{Theorem}
\newtheorem{lemma}[theorem]{Lemma}
\newtheorem*{lemma*}{Lemma}

\newtheorem{proposition}[theorem]{Proposition}
\newtheorem{corollary}[theorem]{Corollary}
\newtheorem{definition}[theorem]{Definition}
\newtheorem*{mainresult*}{Main result}

\usepackage{tikz}
\usetikzlibrary{shapes.misc, positioning, arrows, calc, decorations.markings}
\usepackage{pgfplots}
\makeatletter
\tikzset{use path/.code=\tikz@addmode{\pgfsyssoftpath@setcurrentpath#1}}
\makeatother

\definecolor{statecol}{RGB}{200,210,255}
\definecolor{conccol}{HTML}{ffe6a6}
\definecolor{bcol}{RGB}{230,255,220}
\definecolor{acol}{RGB}{255,230,220}
\definecolor{ccol}{RGB}{250,255,210}
\definecolor{bdark}{RGB}{84, 130, 65}

\pgfmathsetmacro{\width}{512pt}%
\pgfmathsetmacro{\height}{164pt}%
\pgfmathsetmacro{\margw}{3pt}
\pgfmathsetmacro{\rmarg}{19}
\pgfmathsetmacro{\SAWratio}{0.24}
\pgfmathsetmacro{\scalemain}{1/(1+\SAWratio + ((\margw+\rmarg)/\width))}
\pgfmathsetmacro{\bborder}{35pt}
\pgfmathsetmacro{\topShift}{76pt}

\graphicspath{{../../}{tikz/}{../../arxiv/}{../../tikz/}}

\makeatletter
\DeclareFontFamily{OMX}{MnSymbolE}{}
\DeclareSymbolFont{MnLargeSymbols}{OMX}{MnSymbolE}{m}{n}
\SetSymbolFont{MnLargeSymbols}{bold}{OMX}{MnSymbolE}{b}{n}
\DeclareFontShape{OMX}{MnSymbolE}{m}{n}{
	<-6>  MnSymbolE5
	<6-7>  MnSymbolE6
	<7-8>  MnSymbolE7
	<8-9>  MnSymbolE8
	<9-10> MnSymbolE9
	<10-12> MnSymbolE10
	<12->   MnSymbolE12
}{}
\DeclareFontShape{OMX}{MnSymbolE}{b}{n}{
	<-6>  MnSymbolE-Bold5
	<6-7>  MnSymbolE-Bold6
	<7-8>  MnSymbolE-Bold7
	<8-9>  MnSymbolE-Bold8
	<9-10> MnSymbolE-Bold9
	<10-12> MnSymbolE-Bold10
	<12->   MnSymbolE-Bold12
}{}

\let\llangle\@undefined
\let\rrangle\@undefined
\DeclareMathDelimiter{\llangle}{\mathopen}%
{MnLargeSymbols}{'164}{MnLargeSymbols}{'164}
\DeclareMathDelimiter{\rrangle}{\mathclose}%
{MnLargeSymbols}{'171}{MnLargeSymbols}{'171}
\makeatother

\usepackage{tikz}
\usetikzlibrary{shapes.misc, positioning, arrows, calc, decorations.markings}
\usepackage{pgfplots}
\makeatletter
\tikzset{use path/.code=\tikz@addmode{\pgfsyssoftpath@setcurrentpath#1}}
\makeatother

\definecolor{statecol}{RGB}{200,210,255}
\definecolor{conccol}{HTML}{ffe6a6}

\def\TCM{{T.C.M. Group, Cavendish Laboratory, JJ Thomson Avenue, Cambridge CB3 0HE, United Kingdom}}

\def\copenhagen{{Department of Mathematical Sciences, University of Copenhagen, Universitetsparken 5, 2100 Copenhagen, Denmark}}

\def\CQT{{Centre for Quantum Technologies, National University of Singapore, Singapore 117543}}
\def\NUS{{Department of Physics, National University of Singapore, Singapore 117551}}

\begin{document}
	
	\def\papertitle{{Measurement-induced entanglement and complexity in random constant-depth 2D quantum circuits}}

	\title{\papertitle}
	\author{Max McGinley}
	\email{mm2025@cam.ac.uk}
	\affiliation{\TCM}
	\author{Wen Wei Ho}
	\affiliation{\CQT}
	\affiliation{\NUS}
	\author{Daniel Malz}
	\affiliation{\copenhagen}

	\begin{abstract}
		We analyse the entanglement structure of states generated by random constant-depth two-dimensional quantum circuits, followed by projective measurements of a subset of sites. By deriving a rigorous lower bound on the average entanglement entropy of such post-measurement states, we prove that macroscopic long-ranged entanglement is generated above some constant critical depth in several natural classes of circuit architectures, which include brickwork circuits and random holographic tensor networks. This behaviour had been conjectured based on previous works, which utilize non-rigorous methods such as replica theory calculations, or work in regimes where the local Hilbert space dimension grows with system size. To establish our lower bound, we develop new replica-free theoretical techniques that leverage tools from multi-user quantum information theory, which are of independent interest, allowing us to map the problem onto a statistical mechanics model of self-avoiding walks without requiring large local Hilbert space dimension. Our findings have consequences for the complexity of classically simulating sampling from random shallow circuits, and of contracting tensor networks:  First, we show that standard algorithms based on matrix product states which are used for both these tasks will fail above some constant depth and bond dimension, respectively. In addition, we also prove that these random constant-depth quantum circuits cannot be simulated by any classical circuit of sublogarithmic depth.
	\end{abstract}
	
	\maketitle
	
	\section{Introduction}

	Quantum entanglement is a vital resource that can be utilised to achieve tasks that would not be possible in a classical world, such as unconditionally secure cryptography and high-precision sensing. 
	From a computational point of view, highly entangled many-body quantum states are thought to be non-classical resources, in that they are too complex to describe and simulate on a classical computer.
	Accordingly, the generation of macroscopic entanglement is a natural ingredient in any protocol that purports to achieve a quantum computational advantage, i.e.~outperforming classical computers at some task.

	To prepare such complex entangled states using unitary gates alone, 
	deep circuits are typically required. For instance, if we are limited to logarithmic-depth 1D or constant-depth 2D circuits, then the resulting states are simple enough such that expectation values of products of local observables can be efficiently computed classically \cite{Bravyi2021}. It may then seem surprising that the problem of \textit{sampling} from 2D constant-depth circuits---i.e.~generating measurement outcomes sampled according to the Born probabilities encoded by the output state---is known to be hard in the worst case \cite{Terhal2004}. One can gain appreciation for this stark difference by recognising that measurements themselves can convert short-ranged entanglement into long-ranged entanglement---a phenomenon that underpins concepts such as entanglement swapping \cite{Bennett1993, Zukowski1993, Bose1998, Pan1998}, measurement-based quantum computation \cite{Raussendorf2001,Briegel2009}, and the efficient preparation of topologically ordered states \cite{Raussendorf2005,Piroli2021, Lu2022, Iqbal2023, Zhu2023,  Tantivasadakarn2024}. In the worst-case example mentioned above \cite{Terhal2004}, a macroscopic amount of long-ranged \textit{measurement-induced entanglement} (MIE) is generated during the sampling process, which is necessary for such physics to be classically hard to simulate.

	Based on numerical investigations and non-rigorous analytical arguments, it is frequently conjectured that long-ranged MIE is a \emph{typical} phenomenon that occurs with high probability upon measuring a subset of sites in random 2D circuits above some constant critical depth $d_C > d_{\rm crit} = O(1)$ \cite{Napp2022,Liu2022, Google2023, Bao2024, Watts2024}.
	This hypothesised `teleportation transition' \cite{Bao2024} from short- to long-ranged MIE as a function of depth is also thought to coincide with a sharp change in the classical hardness of sampling from such circuits \cite{Napp2022, Watts2024}. However, due to the nonlinear and stochastic relationship between pre- and post-measurement quantum states, the only rigorous results that prove long-ranged MIE in random circuits so far apply to circuits architectures with qudits whose dimension $q$ grows at least polynomially with system size \cite{Hayden2016, Watts2024}, which, when translated to constant-sized qudits, would require circuits with a depth scaling with system size \cite{NoteQDepth}.

	In this work, we prove that extensive long-ranged MIE is generated in random 2D circuits of constant depth $d_C$ and constant qudit dimension $q$ when a subset of qudits are projectively measured (see \hyperlink{thm:mainresult}{Main Result}). These include 2D brickwork circuits as well as plaquette circuits made up of 4-local gates (see Fig.~\ref{fig:2dArch}). By combining our result with previous work relating MIE to the hardness of computing output distributions \cite{Napp2022, Watts2024}, we provide new evidence that sampling from a random constant-depth 2D circuit yields a quantum advantage: First, the existence of a high-MIE phase implies that standard boundary-contraction strategies based on matrix product states for simulating sampling from a random circuit fail with high probability (\cref{thm:ContractInf}). Similarly, boundary-contraction of certain random tensor networks will fail above some critical bond dimension $\chi_{\rm crit} = O(1)$. Second, we prove an unconditional complexity-theoretic separation between random constant-depth 2D circuits and arbitrary sublogarithmic-depth classical circuits (\cref{thm:AdvantageInformal}), thereby generalizing the worst-case result of Ref.~\cite{Bravyi2018} to typical circuit instances.

	To establish our results, we develop a new technique that circumvents the replica theory method, on which almost all analytical studies of measurement-induced dynamics so far were based \cite{Choi2019,Bao2020, Fan2021, Fava2023, Bao2024}. Our approach combines concepts from many-body and statistical physics with tools from multi-user quantum Shannon theory. In particular, we adapt techniques developed in recent works \cite{Colomer2023,Cheng2023} that analyse certain information-theoretic tasks known as \textit{multi-user entanglement of assistance}---a class of problems that have previously been connected to random tensor networks \cite{Hayden2016}.
	In a key intermediate result, we establish a lower bound on the average entanglement of post-measurement states in terms of the free energy of a statistical mechanics model of self-avoiding walks, which are well-studied \cite{Hammersley1962,Chayes1986}. More generally, our method is likely to be applicable to other information-theoretic settings where one works in the thermodynamic limit---in particular without requiring large local Hilbert space dimensions or coherent access to many copies of the same state, as is common in multi-user information theory.
	
	We begin by giving a broad overview of our results and methods in Section \ref{sec:Overview}. Then, in Section \ref{sec:FDS} we formalise our problem, and review previous non-rigorous theoretical approaches to these problems in Section \ref{sec:Prev}. In Section \ref{sec:EntProj}, we prove our lower bound on post-measurement entanglement (Theorem \ref{thm:EntBound}), our main technical result. In Section \ref{sec:Application}, we apply this bound on MIE to specific families of random shallow circuits, which collectively yield our main result, and in turn Corollaries \ref{thm:ContractInf} and \ref{thm:AdvantageInformal} which pertain to the hardness of simulating sampling from constant-depth circuits and contracting random tensor networks. We then discuss the broader implications of our findings in relation to previous work in Section \ref{sec:Discussion}, before concluding in Section \ref{sec:Conclusion}.
	
	\section{Overview of methods and results\label{sec:Overview}}
	
\begin{figure}
		\centering
		\includegraphics[width=\columnwidth]{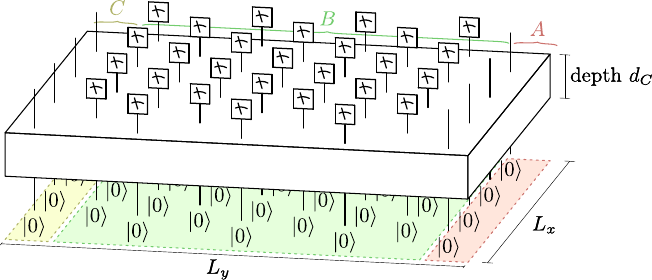}
		\caption{After preparing a state using a geometrically local depth-$d_C$ circuit, we measure degrees of freedom in $B$ and look at the (bipartite) entanglement of the post-measurement state between unmeasured regions $A$ and $C$. This figure shows one particular choice of geometry $ABC$ which we will often refer back to, however for most of this work the choice is arbitrary.}
		\label{fig:ABC}
	\end{figure}

	This paper focuses on states that can be prepared by two-dimensional geometrically local quantum circuits $U$ of constant depth $d_{C}$. Starting from such a state $\ket{\Psi^{ABC}}$, which we divide into regions $A$, $B$, and $C$, we consider the effect of projective measurements on $B$. Conditioned on a particular measurement outcome $s_B$, the unmeasured degrees of freedom collapse into a bipartite post-measurement state $\ket{\phi^{AC}_{s_B}}$ on the unmeasured degrees of freedom $AC$, the entanglement structure of which we wish to study. This setup is illustrated in Fig.~\ref{fig:ABC}. We consider several natural families of random circuits, and rigorously prove that extensive long-ranged measurement-induced entanglement is generated in the regime of constant depth $d_C = O(1)$. This is shown to have consequences for the hardness of simulating sampling from these circuits, and contracting random tensor networks.

	The approach we use to prove our results hinges on two key conceptual ingredients:  1) Interpreting this scenario as a multi-user \textit{entanglement of assistance} problem, and 2) Mapping it onto a statistical mechanics model of self-avoiding walks. Together, these yield our main technical result---Theorem \ref{thm:EntBound}---which is a lower bound on the average entanglement entropy of the post-measurement states in terms of a partition function of these self avoiding walks.

	\begin{figure*}
		\begin{tikzpicture}[x=\scalemain,y=\scalemain,scale=1, every node/.style={transform shape}]
			\tikzset{dot/.style={ circle,fill={#1},inner sep=0pt,outer sep=1pt,minimum size = 6pt}}
			\tikzset{
				>=stealth',
			}

			\newcommand{\measure}{
				\pgfmathsetmacro{\sqsize}{20}
				\pgfmathsetmacro{\meterVshift}{5}
				\pgfmathsetmacro{\meterLength}{8}
				\pgfmathsetmacro{\meterRadius}{12}
				
				\draw[fill=white] (-0.5*\sqsize,0) rectangle (0.5*\sqsize,\sqsize);
				
				\coordinate (base) at  (0,\meterVshift);
				\coordinate (arcl) at ($(base) + (135:\meterLength)$);
				
				\draw (arcl) arc (135:45:\meterLength);

				\draw (base) -- ++(75:\meterRadius);
			}
			
			\coordinate (topLeft) at (0,\height);
			
			\draw[rounded corners, thick] (-\margw,\height) rectangle coordinate[pos=0.5] (eoaMid) (\width - \margw, \height - 97);
			\node[below] at (eoaMid |- topLeft) {\footnotesize \textit{Entanglement of assistance} (Sec.~\ref{subsec:EOA})};

			\node[above right = 0 and 5pt] (prem) at(0,\height - \topShift) {
				\begin{tikzpicture}[x=2.5pt,y=2.5pt,scale=0.8,  every node/.style={transform shape}, anchor=center]

					\fill[acol] (0,0) rectangle (25/4,7);
					\fill[bcol] (25/4,0) rectangle (3*25/4,7);
					\fill[ccol] (3*25/4,0) rectangle (25,7);
					
					\draw (0,0) rectangle (25, 7);
					\node at (12.5,3.5)  {$\Psi^{ABC}$};
					\draw (5, 7) -- ++(0, 5);
					\draw (12.5, 7) -- ++(0, 5);
					\draw (20, 7) -- ++(0, 5);
					
					\node[above] at (5,12) {$A$};
					\node[above] at (12.5,12) {$B$};
					\node[above] at (20,12) {$C$};
					
				\end{tikzpicture}
			};

			\node[above right = 0] (postm) at (125,\height - \topShift) {
				\begin{tikzpicture}[x=2.5pt,y=2.5pt,scale=0.8, every node/.style={transform shape}, anchor=center]
					
					\fill[acol] (0,0) rectangle (25/4,7);
					\fill[bcol] (25/4,0) rectangle (3*25/4,7);
					\fill[ccol] (3*25/4,0) rectangle (25,7);
					
					\draw (0,0) rectangle (25, 7);
					\node at (12.5,3.5)  {$\Psi^{ABC}$};
					\draw (5, 7) -- ++(0, 8);
					\draw (12.5, 7) -- coordinate[pos=1] (B) ++(0, 2);
					\draw (20, 7) -- ++(0, 8);
					
					\draw[double distance = 1] (B) -- ++(0,6);
					
					\begin{scope}[shift={(B)},scale=0.2, every node/.append style={transform shape}]
						\measure
					\end{scope}
					
					
					\node[above, inner sep = 0, rectangle, fill=white, outer sep = 5] at (12.5,14) {$s_B$};
					
				\end{tikzpicture}
			};

			\node[above right = 0] (postd) at (295,\height - \topShift) {
				\begin{tikzpicture}[x=2.5pt,y=2.5pt,scale=0.8,  every node/.style={transform shape}, anchor=center]
					
					\fill[acol] (0,0) rectangle (25/4,7);
					\fill[bcol] (25/4,0) rectangle (3*25/4,7);
					\fill[ccol] (3*25/4,0) rectangle (25,7);
					
					\draw (0,0) rectangle (25, 7);
					\node at (12.5,3.5)  {$\Psi^{ABC}$};
					\draw (5, 7) -- ++(0, 14);
					\draw (12.5, 7) -- coordinate[pos=1] (B) ++(0, 2);
					\draw (20, 7) -- ++(0, 14);
					
					
					\begin{scope}[shift={(B)},scale=0.2, every node/.append style={transform shape}]
						\measure
					\end{scope}

					
					\draw[double distance = 1] (5,15) -- node[above] {\small LOCC} (20,15);
					
					\draw[fill=white] (5,15) circle (2);
					\draw[fill=white] (20,15) circle (2);
					
				\end{tikzpicture}
			};

			\draw[->, thick] (prem.east) -- node[pos=0.5, above] {\footnotesize  measure $B$} (prem.east -| postm.west);
			
			\draw[->, thick] (prem.east -| postm.east) -- node[pos=0.5, above] {\footnotesize  entanglement distillation} node[pos=0.5, below] {\footnotesize (Sec.~\ref{subsec:Distil})}(prem.east -| postd.west);
			
			\node[right = 20] (appr) at (prem.east -| postd.east) {{\Large $\approx$}};
			\node[below = -3] at (appr.south) {(error $\epsilon$)};
			
			\node[above right=-12 and 32] (maxent) at (appr.east) { \begin{tikzpicture}[x=2.5pt,y=2.5pt,scale=0.8,  every node/.style={transform shape}, anchor=center]
					\draw (0,0) -- ++(0,-10) -- node[pos=0.5, fill=black, circle, inner sep = 0, outer sep = 0, minimum size = 4] {} node[pos=0.5,above] {$\ket{\Phi_{d'}}$} ++(15,0) -- ++(0,10);
				\end{tikzpicture}
			};
			
			\node[below = 0] (premlab) at (prem.south) {\footnotesize pre-meas.~state};
			\node[below = -2, fill=white, rounded corners] at (postm.south) {\centering \parbox{100pt}{\footnotesize  post-meas.~state $|\phi^{AC}_{s_B}\rangle$}};
			\node[below = 0] at (postd.south) {\footnotesize post-dist.~state $|\chi_{s_B}^{AC}\rangle$};
			\node[below = 0, text width = 120, align = center] (maxentlab) at (maxent.south) {\footnotesize  max.~entangled \\  state, rank $d'$};
			
			
			
			\node[below right, draw, rounded corners, outer sep = 5, inner sep = 6, text width = 105, align = center] (lem) at ($(premlab.south) + (39,-17)$) 
			{ \footnotesize $E(\phi^{AC}_{s_B}) \geq E(\chi^{AC}_{s_B}) \approx \log d'$ if  $\epsilon$ small (\cref{thm:EntropyQeBound})};
			
			\node[left, draw, rounded corners, outer sep = 5, inner sep = 5, text width = 60, align = center] (eps) at ($(lem.east) + (0,-42)$) 
			{\footnotesize $\epsilon \leq \sqrt{d'}\mathcal{Z}_{\rm SAW}$ (\cref{lem:SAWQeBound}) };
			
			\node[right, draw, rounded corners, outer sep = 5, inner sep = 5, text width = 85, align = center] (thm) at ($(eps.east) + (15,0)$) 
			{\footnotesize $E(\phi^{AC}_{s_B}) \gtrsim -\log \mathcal{Z}_{\rm SAW}$\\   (\cref{thm:EntBound}) };
			
			\node[above right, draw, rounded corners, outer sep = 5, inner sep = 5, text width = 70, align = center] (main) at ($(thm.south east) + (15,0)$) 
			{\footnotesize \textbf{Extensive MIE}\\  $E(\phi^{AC}_{s_B}) \geq \kappa L_x$ (\hyperlink{thm:mainresult}{Main result}) };

			
			\coordinate (origin) at (0,0);
			
			\draw[->, thick] (origin |- lem.west) -- node[pos=0.45,above = -2] {\footnotesize monotonicity of } node[pos=0.45,below=-1] {\footnotesize ent.~under LOCC}(lem.west);
			
			\draw[->, thick] (origin |- eps.west) -- node[pos=0.5,above = -2] {\footnotesize mapping $\epsilon$ onto} node[pos=0.5,below=-1] {\footnotesize  self-avoiding walks (Sec.~\ref{subsec:DWs})}(eps.west);
			
			\draw[->, thick] (lem.east) -| node[pos=0.45,above = -2] {
			} ($(thm.north) + (0,0)$);
			
			\draw[->,thick] (eps.east) -- ($(thm.west) + (0,0)$);
			
			\draw[->,thick] (thm.east) -- (thm.east -| main.west);
			
			\pgfmathsetmacro{\SAWTitleHeight}{26}
			\begin{scope}[shift={(\width - 5,\height - \SAWTitleHeight)}, scale=1]
				\pgfmathsetmacro{\widthSAW}{\SAWratio*\width}%
				\pgfmathsetmacro{\bborder}{5}
				\pgfmathsetmacro{\recth}{48}
				\pgfmathsetmacro{\tmarg}{1}
				\pgfmathsetmacro{\lmarg}{30}
				
				\pgfmathsetmacro{\bmarg}{8}
				\pgfmathsetmacro{\acwid}{13}
				\pgfmathsetmacro{\rectw}{\widthSAW-\lmarg-\rmarg}
				
				\draw[rounded corners, thick] (0.6*\lmarg, \SAWTitleHeight) rectangle coordinate[pos=0.5] (SAWMid) coordinate[pos=0] (SAWTL) (\widthSAW + 0.45*\rmarg, -\height - \bmarg);

				\node[below=-1] at (SAWMid |- SAWTL) {\parbox{60pt}{\centering \footnotesize \textit{Self-avoiding walks}}};
				
				\newcommand{\tstar}[5]{
					\pgfmathsetmacro{\starangle}{360/#3}
					\draw[#5] (#4:#1)
					\foreach \x in {1,...,#3}
					{ -- (#4+\x*\starangle-\starangle/2:#2) -- (#4+\x*\starangle:#1)
					}
					-- cycle;
				}
				
				\newcommand{\ngram}[4]{
					\pgfmathsetmacro{\starangle}{360/#2}
					\pgfmathsetmacro{\innerradius}{#1*sin(90-\starangle)/sin(90+\starangle/2)}
					\tstar{\innerradius}{#1}{#2}{#3}{#4}
				}
				
				\tikzset{saw/.style={line width = 1}}

				\path[use as bounding box] (0,-\height) rectangle (\widthSAW,0);
				
				\fill[acol] (\lmarg,-\recth-\tmarg) rectangle node[pos=0.5] {\textcolor{acol!50!black}{$A$}} ++(\acwid,\recth);
				\fill[bcol] (\lmarg+\acwid,-\recth-\tmarg) rectangle ++(\rectw-2*\acwid,\recth);
				\fill[ccol] (\lmarg+\rectw-\acwid,-\recth-\tmarg) rectangle  node[pos=0.5] {\textcolor{ccol!50!black}{$C$}} ++(\acwid,\recth);
				
				\coordinate (tr1) at (\widthSAW-\rmarg,-\tmarg);
				\coordinate (tr2) at (\widthSAW-\rmarg,-\height+\tmarg+\recth);
				
				\coordinate (bl1) at (\lmarg,-\tmarg-\recth);
				\coordinate (bl2) at (\lmarg,-\height+\tmarg);
				
				\coordinate (br1) at (\widthSAW-\rmarg,-\tmarg-\recth);
				\coordinate (br2) at (\widthSAW-\rmarg,-\height+\tmarg);
				
				\coordinate (tl1) at (\lmarg,-\tmarg);
				\coordinate (tl2) at (\lmarg,-\height+\tmarg+\recth);
				
				\fill[acol] (\lmarg,-\height+\tmarg) rectangle node[pos=0.5] {\textcolor{acol!50!black}{$A$}} ++(\acwid,\recth);
				\fill[bcol] (\lmarg+\acwid,-\height+\tmarg) rectangle ++(\rectw-2*\acwid,\recth);
				\fill[ccol] (\lmarg+\rectw-\acwid,-\height+\tmarg) rectangle  node[pos=0.5] {\textcolor{ccol!50!black}{$C$}} ++(\acwid,\recth);
				\draw[thick] (\lmarg,-\height+\tmarg) rectangle ++(\rectw,\recth);
				
				\draw[bcol!70!bdark,decoration={random steps,segment length=2,amplitude=1}, decorate, rounded corners = 0.3,saw] (\lmarg+0.5*\rectw,-\tmarg) to[out=210, in=80] ++(-20,-\recth);
				
				\draw[bcol!58!bdark,decoration={random steps,segment length=2,amplitude=1}, decorate, rounded corners = 0.3,saw] (\lmarg+0.5*\rectw,-\tmarg) to[out=250, in=80] ++(-15,-\recth);
				
				\draw[bcol!45!bdark,decoration={random steps,segment length=2,amplitude=1}, decorate, rounded corners = 0.3,saw] (\lmarg+0.5*\rectw,-\tmarg) to[out=255, in=80] ++(-7,-\recth);
				
				\draw[bcol!45!bdark,decoration={random steps,segment length=2,amplitude=1}, decorate, rounded corners = 0.5,saw] (\lmarg+0.5*\rectw,-\tmarg) to[out=280, in=120] ++(3,-\recth);

				\draw[bcol!58!bdark,decoration={random steps,segment length=2,amplitude=1}, decorate, rounded corners = 0.3,saw] (\lmarg+0.5*\rectw,-\tmarg) to[out=310, in=80]  ++(10,-0.35*\recth) to[out=210, in=80] ++(-4,-0.65*\recth);
				
				\draw[bcol!70!bdark,decoration={random steps,segment length=2,amplitude=1}, decorate, rounded corners = 0.3,saw] (\lmarg+0.5*\rectw,-\tmarg) to[out=350, in=40] ++(15,-\recth);
				
				\fill[white] (\lmarg,-\tmarg) rectangle (\widthSAW-\rmarg,2-\tmarg);
				\draw[thick] (\lmarg,-\recth-\tmarg) rectangle (\widthSAW-\rmarg,-\tmarg);
				
				\draw[bcol!0!bdark,decoration={random steps,segment length=2,amplitude=1}, decorate, rounded corners = 0.3,saw] (\lmarg+0.5*\rectw,-\height + \tmarg + \recth) to[out=270, in=90] ++(0,-\recth);
				
				\node[right] at ($(tr1) + (0,-0.5*\recth)$) {};
				
				\node[below = 2 , text width = 80, align = center] at ($(bl1)!0.5!(br1)$) {\footnotesize  disordered\\ $\mathcal{Z}_{\rm SAW} \rightarrow \infty$};
				\node [above = 2, text width = 80, align = center] (order) at ($(tl2)!0.5!(tr2)$)  {\footnotesize  ordered\\ $\mathcal{Z}_{\rm SAW} \leq e^{-\kappa L_x}$};

				\draw[-{Stealth[scale=1.6]}] ($(tr1) + (8,-1)$) -- node[above, rotate=-90] {\footnotesize increasing pre-meas.~entanglement} ($(tr2) + (5,-\recth)$);
				\coordinate (tp) at (\widthSAW-\rmarg + 6,-0.5*\height);
				
				\draw[dashed] (\lmarg,-0.5*\height) -- (tp);

				\tstar{2}{4}{7}{0}{fill=red,draw=black,shift=(tp)}
			\end{scope}
			
			\draw[->,thick] (order.west) -| (main.north);

		\end{tikzpicture}
		\caption{ Flow of logic leading to our primary technical finding, \cref{thm:EntBound}, and in turn our main result---that the post-measurement states $\ket{\phi^{AC}_{s_B}}$ possess extensive measurement-induced entanglement (scaling with the linear system size $L_x$). Top row: After measuring on $B$, we consider the effect of some distillation protocol, whose aim is to generate a state that is close to the maximally entangled state $\ket{\Phi_{d'}}$ of rank $d'$. Overall, this procedure constitutes an entanglement of assistance task. Right panel: The self-avoiding walk partition function is known to exhibit an ordering transition driven by the temperature $\beta^{-1}$. Here, $\beta$ is set by the amount of entanglement in the pre-measurement state, which is determined by circuit depth and/or qudit dimension. As a result, we find regimes where we can lower bound MIE in the form  \eqref{eq:LinearMIE}.
		}
		\label{fig:Flow}
	\end{figure*}
	
	The overall logic we use to obtain Theorem \ref{thm:EntBound} is shown in Fig.~\ref{fig:Flow}. Here, we sketch the main arguments, which are described in more detail in Section \ref{sec:EntProj}.
	
	First, we relate the problem of quantifying post-measurement entanglement to a family of information-theoretic tasks known as entanglement of assistance \cite{Cohen1998,DiVincenzo1999}, which is closely related to the concept of localizable entanglement \cite{Popp2005}. While several variations exist, we focus on the task of preparing a maximally entangled state $\ket{\Phi_{d'}^{AC}} = (d')^{-1/2}\sum_{j=1}^{d'}\ket{j}_{A}\otimes \ket{j}_{C}$ of some size $d'$ between two parties $A$ and $C$ by acting with local operations and classical communication (LOCC) on a pre-specified tripartite `resource' state $\rho^{ABC}$. The actions of $B$ help $A$ and $C$ to prepare this entangled state, hence the terminology. 
	
	A natural strategy to achieve this task is to perform measurements on $B$ (step 1), followed by some LOCC operations between $A$ and $C$, conditioned on those outcomes (step 2).
	The operations in step 2 cannot increase the entanglement between $A$ and $C$, so if we succeed, then the state immediately after step 1 must have had entanglement entropy $E(A:C)$ at least as large as that of the target state $\ket{\Phi_{d'}^{AC}}$, namely $\log d'$. More generally, if $A$ and $C$ end up in a state that is $\epsilon$-close to the target state, then we get a correspondingly close lower bound.
	
	We apply this logic to our setup, with $\ket{\Psi^{ABC}}$ as the resource state and the measurements on $B$ being in the computational basis (see the top row of Fig.~\ref{fig:Flow}); this gives us an entanglement lower bound for the post-measurement states $\ket{\phi^{AC}_{s_B}}$ in terms of the chosen dimension $d'$ and protocol error $\epsilon$ (Lemma \ref{thm:EntropyQeBound}). The problem then becomes quantifying $\epsilon$ as a function of $d'$, for this class of entanglement of assistance protocols. In particular, because $B$ is made up of many parties $B = B_1 B_2 \ldots$, each of which act locally, and the operations involve only use a single copy of the resource state at a time, this corresponds to a \textit{one-shot, multi-user} version of entanglement of assistance \cite{dutil2010,Dutil2011}. As we shall see later (Sec.~\ref{subsec:MultipartySplit}), it will be possible to leverage certain tools that were developed to study these kind of problems \cite{Colomer2023,Cheng2023}. 

	The second ingredient in our proof is a mapping of the error $\epsilon$ onto a statistical mechanics model of self-avoiding walks. We start by adopting a particular `multiparty splitting method' introduced in Refs.~\cite{Colomer2023, Cheng2023}, which allows one to decompose the measurement channel on $B$ into $2^{|B|}$ terms, each of which can be associated with a configuration of Ising spins $\sigma_i = \pm 1$ on each cell. (These $\mathbbm{Z}_2$-valued variables are reminiscent of, but distinct from, the Ising spins that arise in replica-based statistical mechanics descriptions of random quantum circuits~\cite{Hayden2016, Nahum2017, Keyserlingk2018}, as we discuss in Section \ref{subsec:StatMechInterp}.) 
	Accordingly, the overall error of the distillation protocol can be expressed in terms of a (operator-valued) sum over these spin configurations.
	
	By invoking the local structure of correlations in the pre-measurement state, along with certain properties of the random gates used to prepare it, we demonstrate how to group these Ising configurations into subsets that feature certain domain wall configurations $W$. Each such subset of terms is shown to be collectively suppressed by a factor that is exponentially small in the entanglement entropy of the resource state $\ket{\Psi^{ABC}}$ across the cut $W$. The result is an expression for the error in terms of a partition function of self-avoiding walks $\mathcal{Z}_{\rm SAW}$ with an effective inverse temperature set by the entanglement structure of $\ket{\Psi^{ABC}}$ (Lemma \ref{lem:SAWQeBound}). Combining Lemmas \ref{thm:EntropyQeBound} and \ref{lem:SAWQeBound}, and optimizing over the choice of $d'$ gives us Theorem \ref{thm:EntBound}.

	Thanks to longstanding rigorous results on these statistical mechanical models \cite{Hammersley1962,Chayes1986}, the partition function  $\mathcal{Z}_{\rm SAW}$ is known to undergo a sharp transition below a critical temperature, where it becomes exponentially suppressed in system size $\mathcal{Z}_{\rm SAW} \leq e^{-\kappa L_x}$ (see right panel of Fig.~\ref{fig:Flow}). The distillation error $\epsilon$ then becomes small even for large values of $d'$, and in turn we infer that the MIE is large. In summary, we arrive at the following.
	
	\begin{mainresult*}[\hypertarget{thm:mainresult}{Informal}]
		\label{thm:informal}
		There exist ensembles of unitaries $\U$ obtained from natural 2D circuit architectures on $q$-dimensional qudits of variable depth $d_C$, for which there are values $(q,d_C) = O(1)$, such that the average measurement-induced entanglement between regions $A$ and $C$  in the setup shown in Fig.~\ref{fig:ABC} satisfies
		\begin{align}
			\mathbbm{E}_{U\sim\U}\,\mathbbm{E}_{s_B} E(A:C)_{\phi^{AC}_{s_B}} \geq \kappa L_x.
			\label{eq:LinearMIE}
		\end{align}
		for some constant $\kappa > 0$ for sufficiently large $L_x$, where $L_x$ is the linear system size. 
	\end{mainresult*}
	\noindent The families of circuits we consider are:
	\begin{enumerate}
		\item Random 2D brickwork circuits [Fig.~\ref{fig:2dArch}(a); \cref{thm:Brickwork}]
		\item Circuits with random 4-local gates [Fig.~\ref{fig:2dArch}(b); \cref{thm:4local}]
		\item Holographic random tensor network states [Section \ref{subsec:Holog}; Corollary \ref{thm:Holog}] 
	\end{enumerate}
	
	The fact that we can obtain nontrivial lower bounds on MIE in the regime of constant Hilbert space dimension and depth improves on what had previously been proved, which relied on taking $ q = \text{poly}(N)$ \cite{Hayden2016,Watts2024}; physically, such a scenario could only be implemented by $\omega(1)$-depth  circuits on qudits of constant dimension \cite{NoteQDepth}. This confirms the conjecture that there is a `finite-time teleportation transition' in random 2D circuits, as argued in Refs.~\cite{Napp2022, Bao2024} based on non-rigorous replica-theoretic calculations. Our techniques can also be applied to different geometries of the regions $ABC$---these simply correspond to different boundary conditions being applied to the self-avoiding walk model.

	The above result has consequences for the hardness of two related computational problems: simulating random constant-depth 2D quantum circuits, and contracting random 2D tensor networks. (An explicit correspondence between these two problems has been established in Refs.~\cite{Hayden2016, Vasseur2019, Gonzalez2024}.) Specifically, the main algorithms used in both cases sweep through the system one row at a time, and at each step one stores a matrix product state (MPS) representation of a 1D wavefunction that is precisely of the same form as the post-measurement states $\phi^{AC}_{s_B}$ (with a slightly different geometry of $A$, $B$, and $C$). If MIE is large, then an efficient MPS representation of the state cannot exist. By formalising this argument, we show that the `sideways evolving block decimation' (SEBD) algorithm for simulating sampling \cite{Napp2022} will fail in regimes above the threshold.
	\begin{corollary}[Informal]
		\label{thm:ContractInf}
		If one uses the SEBD method to simulate sampling from a random circuit for values of $(q, d_C)$ where our \hyperlink{thm:mainresult}{main result} applies, the algorithm will abort with high probability $1 - e^{-\Omega(\sqrt{N})}$.
	\end{corollary}
	\noindent The same behaviour will also be seen when using the related boundary matrix product state (bMPS) algorithm to contract random tensor networks in analogous regimes \cite{Verstraete2004,Jordan2008}. For a particular natural family of random tensor networks (defined in Section \ref{subsec:Holog}), we find that bMPS fails for bond dimension $\chi \geq 7$.

	Finally, we combine our main result with the results of Ref.~\cite{Watts2024} to prove a certain kind of complexity-theoretic quantum advantage exhibited by random constant-depth circuits.
	\begin{corollary}[Informal]
		\label{thm:AdvantageInformal}
		No classical circuit of sub-logarithmic depth can simulate sampling from the output distribution of certain families of random constant-depth quantum circuits.
	\end{corollary}
	The above result implies that the random circuit sampling problems considered here are not contained in the classical computational complexity class $\textsf{NC}^0$ consisting of constant-depth classical circuits, even though these problems themselves lie in the corresponding class $\textsf{QNC}^0$ consisting of constant-depth quantum circuits (or, more precisely, the sampling versions of these decision classes). This demonstrates that the kind of quantum advantage established for worst-case instances in Ref.~\cite{Bravyi2018} is actually exhibited in \textit{typical} instances of constant-depth circuits. While this notion of quantum advantage does not imply the long sought-after result that polynomial quantum computations are stronger than polynomial classical computations, which itself would be a milestone achievement in complexity theory, this unconditional separation between two naturally corresponding classical and quantum computational problems is a strong statement, which complements our algorithm-specific hardness results.\\

	We now turn to the technical part of the paper, where we formalise the approach described above, state our claims in more detail, and prove  the main theorems.
	
	\section{Setup\label{sec:FDS}}
	
	\subsection{Sampling and measurement-induced entanglement}
	
	Starting from a state $\ket{\Psi}$ made up of $N$ qudits of dimension $q$, the problem of \textit{sampling} involves projectively measuring each qudit in some local basis $\{\ket{s_i}\}_{s_i = 1, \ldots q}$. This is a stochastic process that generates a string of dits $s = (s_1, \ldots, s_N)$ distributed according to the Born probabilities $p(s) = |\braket{s|\Psi}|^2$. It will be convenient to represent the outcome distribution as a classical (i.e.~diagonal) density matrix $\rho^M = \sum_s p(s)\proj{s}_M$, which describes the state of a classical measurement register $M = M_1 \cdots M_N$.  The physical measurement process itself can be described by a channel $\mathcal{T}^{Q \rightarrow M}$, which is a completely-positive trace-preserving (CPTP) map, that relates $\rho^M$ to the pre-measurement state $\ket{\Psi}$. Because measurements are local, $\mathcal{T}^{Q \rightarrow M}$ has a tensor product structure,
	\begin{align}
		\rho^{M} = \left(\bigotimes_{i=1}^N \mathcal{T}^{Q_i \rightarrow M_i}\right)[\Psi],
		\label{eq:MeasurementDistChannel}
	\end{align}
	where $\mathcal{T}^{Q_i \rightarrow M_i}[\sigma^{Q_i}] = \sum_{s_i} \proj{s_i}  \braket{s_i|\sigma^{Q_i}|s_i}$ defines the local measurement channel on each site $i$. Where obvious from context, we use $\Psi \equiv \proj{\Psi}$ for projectors onto pure states throughout.
	
	Insight into this problem can be gained by partitioning the state $\ket{\Psi^{ABC}}$ into regions $A$, $B$, $C$, and considering the state of the system after sites in $B$ have been measured. Conditioned on the measurement outcome $s_B$, which occurs with probability $p_{s_B} = \braket{s_B|\rho^B|s_B}$, where $\rho^B = \Tr_{AC}[\Psi]$ is the reduced density matrix on $B$, the remaining unmeasured degrees of freedom $AC$ end up in a post-measurement quantum state
	\begin{align}
		\ket{\phi_{s_B}^{AC}} = \frac{1}{\sqrt{p_{s_B}}} \big(\bra{{s_B}} \otimes I_{AC}\big)\ket{\Psi}.
		\label{eq:PostMeasurementState}
	\end{align}
	Together, the post-measurement states along with their corresponding probabilities constitute a statistical ensemble of quantum states $\mathcal{E}_{AC} = \{(p_{s_B},\ket{\phi_{s_B}^{AC}} ) \}$, which has recently been referred to as the \textit{projected ensemble} \cite{Ho2022,Cotler2021,Claeys2022}.
	
	The \textit{measurement-induced entanglement} (MIE) is the average entanglement between $A$ and $C$ in the projected ensemble, 
	\begin{align}
		\overline{E(A:C)} \coloneq \sum_{s_B}p_{s_B} E(A:C)_{\phi^{AC}_{s_B}} \equiv \sum_{s_B}p_{s_B} S(\rho^A_{s_B})
		\label{eq:AvgMIE}
	\end{align}
	where $\rho^A_{s_B} = \Tr_C \phi^{AC}_{s_B}$ is the reduced density matrix of the post-measurement state on $A$, and $S(\sigma) = -\Tr[\sigma \log \sigma]$ is the von Neumann entropy.
	It is sometimes convenient to write the above as a conditional entropy of the quantum-classical state $\rho^{AM_BC}$
	consisting of a classical measurement register $M_B$ with $|B|$ dits and a quantum register $AC$ which holds the state of the unmeasured qudits
	\begin{align}
		\rho^{AM_BC} &= \sum_{s_B} p_{s_B} \proj{s_B}_{M_B} \otimes \proj{\phi_{s_B}^{AC}}\\
		&=
		\left(\bigotimes_{i \in B} \mathcal{T}^{Q_i \rightarrow M_i} \otimes \bigotimes_{i\in AC} \text{id}_{Q_i} \right)[\Psi].
		\label{eq:CQState}
	\end{align}
	Here, and throughout, we use $\text{id}_X$ to denote the identity channel $\text{id}_X[\sigma^X] = \sigma^X$ on some set of degrees of freedom $X$.
	Specifically, we can write
	\begin{align}
		\overline{E(A:C)} = S(A|M_B)_{\rho^{AM_BC}} \coloneqq S(AM_B) - S(M_B),
		\label{eq:CondEntr}
	\end{align}
	where $S(M_B)$ is a shorthand for $S(\rho^{M_{B}})$, and similar for $S(AM_B)$.

	\subsection{Constant-depth 2D Circuits \label{subsec:TNS}}
	
\begin{figure}
		\includegraphics[width=246pt]{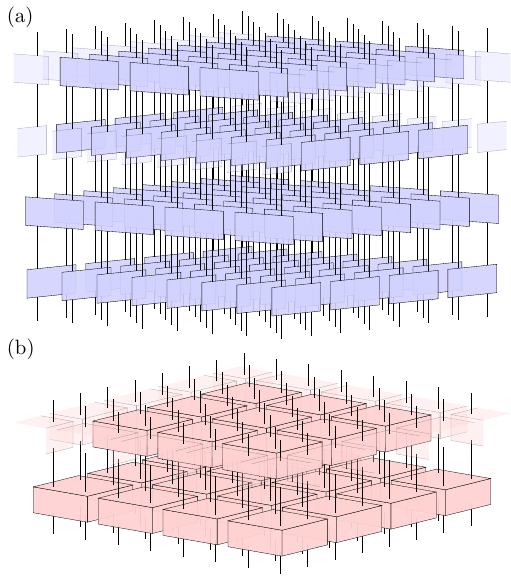}
		\caption{Circuit architectures that we explicitly consider in this work. Here, the order of gates runs from bottom to top. (a) A 2D brickwork circuit, where  2-local gates (blue rectangles) are applied between pairs of qudits in an arrangement that depends on $t \mod 4$, where $t = 1, \ldots, d_C$ is a discrete time index labeling each layer. (b) 4-local plaquette circuit. By combining groups of 4 two-qudit gates in the 2D brickwork architecture, depth-$2d_C$ brickwork circuits can be thought of as special cases of depth-$d_C$ 4-local circuits.} 
		\label{fig:2dArch}
	\end{figure}

	We are specifically interested in sampling from states $\ket{\Psi_\Lambda}$ that can be prepared by constant-depth geometrically local circuits $U$ on a two-dimensional (2D) lattice $\Lambda$, starting from a product state, i.e., $\ket{\Psi_\Lambda} = U\ket{0}^{\otimes N}$, where $U$ is a 2D local circuit whose depth $d_C$ (i.e.~the number of layers) is an $O(1)$ constant, independent of the number $|\Lambda|=N$ of sites in the lattice.
	For concreteness, we often refer to square lattices with either a 2D brickwork layout of 2-local gates, as depicted in Fig.~\ref{fig:2dArch}(a), or an arrangement of 4-local gates around square plaquettes shown in Fig.~\ref{fig:2dArch}(b). In the following we use $d_C$ to denote the depth with respect to the latter arrangement, which can be straightforwardly seen to include brickwork circuits of depth $2d_C$.
	
\begin{figure}
		\centering
		\includegraphics[width=\columnwidth]{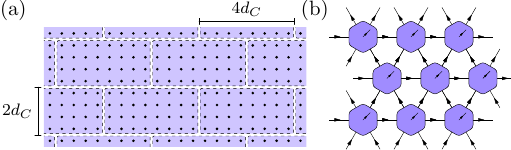}
		\caption{(a) Blocking sites into rectangular cells of dimension $4d_C \times 2d_C$. (b) Treating all the qudits within each cell as a single degree of freedom, any state prepared by $d_C$ layers of gates in the architecture shown in Fig.~\ref{fig:2dArch}(b) can be written as a strictly local isoTNS on the triangular lattice.}
		\label{fig:Tessellation}
	\end{figure}

	States generated by constant-depth circuits have strictly short-ranged correlations (strict light cone) and entanglement, which will play a role in our analysis later on. Blocking qudits into rectangular cells of size $2d_C \times 4d_C$, which tessellate the plane as shown in Fig.~\ref{fig:Tessellation}(a), gives us a new lattice (triangular) lattice $\Lambda'$, with qudits of dimension $q' = q^{8d_C^2}$ and with strictly nearest-neighbour correlations and entanglement.
	Except when referring to specific circuit architectures (Section \ref{sec:Application}), in the following we use $q$, $\Lambda$ to denote the blocked Hilbert space dimension and lattice.
	
	We represent the state $\ket\Psi$ on the blocked lattice as a projected-entangled pair state (PEPS)~\cite{Verstraete2004}, a form of tensor network states (TNS).
	Generally, a TNS is specified by a collection of tensors $T^{s_i}_{\alpha_1, \ldots, \alpha_{z_i}}$ at each site $i$, which have $z_i$ `virtual' indices $\alpha_a$, where $z_i$ is the number of neighbours of site $i$, and one physical index $s_i \in [q]$. Given a local basis $\{\ket{s_i}\}_{s_i = 1, \ldots q}$, the components of the state $\braket{s\,|\Psi_\Lambda}$ are computed by fixing the physical index of each tensor to $s_i$, and contracting all the remaining indices according to the links of the lattice. We define the bond dimension $\chi$ of the TNS as the largest of the dimensions of all the virtual indices. 
	
	A state produced by a local quantum circuit of depth $d_C$ can always be written as a TNS with $\chi\leq\exp(\log(q)d_C)$ by ``flattening'' the circuit.
	In Ref.~\cite{Soejima2020} it was shown that this can be done such that the tensors of the resulting state obey the an isometry condition, making it an instance of \emph{isometric TNS} as introduced in Ref.~\cite{Zaletel2020}.
	
	\begin{definition}[IsoTNS~\cite{Zaletel2020}]
		An isoTNS is a TNS on a simple directed graph (i.e., the flow has no loops), for which each tensor is an isometry from its input legs (virtual bonds point toward the tensor) to its output legs (bonds pointing away, including by definition the physical leg).
	\end{definition}
	When depicting isometries and isoTNSs diagrammatically, arrows will be used to indicate which legs are inputs vs.~outputs [see e.g.~Fig.~\ref{fig:Tessellation}(b)].
	
	While the construction of Ref.~\cite{Soejima2020} demonstrates that all finite-depth states have isoTNS representations, the converse is not true, as the family includes, for example, the Greenberger-Horne-Zeilinger (GHZ) state and string-net states~\cite{Soejima2020}.
	We therefore need an extra property which reflects the locality of correlations in these states. 
	\begin{definition}[Strictly local correlations] \label{def:SimpleTNS}
		A state $\rho^\Lambda$ on a graph $\Lambda$ is strictly local if for any tripartition $\Lambda = A \cup B \cup C$ for which $B$ screens $A$ from $C$ (i.e.~there is no edge connecting $A$ and $C$), we have
		\begin{align}
			\Tr_{B}(\rho^{ABC}) = \rho^A \otimes \rho^C
			\label{eq:LocalCorr}
		\end{align}
		A tensor network state is a strictly local isoTNS if it is an isoTNS and also has strictly local correlations.
	\end{definition}

	In Appendix \ref{app:isoTNS}, we explicitly establish that one can find a TNS representation satisfying both of the above conditions:
	\begin{proposition} \label{thm:SimpleFiniteDepth}
		By blocking sites into cells of size at most $8d_C^2$, any state preparable by a depth-$d_C$ circuit of 4-local gates in the geometry shown in Fig.~\ref{fig:2dArch}(b) can be represented as a strictly local isoTNS on the triangular lattice, with bond dimension $\chi \leq \exp(O(d_C^2))$, independent of system size $N$.
	\end{proposition}
	The reason for taking the blocked lattice to be triangular, rather than square, is that if we choose to tile the plane using the standard square tessellation, there will generically be correlations between degrees of freedom within cells that share a corner, even though these cells do not share an edge in the square lattice. In general, we expect that for any geometry, we will be able to find a strictly local isoTNS on some graph that forms a triangulation of the plane. Note also that the above result includes 2D brickwork circuits of depth $2d_C$, as can be seen by combining gates around square plaquettes (Fig.~\ref{fig:2dArch}).
	
	When analysing the physics of sampling from finite-depth states in Section \ref{sec:EntProj}, we will work with these strictly local isoTNS representations as much as possible, in order to abstract from the particular structure of the circuit used to prepare the state.

	\subsection{Random holographic tensor networks and holographic circuits\label{subsec:Holog}}
	
\begin{figure}
		\includegraphics[width=\columnwidth]{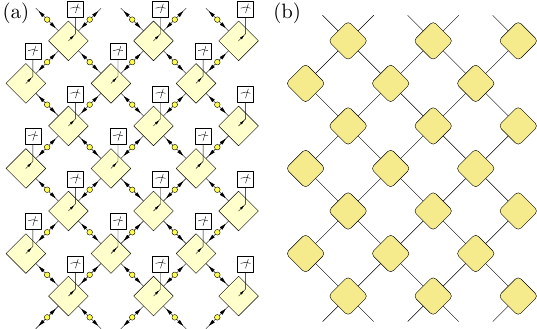}
		\caption{(a) In the holographic circuit architecture, a bipartite entangled state $\ket{\omega_e}$, denoted by dark yellow circles, is prepared on each bond, which is usually taken to be maximally entangled. Then, at each site, a random unitary is applied (light yellow squares). When projective measurements are made in the bulk, one obtains a random holographic state (b), where the physical legs of each bulk tensor have been projected out. In this geometry, the physical legs of the holographic tensor network reside on the top and bottom boundaries. The distribution of holographic tensor network states generated in this way is equivalent to Eq.~\eqref{eq:HologDist}.}
		\label{fig:Holog}
	\end{figure}

	As a final remark before going into the main technical part of this paper, we highlight that random holographic tensor network states (see Refs.~\cite{Hayden2016, Vasseur2019}) fall within the scope of our analysis. In that setting, a quantum state in $D$ spatial dimensions (the boundary) is constructed from a tensor network in $(D+1)$ dimensions (the bulk, often taken to be in curved space). An example is shown in Fig.~\ref{fig:Holog}(a), in the square lattice. The bulk tensors have no physical leg, and are fully contracted with each other, while at the boundary there are some uncontracted legs, which represent the components of a boundary state.
	
	A natural way to generate random holographic tensor network states using a constant-depth circuit was presented in Ref.~\cite{Hayden2016}. 
	We start by generating a bipartite entangled state $\ket{\omega_e}$ on each edge $e \in E$ of a graph $\Lambda = (V, E)$, usually taken to be the maximally entangled state of dimension $\chi$. The corresponding TNS is represented by tensors that are just the identity map from the virtual to the physical space, $T_{[i],\alpha_1\dots\alpha_{z_i}}^{s_i}=\delta^{s_i}_{\alpha_1\dots\alpha_z}$ (where we think of both $s_i$ and $\alpha_1\dots\alpha_{z_i}$ as $z_i$-digit numbers in base $\chi$, where $z_i$ is the degree of vertex $i \in V$).
	We then apply a Haar-random unitary $U_i$ of dimension $\chi^{z_i}$ to each site.
	This overall procedure can be implemented by a constant-depth circuit, and we will refer to this particular circuit architecture as the `holographic circuit' model. The resulting state $\ket{\Psi_\Lambda}$ is manifestly a strictly local isoTNS (even for non-maximally entangled bond states; see Appendix \ref{app:isoTNS}).
	
	We then perform projective measurements on all sites in the bulk [Fig.~\ref{fig:Holog}(b)], whose outcomes $s_B = (s_1, \ldots, s_N)$ are distributed according to Born probabilities. The conditional state on the boundary is given by a projection of $\ket{\Psi_\Lambda}$ onto $\bra{s_B}$. This is evidently a holographic tensor network state, and the components of each bulk tensor $T_{[i],\alpha_1 \ldots \alpha_{z_i}}$ are given by the components of $\bra{s_i}U_i$. Up to normalization, for fixed $s_B$ the distribution of $T = \{T_{[i]}\}$ over the Haar-random unitaries $U_i$ is identical to that of random complex $\chi^{z_i}$-dimensional vectors, where each component of every tensor is independently Gaussian-distributed. Combining this with the Born probabilities, which are given by the norm of the corresponding boundary state $\ket{\Psi[T]}$, we get a distribution of holographic tensor network states for which the probability density of bulk tensors is simply 
	\begin{align}
		p(T) = p_{\rm Gauss}(T) \times \| \ket{\Psi[T]}\|^2 
		\label{eq:HologDist}
	\end{align}
	where $p_{\rm Gauss}(T)$ is the probability density for independent Gaussian-random tensors $T$. Our methods will allow us to evaluate entanglement properties of holographic tensor network states averaged over the distribution \eqref{eq:HologDist}. Note that in the limit of large bond dimension $\chi$, we can neglect the Born probabilities and approximate $p(T) \approx p_{\rm Gauss}(T)$ \cite{Cheng2023a}.

	\section{Review of previous approaches\label{sec:Prev}}
	
	To elucidate the challenges of computing MIE \eqref{eq:AvgMIE} in many-body settings, and to help contextualize our new results, we review some previous methods that have been used to analyse similar problems. Readers familiar with such prior works can skip directly to Section \ref{sec:EntProj}.
	
	\subsection{Quantifying post-measurement entanglement \label{subsec:PMEnt}}
	
	The problem of computing average MIE for any specific circuit will generically be an intractable problem if it is extensive. Nevertheless, we might hope that for an ensemble of random circuits $\U$, a calculation of some form of circuit-averaged behaviour may be possible, the results of which would then tell us about the behaviour of typical instances of such circuits. Indeed, this is the case for Haar-random unitary circuits without measurements \cite{Nahum2017,Keyserlingk2018}. Specifically, one can calculate averages of the form $\mathbbm{E}_{U \sim \U} \Tr[(\rho^A)^\alpha]$ for some subregion $A$ and integer $\alpha = 2, 3, \ldots$, by virtue of the fact that the quantities $\Tr[(\rho^A)^\alpha]$ are polynomial functions of the random state $\Psi$, and that integer moments of the Haar ensemble can be computed analytically. This family of quantities are related to $\alpha$-R{\'e}nyi entropies $S^{(\alpha)}(\rho) = (1-\alpha)^{-1}\log \Tr[\rho^\alpha]$, which generalize the von Neumann entropy $S(\rho)$ in the sense that $\lim_{\alpha \rightarrow 1}S^{(\alpha)}(\rho) = S(\rho)$. Since these calculations typically proceed by working with $\alpha$ copies of the state in question, such techniques go under the collective name of the \textit{replica method}.

	Unfortunately, computing entanglement entropies of post-measurement states is qualitatively harder, even for random circuits. This is because each individual post-measurement state $\phi^{AC}_{s_B}$ is a nonlinear function of $\Psi^{ABC}$, due to the need to re-impose normalization after projecting onto $\proj{s_i}$ [Eq.~\eqref{eq:PostMeasurementState}]. 
	As a workaround, in previous studies of measurement-induced entanglement in random circuits, a proxy quantity, sometimes referred to as quasientropy, has been introduced. We give its definition here.
	Let $\Psi^{ABC}_U$ denote the pre-measurement state constructed for a unitary $U$, with reduced density matrix $\rho^{AB}_U$.
	Define the subnormalized post-measurement states on $A$ (distinguished with a tilde) as
	\begin{align}
		\tilde{\rho}^A_{s_B, U} \coloneqq& (I_A \otimes \bra{s_B} )\rho^{AB}_U(I_A \otimes \ket{s_B})
		\label{eq:rhoASubnorm}
	\end{align}
	such that the outcome probabilities can be written $p_{s_B} = \Tr[\tilde{\rho}^A_{s_B}]$. Whereas the $\alpha$-R{\'e}nyi entropy of the post-measurement states averaged over outcomes $s_B$ and unitaries $U$ distributed according to $\U$ can be written as
	\begin{align}
		&\mathbbm{E}_{s_B, U \sim \U} S^{(\alpha)}(\rho^A_{s_B, U}) \nonumber\\ =& \mathbbm{E}_{U \sim \U} \sum_{s_B} 
		\frac{1}{1-\alpha} \Tr[\tilde{\rho}^A_{s_B, U}] \log\left(\frac{\Tr[(\tilde{\rho}^A_{s_B, U})^\alpha]}{\Tr[\tilde{\rho}^A_{s_B, U}]^\alpha}\right),
		\label{eq:AvgRenyi}
	\end{align}
	in contrast, the quasi-entropy with corresponding index $\alpha$ is defined as
	\begin{align}
		Q^{(\alpha)} = \frac{1}{1-\alpha}\log \left(\frac{\mathbbm{E}_{U \sim \U}\sum_{s_B}
			\Tr[(\tilde{\rho}^A_{s_B, U})^\alpha]}{\mathbbm{E}_{U \sim \U}\sum_{s_B} 
			\Tr[\tilde{\rho}^A_{s_B, U}]^\alpha}\right).
		\label{eq:QuasiEntropyAlt}
	\end{align}
	We can see the averages over outcomes $s_B$ and unitaries $U$ are evaluated separately in the numerator and denominator, and additionally the logarithm is taken after averaging. 
	
	Although not entirely physical, the advantage of quasientropy is that when we pick $\alpha$ to be a fixed integer greater than or equal to 2, we can in principle evaluate the averages in Eq.~\eqref{eq:QuasiEntropyAlt} using Haar calculus, as we will see in the next section. Then, mathematically, the desired average entanglement entropy \eqref{eq:CondEntr} can be related to quasientropy via the so-called \textit{replica limit}
	\begin{align}
		\mathbbm{E}_{U \sim \U} S(A|M)_{\rho^{AM}} = \lim_{\alpha \rightarrow  1} Q^{(\alpha)}
		\label{eq:QuasiEntropyLimit}
	\end{align}
	Unfortunately, even if we are able to evaluate $Q^{(\alpha)}$ for some fixed values of integer $\alpha$, the limit cannot be evaluated unless we have a closed-form expression valid for all $\alpha = 2, 3, \ldots$, in which case an analytic continuation may be possible, akin to the `replica limit' in the study of disordered systems \cite{Dotsenko2000}. This procedure has many subtleties and is generally only possible in the simplest of scenarios, e.g.~circuits without locality \cite{Fava2023,Giachetti2023,Bulchandani2024, Deluca2024}. We also note that unlike the case of von Neumann entropy [Eq.~\eqref{eq:CondEntr}], the difference of R{\'e}nyi entropies $S^{(\alpha)}(\rho^{AM_B}) - S^{(\alpha)}(\rho^M)$ is not equal to the ensemble-averaged R{\'e}nyi entropy for $\alpha \neq 1$, and so this cannot be used as an alternative.
	
	Thus, in general there is no concrete means to obtain $\mathbbm{E}_{U \sim \U} S(A|M)_{\rho^{AM}}$ or any correctly-averaged entropic quantity from the value of quasientropy $Q^{(\alpha)}$ for a fixed value of $\alpha \neq 1$. Nevertheless, in light of Eq.~\eqref{eq:QuasiEntropyLimit}, many works have been dedicated to evaluating $Q^{(\alpha)}$ for $\alpha = 2$, in the hope that a qualitative picture of the underlying physics may be obtained. We review some of these approaches in the next section.
	
	\subsection{Replica-theoretic calculation of quasientropy \label{subsec:PrevQuasiEnt}}
	
	Here we show how quasientropy can be directly computed using replica techniques. For pedagogical reasons, here we specifically discuss random holographic tensor network states, focusing  on the case $\alpha = 2$ for simplicity, which was first analysed in Ref.~\cite{Hayden2016}. We will find it useful to refer back to this calculation to help our intuition when calculating the correctly-averaged entanglement entropy \eqref{eq:CondEntr} in generic shallow circuits later on.
	
	In holographic circuits, a random local unitary $U_i$ is applied to each site $i$ in the bulk immediately before the qudits are measured. In this section and throughout this paper, we will find it convenient to absorb the final layer of random unitaries into the measurement channel, which we can do by defining a generalized measurement process as follows. Any generalized measurement can be specified by a positive operator-valued measure (POVM) $F = \{F_{r}\}$ \footnote{Here we explicitly deal with discrete POVMs, which describe the case where the random unitaries are drawn from a discrete ensemble, though all our arguments are readily generalized to include continuous POVMs.}---a collection of positive operators defined such that the probability of obtaining outcome $r$ given a pre-measurement state $\rho^{Q}$ is given by $\text{Pr}(r|\rho^{Q}) = \Tr[\rho^{Q}F_{r}]$. Let $\{(q_{U_i}, U_i)\}$ be the ensemble of random unitaries, where $U_i$ is applied on site $i$ with probability $q_{U_i}$. The POVM describing this random unitary followed by a projective measurement in the basis $\{\proj{s_i}\}$ is
	\begin{align}
		F_{r_i} &= \proj{\tilde{\chi}_{r_i}} & r_i = (s_i, U_i)\\
		\text{where }\ket{\tilde{\chi}_{r_i}} &= \sqrt{q_i}U_i^\dagger \ket{s_i}.
		\label{eq:POVMRandomUnitary}
	\end{align}
	Here, $r_i$ is a multi-index labelling both the unitary applied and the measurement outcome. We use tildes to emphasise that the states $\ket{\tilde{\chi}_{r_i}}$ are sub-normalized. The measurement channel $\mathcal{T}^{Q_i \rightarrow M_i}$ corresponding to a given POVM is
	\begin{align}
		\mathcal{T}^{Q_i \rightarrow M_i}[\rho^{Q_i}] = \sum_{r_i} \Tr[F_{r_i} \rho^{Q_i}] \proj{r_i}_{M_i}.
		\label{eq:MeasurementPOVM}
	\end{align}
	
	For holographic tensor network states, the ensemble of pre-measurement unitaries is chosen to be a unitary 2-design. This is naturally reflected in the second moments of the POVM $F$, namely the operator
	\begin{align}
		\mathcal{E}^{(2)}_F \coloneqq \sum_{r_i} \Tr[F_{r_i}] \left(\frac{F_{r_i}}{\Tr[F_{r_i}]}\right)^{\otimes 2},
	\end{align}
	which lives on a twofold replicated Hilbert space (see e.g.~Ref.~\cite{Lancien2013}). For a 2-design POVM such as the one considered here, one has
	\begin{align}
		\mathcal{E}^{(2)}_{F:\text{2-design}} = d_i \int \dif\mu_H(\phi) \Big(\proj{\phi}\Big)^{\otimes 2} = \frac{\mathcal{I}_i + \mathcal{F}_i}{d_i + 1},
		\label{eq:POVM2Design}
	\end{align}
	where $\dif \mu_H(\phi)$ is the uniform (Haar) measure over all normalized states of dimension $d_i$; $\mathcal{I}_i = I^{\otimes 2}_{Q_i}$, and $\mathcal{F}_i$ is the swap operator on $Q_i \otimes Q_i$.

	The quasientropy can be now computed. 
	By construction, quasientropy satisfies the convenient property that the summands in side the logarithm in \eqref{eq:QuasiEntropyAlt} are polynomial functions of the unnormalized post-measurement states \eqref{eq:rhoASubnorm}, which are themselves linear in the pre-measurement state. Accordingly, using a `swap trick', one can write the numerator as a simple trace
	\begin{align}
		\sum_{s_B} q_{s_B} \Tr[(\tilde{\rho}^A_{s_B})^2] = \Tr[(\Psi^{ABC})^{\otimes 2} \cdot (\mathcal{F}_A \otimes \mathcal{E}^{(2)}_B \otimes \mathcal{I}_C)],
		\label{eq:QuasiNumerator}
	\end{align}
	where
	\begin{align}
		\mathcal{E}^{(2)}_{B} = \bigotimes_{i \in B}\mathcal{E}^{(2)}_{F_i} = \bigotimes_{i \in B}\frac{\mathcal{I}_i + \mathcal{F}_i}{d_i + 1}.
	\end{align}
	Expanding out the tensor product on the right hand side, one sees that there are $2^{|B|}$ terms, which  can be associated with configurations of spin variables $\sigma_i = \pm 1$. The choice $\sigma_i = -1$ corresponds to the term where the swap $\mathcal{F}_i$ is chosen, and $\sigma_i = +1$ for the identity operator $\mathcal{I}_i$. Looking at the right hand side of Eq.~\eqref{eq:QuasiNumerator}, one can also define fixed spins $\sigma_{A} = -1$, $\sigma_C = +1$ which indicate the relevant operators on regions $A$ and $C$. This gives a convenient representation in terms of a statistical mechanics partition function
	\begin{align}
		\sum_{s_B} q_{s_B} \Tr[(\tilde{\rho}^A_{s_B})^2] =& \left(\prod_{i\in B}\frac{1}{d_i+1}\right)\mathcal{Z}_{-+} \nonumber\\
		\text{where }\mathcal{Z}_{\tau_A, \tau_C} \coloneqq& \sum_{\{\sigma_i\}} \delta_{\sigma_A, \tau_A}\delta_{\sigma_C, \tau_C} e^{-\mathcal{H}[\{\sigma_i\}]}
		\label{eq:QuasiNumeratorPartition}
	\end{align}
	with the `free energy' defined as
	\begin{align}
		\mathcal{H}[\{\sigma_i\}] \coloneqq S^{(2)}\left[\rho^{I[{\sigma_i}]} \right]
		\label{eq:FreeEnergy}
	\end{align}
	where $I[\sigma_i] = \bigcup_{i : \sigma_i = -1} \{i\}$. This evidently depends on the entanglement structure of the pre-measurement state $\Psi^{ABC}$. In the canonical holographic model, this consists entirely of entangled pairs $\ket{\omega_e}$ at each bond, and hence the free energy simplifies to that of a ferromagnetic Ising model
	\begin{align}
		\mathcal{H}[\{\sigma_i\}] = \sum_{\braket{ij}} S^{(2)}[\omega_{\braket{ij}}] \frac{1 - \sigma_i \sigma_j}{2}
		\label{eq:IsingHamiltonian}
	\end{align}
	where $S^{(2)}[\omega_{\braket{ij}}]$ is the R{\'e}nyi entanglement entropy across the bond $\braket{ij}$. Assuming all bonds host the same entangled pair state, this entropy sets the effective temperature of the statistical mechanics model $T^{-1} = S^{(2)}(\omega)$.
	
	Applying the same logic with the denominator gives an exact expression for the quasientropy
	\begin{align}
		Q^{(2)} = \log\left(\frac{\mathcal{Z}_{++}}{\mathcal{Z}_{-+}}\right)
		\label{eq:QuasiRatioIsing}
	\end{align}
	The argument of the logarithm is a ratio of partition functions with different boundary conditions on $A$ and $C$. If the bond entanglement is large enough such that the corresponding Ising model is in its ferromagnetic phase, then one expects $\mathcal{Z}_{-+}$ to be suppressed by a factor exponential in $L_x$, since all configurations host at least one domain wall separating $A$ from $C$. Using the exact solution of the 2D Ising model, this ratio can be shown to behave asymptotically as (provided $L_y \sim \text{poly}(L_x)$)
	\begin{align}
		Q^{(2)} \underset{L_{x,y} \rightarrow \infty}{\longrightarrow} \zeta L_x - \log L_y
	\end{align}
	where $\zeta$ is the Ising model surface tension, which is a known function of the temperature \cite{Abraham1973}. (The $-\log L_y$ is due to the choices of location for the domain wall \cite{Li2021a}.)
	
	In studies of holography, the primary interest is in the limit of infinite bond state dimension $\chi \rightarrow \infty$. In fact, if one takes $\chi \rightarrow \infty $ \textit{before} the thermodynamic limit $L_{x,y} \rightarrow \infty$ is taken, then the discrepancy between the quasientropy and the properly averaged R{\'e}nyi entropy \eqref{eq:AvgRenyi} becomes provably negligible \cite{Hayden2016}, which means that Eq.~\eqref{eq:QuasiRatioIsing} also serves as an expression for the correctly averaged entropy in that case. In this limit, both partition functions are dominated by their lowest-energy configurations, and so the average entanglement between $A$ and $C$ is determined by the minimal cut between them (a manifestation of the Ryu-Takanayagi formula \cite{Ryu2006}). In contrast, our interest will be in the regime where $\chi$ and $q$ are fixed and finite, i.e.~we do not let $q$ scale with $L_{x,y}$. We cannot rely on calculations of quasientropy such as this to prove results about the averaged entropy. For this, we will need a different approach, as described in the next section.
	
	\section{Rigorous bounds on measurement-induced entanglement  \label{sec:EntProj}}
	
	In this section, we explain our new method for rigorously characterizing measurement-induced entanglement in random circuits, which circumvents the replica technique. The main result of this section is Theorem \ref{thm:EntBound}: a lower bound on the average post-measurement entanglement entropy \eqref{eq:AvgMIE} in constant-depth quantum circuits. We again refer readers to Figure \ref{fig:Flow}, which maps out the overall flow of logic in proving this result.\\
	
	\noindent Our first key insight is to identify a connection between the problem of quantifying MIE with the success of a particular information-theoretic task known as one-shot multi-user entanglement of assistance (EOA).
	
	\subsection{One-shot multi-user entanglement of assistance \label{subsec:EOA}}
	In quantum Shannon theory, one-shot entanglement of assistance (or sometimes entanglement of collaboration \cite{Gour2006} or assisted entanglement distillation \cite{Dutil2011a}) relates to the following task: Using one copy (hence one-shot) of a tripartite state $\rho^{ABC}$ as a resource, one attempts to generate a maximally entangled state $\ket{\Phi_{d'}} = \frac{1}{\sqrt{d'}}\sum_{j=1}^{d'}\ket{j} \otimes \ket{j}$ of some specified dimension $d'$ between $A$ and $C$ using local operations and classical communication (LOCC) between all three parties $A$, $B$, $C$ \cite{DiVincenzo1999,Cohen1998,Lausten2003, Smolin2005, Gour2006a}. In the multi-user version of this problem, $B$ is itself made up of $n_B$ subregions $B = B_1\cdots B_{n_B}$, and the allowed operations are LOCC between all $(n_B+2)$ parties.
	
	To relate this to MIE in our setting, we consider the local measurements on each qudit in $B$ as the first step of a one-shot multi-user EOA protocol that uses $\ket{\Psi^{ABC}}$ as the resource state.
	The measurement on $B$ produces an ensemble of pure states $\{p_{s_B}, \ket{\phi_{s_B}^{AC}}\}$.
	Dependent on the particular outcome $s_B$, we then apply an LOCC channel $\mathcal{T}^{AC}_{s_B}$ to the state on $AC$ in an attempt to distill $\ket{\Phi_{d'}}$ for some $d'$.
	Because entanglement is non-increasing under LOCC, if we can find such a channel $\mathcal{T}^{AC}_{s_B}$ that achieves this, we can conclude that $\phi_{s_B}^{AC}$ contains at least as much entanglement as $\ket{\Phi_{d'}}$, namely $\log_2 d'$ bits' worth. 
	
	\subsection{Universal distillation based on random measurements \label{subsec:Distil}}
	We now specify a particular protocol for distilling maximally entangled pairs from the post-measurement states, which we adapted from prior work on few-party information theory \cite{Horodecki2006}. This protocol requires very little information about the states $\ket{\phi^{AC}_{s_B}}$ to analyse, and is in this sense universal.
	
	Starting from the post-measurement state $\ket{\phi^{AC}_{s_B}}$, we perform random generalized measurements of rank $d'$ to the state on $A$, whose outcome we denote $r_A$. This is most easily understood in the special case where $d' = q^m$ for some integer $m$. Then, the procedure involves drawing a random unitary $V^A$ from a 2-design, applying it to $A$, and then projectively measuring $(|A| - m)$ qudits. The remaining $m$ unmeasured qudits constitute a new, smaller quantum register $A'$ of dimension $d'$. The final step is a local operation on $C$ such that the final state $\ket{\chi^{A'C'}_{r_A, s_B}}$ is as close as possible to $\ket{\Phi^{A'C'}_{d'}} = \frac{1}{\sqrt{d'}}\sum_{j=1}^{d'}\ket{j}_{A'}\otimes \ket{j}_{C'}$, where $C'$ is also a register of dimension $d'$.
	
	Since we wish to obtain a bound for arbitrary $d'$, we introduce a more general distillation protocol which has an equivalent effect. To specify the channel, we start with a fixed isometry $L_0$ from a space $A'$ of dimension $d'$ to the space $A$ of dimension $d_A$. This matrix satisfies $L_0^\dagger L_0 = I_{A'}$, while $L_0L_0^\dagger$ will be a projector on $A$.
	Now pick an ensemble of unitaries $V_{r_A}$ on $A$ with corresponding probabilities $q_{r_A}$ that form a unitary 2-design (see e.g.~Ref.~\cite{Gross2007}). The object $L_{r_A}^\dagger \coloneqq L_0^\dagger V_{r_A}$ then constitutes our random measurement operator. Since the unitaries are drawn from a 2-design, we have that $\sum_{r_A}q_{r_A} V_{r_A}X V_{r_A}^\dagger = \Tr[X]\pi^A$ for all operators $X$, where $\pi^A = I_A/d_A$ is the maximally mixed state on $A$. This in turn implies that the map
	\begin{align}
		\mathcal{T}^{A \rightarrow A'M_A}[\rho^A] = \frac{d_A}{d'} \sum_{r_A}  q_{r_A} L_{r_A}^\dagger\rho^A L_{r_A}  \otimes \proj{{r_A}}_{M_A}
		\label{eq:ChannelRandomProj}
	\end{align}
	is a quantum channel from $A$ to $A'M_A$, where here $M_A$ is a classical measurement register. When applied to the state $\phi^{AC}_{s_B}$, we obtain a classical-quantum state $\rho^{A'CM_A}_{s_B} = (\mathcal{T}^{A \rightarrow A'M_A} \otimes \textrm{id}^C)[\phi^{AC}_{s_B}]$ which describes the ensemble of states
	\begin{align}
		\ket{\chi^{A'C}_{r_A s_B}} &= \frac{L_{r_A}^\dagger \ket{\phi^{AC}_{s_B}} }{\sqrt{p_{r_A|s_B}}}
		\label{eq:ChannelRandomProjStoch}
	\end{align}
	with probability $p_{r_A|s_B} = \braket{\phi^{AC}_{s_B}|L_{r_A}L_{r_A}^\dagger |\phi^{AC}_{s_B}}$. 
	
	
	The next step is to perform local operations on $\ket{\chi^{A'C}_{r_A s_B}}$ to get as close as possible to $\ket{\Phi^{A'C'}_{d'}}$.
	A useful way to quantify how close we can get is to measure how far the reduced states on $A'$ are from the maximally mixed state $\pi^{A'} = I^{A'}/d'$,
	\begin{align}
		\epsilon_{s_B} &\coloneqq \sum_{r_A} p_{r_A|s_B} \big\|\Tr_{C}[\chi^{A'C}_{r_A s_B}] - \pi^{A'}\big\|_1
		\nonumber\\ &= \left\| \rho^{A'M_A}_{s_B} - \pi^{A'}\otimes \rho^{M_A}_{s_B} \right\|_1.
		\label{eq:EpsilonDefSB}
	\end{align}
	Here, $\|X\|_1=\tr\sqrt{X^\dagger X}$ is the trace norm.
	We can also consider averaging over measurement outcomes on $B$, thereby obtaining the overall average trace distance
	\begin{align}
		\overline{\epsilon} \coloneqq \sum_{s_B} p_{s_B}\epsilon_{s_B}.
		\label{eq:EpsilonDef}
	\end{align}

	As demonstrated in Ref.~\cite{Horodecki2006}, based on the value of $\overline{\epsilon}$, one can infer the existence of a local operation on $C$ that converts $\ket{\chi^{A'C}_{r_A s_B}}$ into states that have an average overlap of at least $1 - 2\sqrt{\overline{\epsilon}}$ with the maximally entangled state. Thus, $\overline{\epsilon}$ constitutes a figure of merit for the overall EOA task. Because $\overline{\epsilon}$ can be computed without needing to know what the final operation on $C$ actually is, and the channel \eqref{eq:ChannelRandomProj} applied to $A$ does not explicitly depend on $s_B$, this quantity will be particularly convenient to work with later on.\\

	As an aside, we note that the particular channel $\mathcal{T}^{A \rightarrow A'M_A}$, which already has the advantage of being universal, is also known to be a near-optimal choice for distilling entanglement from pure states, among other information-theoretic tasks (see e.g.~Refs.~\cite{Horodecki2006,Dupuis2014}). Intuitively, we can see that $\mathcal{T}^{A \rightarrow A'M_A}$ projects $\phi^{AC}_{s_B}$ onto randomly chosen subspaces of dimension $d'$, which converts the potentially highly structured Schmidt spectrum of $\phi^{AC}_{s_B}$ into a new spectrum that is smaller but flatter, i.e.~closer to maximally mixed. In fact, from Ref.~\cite{Horodecki2006} we have an \textit{a priori} guarantee that if $\phi^{AC}_{s_B}$ is sufficiently entangled, then the final state on $A'$ will indeed be close to maximally mixed. Specifically,
	\begin{align}
		\epsilon_{s_B} \leq \exp\left(\frac{1}{2}\big[\log d' - S^{(2)}(\rho^A)\big]\right)
		\label{eq:DistGuarantee}
	\end{align}
	Eq.~\eqref{eq:DistGuarantee} assures us that the inequalities applied in this step are essentially tight, assuming that the von Neumann and R{\'e}nyi entropies of the post-measurement states behave similarly.
	
	\subsection{Bounding measurement-induced entanglement in terms of $\overline{\epsilon}$}
	
	Since direct computations of entanglement entropies of the post-measurement states $\phi^{AC}_{s_B}$ are not possible (see Section \ref{sec:Prev}), our alternative approach will be instead to compute an upper bound on the average error $\overline{\epsilon}$. From this, we can obtain the following lower bound on the entanglement in the post-measurement states:
	\begin{lemma}\label{thm:EntropyQeBound}
		Let $\{(p_{r_{A}|s_B}, \chi^{A'C}_{s_B r_{A}})\}_{r_A}$ be the ensemble of states that arises from applying the channel $\mathcal{T}^{A \rightarrow A'M_A}$ defined in Eq.~\eqref{eq:ChannelRandomProj} to to the state $\phi^{AC}_{s_B}$, and define $\epsilon_{s_B}$ as in Eq.~\eqref{eq:EpsilonDefSB}. The entanglement entropy $\phi^{AC}_{s_B}$ is lower bounded as
		\begin{align}
			S(\rho^A_{s_B}) \geq \left(1 - \frac{1}{2}\epsilon_{s_B}\right)\log d' - h_2\left(\frac{\epsilon_{s_B}}{2}\right),
			\label{eq:SBEntropyBound}
		\end{align}
		where $h_2(x) = -x\log x - (1-x)\log(1-x)$ is the binary entropy.
		Averaged over all measurement outcomes $s_B$, we get
		\begin{align}
			\mathbbm{E}_{s_B} S(\rho^{A}_{s_B}) \geq \left(1 - \frac{1}{2}\overline{\epsilon}\right)\log d' - h\left(\frac{\overline{\epsilon}}{2}\right),
			\label{eq:MeanEntropyBound}
		\end{align}
		
	\end{lemma}
	
	\textit{Proof.---}Because the distillation channel is LOCC, and the states before and after are pure, the entanglement entropy cannot increase on average \cite{Vidal2000}, and thus
	\begin{align}
		S(\rho^A_{s_B}) \geq \sum_{r_A}p_{r_A|s_B}S(\rho^{A'}_{r_As_B})
		\label{eq:Monotonicity}
	\end{align}
	where $\rho^{A'}_{r_As_B} = \Tr_C \chi^{A'C}_{r_As_B}$. Next, we apply the Audenaert-Fannes inequality: for two states $\rho$, $\sigma$ of dimension $d$ whose trace distance is $T = \frac{1}{2}\|\rho - \sigma\|_1$, we have \cite{Audenaert2007}
	\begin{align}
		|S(\rho) - S(\sigma)| \leq T \log d + h_2(T).
	\end{align}
	We take $\rho = \rho^{A'}_{r_As_B}$ and $\sigma = \pi^{A'}$, for which $S(\sigma) = \log d'$, which gives $S(\rho^{A'}_{r_As_B}) \geq (1 - T)\log d' - h_2(T)$. Taking the average with respect to $r_A$,  because $h_2$ is concave we have $\mathbbm{E}_{r_A}h_2(T) \leq h_2(\mathbbm{E}_{r_A}T)$, and by \eqref{eq:EpsilonDefSB} we have $\mathbbm{E}_{r_A} T = \epsilon_{s_B}/2$. Together with \eqref{eq:Monotonicity}, this yields \eqref{eq:SBEntropyBound}. Applying the same argument to the average over $s_B$, $r_A$, and $U$ gives Eq.~\eqref{eq:MeanEntropyBound}. \hfill $\square$\\
	
	Later on, we will also require statistical properties of post-measurement entanglement beyond the mean entropy \eqref{eq:MeanEntropyBound}, specifically in the form of concentration inequalities. A relatively simple bound of this kind which will be sufficient for our purposes can be obtained by applying apply Markov's inequality to the variable $\epsilon_{s_B}$, combined with our pre-averaged bound Eq.~\eqref{eq:SBEntropyBound}.	
	\begin{align}
		&\text{Pr}_{{\rm s_B}, U\sim \U}\Big(S(\rho^A_{\rm s_B}) < \log d' - \delta \Big) \nonumber\\ \leq&\; \text{Pr}_{{\rm s_B}, U \sim \U}\Big({\textstyle \frac{1}{2}}\epsilon_{s_B}\log d' >  \delta - \log 2\Big) \nonumber\\
		\leq&\;  \mathbbm{E}_{U \sim \U} \frac{1}{2}\frac{\overline{\epsilon}\log d'}{\delta -  \log 2}
		\label{eq:EntropyConcDelta}
	\end{align}
	where we have used the fact that $h_2(\epsilon/2) \leq \log 2$ over the domain $0 \leq \epsilon \leq 2$. As a concrete case,
	\begin{align}
		\text{Pr}_{{\rm s_B}, U \sim \U }\Big(S(\rho^A_{\rm s_B}) < \log d' - 2 \log 2\Big) \leq  \mathbbm{E}_{U \sim \U} \frac{1}{2} \overline{\epsilon} \log_2 d'.
		\label{eq:EntropyConcO1}
	\end{align}

	\subsection{Multiparty splitting technique\label{subsec:MultipartySplit}}
	
	In this section, we introduce a convenient trick that was developed for understanding one-shot multi-user quantum information tasks, including entanglement of assistance, which will bring us closer to a usable upper bound for $\overline{\epsilon}$. The following approach was referred to as a multiparty mean-zero splitting method in Ref.~\cite{Cheng2023}, or the telescoping trick in Ref.~\cite{Colomer2023}.
	
	We start by writing out the classical-quantum state $\rho^{A'CM_AM_B}$ that describes the ensemble of states after both the measurements on $B$ and the distillation channel $\mathcal{T}^{A\rightarrow A'M_A}$ [Eq.~\eqref{eq:ChannelRandomProj}] on $A$ have been applied. By virtue of the fact that $\mathcal{T}^{A\rightarrow A'M_A}$ is independent of $s_B$, we can write
	\begin{align}
		\rho^{A'CM_AM_B} = \big(\mathcal{T}^{A \rightarrow A'M_A} \otimes \mathcal{T}^{B \rightarrow M_B} \otimes \text{id}^C\big) [\Psi^{ABC}]
	\end{align}
	The average error \eqref{eq:EpsilonDef} can then be written as $\overline{\epsilon} = \|\rho^{A'M_AM_B}-\pi^{A'}\otimes \rho^{M_AM_B}\|_1$, which in terms of the channels is
	\begin{align}
		\overline{\epsilon} = \left\|\Big(\big(\text{id}^{A'} - \mathcal{D}^{A'}\big)\circ \mathcal{T}^{A \rightarrow A'M_A}  \Big) \otimes \mathcal{T}^{B \rightarrow M_B} [\rho^{AB}]\right\|_1,
		\label{eq:QeChannels}
	\end{align}
	where $\circ$ denotes composition of channels. Throughout, we use the notation $\mathcal{D}^{X}[\rho^{X}] = \pi^{X}\Tr[\rho^{X}]$ for the depolarizing channel on some set of degrees of freedom $X$, with $\pi^X = I_X/d_X$ the maximally mixed state. We have also temporarily suppressed the average over circuits $\mathbbm{E}_{U \sim \U}$ for notational convenience. 
	
	The technique that we take from Refs.~\cite{Colomer2023, Cheng2023} is to use a particular decomposition of the measurement channel $\mathcal{T}^{B \rightarrow M_B}$, namely
	\begin{align}
		\mathcal{T}^{B \rightarrow M_B} &= \bigotimes_{i \in B} (\Theta_i + \hat{\mathcal{D}}_i) = \sum_{I \subseteq B} \Theta_I \otimes \hat{\mathcal{D}}_{I^c} \label{eq:TDecompSum} \\
		\text{where }\Theta_i &\coloneqq \mathcal{T}^{Q_i \rightarrow M_i} \circ (\text{id} - \mathcal{D}^{Q_i}) \label{eq:ThetaDef}
	\end{align}
	where $I^c \coloneqq B \backslash I$, and we use the notation $\Theta_I \coloneqq \bigotimes_{i\in I}\Theta_i$. The channel $\hat{\mathcal{D}}_i[\rho^{Q_i}] = \mathcal{T}^{Q_i \rightarrow M_i} \circ \mathcal{D}^{Q_i}$, is made up of the depolarizing channel on site $i$, with the output being sent through $\mathcal{T}^{Q_i \rightarrow M_i}$ to ensure the correct matrix dimensions (this is only necessary for generalized measurements). Note that the map $\Theta_i$ is a difference of two channels, and hence is not completely positive. Its action on some operator $X$ is the same as applying $\mathcal{T}^{Q_i \rightarrow M_i}$ to the traceless part of $X$, namely $\tilde{X} = X - \pi^{Q_i}\Tr[X]$. Thus, the sum in \eqref{eq:TDecompSum} separates out operators on $B$ according to those that are traceless on subregions $I$ and proportional to identity on $I^c$.
	
	This splitting into regions $I$ and $I^c$ is reminiscent of the Ising spins that we encountered in our calculation of quasientropy in Section \ref{subsec:PrevQuasiEnt}: To each choice of $I$ we can associate variables $\sigma_i = \pm 1$ living on each site, where $\sigma_i = +1$ if $i \notin I$, and $\sigma_i = -1$ if $i \in I$; these variables signify which of the maps $\hat{\mathcal{D}}_i$ and $\Theta_i$ is applied to site $i$. The sum over $I$ in Eq.~\eqref{eq:TDecompSum} then becomes a sum over all spin configurations.
	
	We can build on this interpretation by observing that the overall map applied to region $A$ in Eq.~\eqref{eq:QeChannels} bears a resemblance to $\Theta_i$. Indeed, $\mathcal{T}^{Q_i \rightarrow M_i}$ can be thought of as an instance of the distillation channel \eqref{eq:MeasurementDistChannel} in the special case of fully destructive measurements $d'$. It is therefore natural to define the analogous map
	\begin{align}
		\Theta_A &\coloneqq (\text{id}^{A'} - \mathcal{D}^{A'}) \circ \mathcal{T}^{A \rightarrow A'M_A}.
		\label{eq:ThetaADef}
	\end{align}
	Noticing also that region $C$ is traced out in \eqref{eq:QeChannels}, it is helpful to view $A$ and $C$ as two extra sites in addition to those in $B$, which host Ising spins fixed as $\sigma_A = -1$, $\sigma_C = +1$. This again reminds us of the calculation of quasientropy. Let $\mathcal{I}$ be the set of spin configurations on $ABC$ with $\sigma_{A,C}$ fixed in this way. We then have
	\begin{align}
		\overline{\epsilon} =  \left\| \sum_{I \in \mathcal{I}} \Theta_I \otimes \hat{\mathcal{D}}_{I^c}[\Psi^{ABC}]\right\|_1
		\label{eq:ErrorSumI}
	\end{align}
	In Ref.~\cite{Colomer2023, Cheng2023}, having obtained an analogous expression, the next step was to apply the triangle inequality in the form
	\begin{align}
		\Big\|\textstyle \sum_I X_I\Big\|_1 \leq \sum_I \|X_I\|_1,
		\label{eq:TriangleIneqXI}
	\end{align}
	where $X_I = \Theta_I \otimes \hat{D}_{I^c}[\Psi^{ABC}]$ is the summand in \eqref{eq:ErrorSumI}, to obtain a sum over scalars. Upon bounding each norm separately, one arrives at an upper bound for $\overline{\epsilon}$. Unfortunately, if we take this approach here, the bounds we find become vacuous in the thermodynamic limit, unless we take $q$ to scale with $L_B$. As we show in Appendix \ref{app:HolographicBoundVac}, for the holographic model the bound we get on $\overline{\epsilon}$ is at least as large as $\sqrt{d' \mathcal{Z}_{-+}}$, where $\mathcal{Z}_{-+}$ is the partition function defined in Eq.~\eqref{eq:QuasiNumeratorPartition} (without the denominator $\mathcal{Z}_{++}$). Since the free energy density of the Ising model is finite for any nonzero temperature (here given by the bond entanglement entropy), this partition function must scale exponentially with $N$, leading to a vacuous bound. Thus, we will have to take a different approach.

	\subsection{Mapping to self-avoiding walks\label{subsec:DWs}}

	Instead of applying the triangle inequality as in \cref{eq:TriangleIneqXI}, in this section we analyse the expression \cref{eq:ErrorSumI} more carefully. This will result in a nontrivial upper bound on $\epsilon$ in terms of a different statistical mechanics partition function, which describes self-avoiding walks, \cref{lem:SAWQeBound}. In turn, we will arrive at our main technical result, \cref{thm:EntBound}, a lower bound on MIE.\\

\begin{figure}
		\includegraphics[width=\columnwidth]{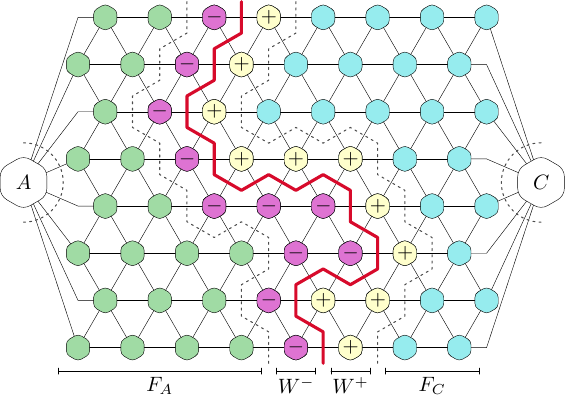}
		\caption{Partitioning of sites for a particular separating domain wall $W$ (thick red line), here for the geometry depicted in Fig.~\ref{fig:ABC}. The unmeasured sites in $A$ and $C$ are  grouped together at opposite ends, and have spins fixed to $\sigma_i = -1$ and $+1$, respectively. The sites immediately on the $A$ side of the domain wall must all have $\sigma_i = -1$, and these define the region $W^-$ (dark purple tiles). Similarly, $W^+$ is the set of sites immediately on the opposite side (light yellow tiles). The remaining sites between $A$ and $W^-$ ($C$ and $W^+$) have free spins. These regions are denoted $F_A$ ($F_C$), and are shaded green (light blue).}
		\label{fig:DomainWall}
	\end{figure}

	\noindent The Ising spins defined in the previous section are reminiscent of those that appeared in the calculation of quasientropy. As before, they are subject to boundary conditions on $A$ and $C$. These ensure that every configuration has at least one domain wall separating $A$ from $C$ with $\sigma_i = +1$ on the $C$ side and $\sigma_i = -1$ on the $A$ side, and which terminates on the boundaries of the system.  We refer to these domain walls as \textit{separating domain walls}, an example of which is depicted in Fig.~\ref{fig:DomainWall}. The intuition we have developed so far indicates that separating domain walls are associated with entanglement, and thus it makes sense to treat them more explicitly. In the following, we will consider pre-measurement states on the triangular lattice, understanding that the square lattice corresponds to the special case where we set the virtual dimension of certain bonds to 1 (those that are horizontal in Figure \ref{fig:DomainWall}).
	
	Let us start by picking one particular separating domain wall $W$, which lives on the bonds of the lattice, and write $\mathcal{I}_W \subset \mathcal{I}$ for the set of all configurations that have this domain wall in their configuration. We write $W^- \subset B$ for the subset of sites that immediately neighbour the domain wall on the $A$-side, and $W^+$ for the same on the $C$ side. We can characterize $\mathcal{I}_W$ as the set of all spin configurations for which $\sigma_i = \pm 1$ for all $i \in W^{\pm}$. All other spins apart from $\sigma_{A,C}$ are unconstrained, and we can write $F_A$, $F_C$ for the sets of `free' sites that are on the $A$- and $C$-side of the domain wall, respectively. The sum over $I \in \mathcal{I}_W$ then becomes
	\begin{align}
		\sum_{I \in \mathcal{I}_W} (\cdots)  = \sum_{\{\sigma_i\}} \left(\prod_{i \in W^+}\delta_{\sigma_i, +1} \right) \left(\prod_{i \in W^-}\delta_{\sigma_i, -1} \right) (\cdots)
	\end{align}
	Now, for the free sites in $F_{A,C}$ we can sum over each $\sigma_i = \pm 1$ which involves recombining the maps $\Theta_i$ and $\hat{D}_i$ back into the original physical channel $\mathcal{T}^{Q_i \rightarrow M_i}$. Altogether, the subset of terms $I \in \mathcal{I}_W$ in the sum in Eq.~\eqref{eq:ErrorSumI} can be written
	\begin{align}
		X(\mathcal{I}_W) &\coloneqq \sum_{I \in \mathcal{I}_W} \Theta_I \otimes \hat{\mathcal{D}}_{I^c}[\Psi^{ABC}] = \mathcal{L}_{W}[\Psi^{ABC}]
		\label{eq:XWDef}
	\end{align}
	where we define the map
	\begin{align}
		\mathcal{L}_W &\coloneqq \mathcal{T}^{F_{A}} \otimes \mathcal{T}^{F_C} \otimes \hat{\mathcal{D}}_{W^+} \otimes \Theta_{W^-}  \otimes \Theta_A \otimes \mathcal{D}_C. 
		\label{eq:LMap}
	\end{align}
	The above expression specifies which operations we should apply to each site of the lattice: Depending on which region we are considering, we apply the corresponding measurement channel $\mathcal{T}$, depolarizing channel $\mathcal{D}$, or difference channel $\Theta$. Since the maps corresponding to each site have non-overlapping support, we are free to apply them to $\Psi^{ABC}$ in any order we wish.
	
	The considerations so far have been fully general. However, at this point, we will need to invoke properties of 2D finite-depth states that we derived in Section \ref{subsec:TNS}. In particular, we take $\ket{\Psi}$ to be a state that satisfies the strictly local property (\cref{def:SimpleTNS}) on a lattice $\Lambda$ which forms a triangulation of the plane, i.e.~the dual lattice $\Lambda^*$ is 3-valent [for instance the triangular lattice isoTNS shown in Fig.~\ref{fig:Tessellation}(b)]. Note that we can turn any 2D lattice into a triangulation by adding bonds, so this condition is not a limitation. In particular, by Proposition \ref{thm:SimpleFiniteDepth}, this can always be guaranteed for finite-depth states by blocking. Observe that domain walls, which live on the dual lattice, must obey a one-in-one-out rule on a triangulation, and thus there are no intsersections or loops. Thus, there are no edges that connect the region $AF_AW^-$ to $F_CC$ (see Fig.~\ref{fig:DomainWall}). According to the strictly local condition [\cref{eq:LocalCorr}], when we trace out $W^+$, the resulting state factorizes. Specifically,
	\begin{align}
		\hat{\mathcal{D}}_{W^+}[\Psi^{ABC}] = \rho^{A F_A W^-} \otimes \rho^{C F_C} \otimes \sigma^{W^+},
		\label{eq:DWStateFact}
	\end{align}
	where $\sigma^{W^+} = \mathcal{T}^{W^+}[\pi^{W^+}]$.
	
	The channel $\mathcal{T}^{F_A}$ is also CPTP, and implements the measurement process on $F_A$. Therefore, we can write
	\begin{align}
		&(\mathcal{T}^{F_A} \otimes \text{id}_{A W^-})[\rho^{W^-F_AA}] 
		=\sum_{r} p_r \proj r_{M_{F_A}} \otimes \sigma^{AW^-}_r
		\label{eq:FAMeasureCQ}
	\end{align}
	which is a quantum-classical state representing an ensemble of mixed states on $A W^-$, with classical labels $r$.
	
	It turns out that all $\sigma^{AW^-}_r$ have the same spectrum and thus the same entropy, independent of $r$.
	To see this, consider applying the measurement first,
	\begin{align}
		&(\T^{F_A}\otimes\id_{AW^-W^+F_CC})[\Psi^{ABC}] \nonumber\\
		&= \sum_rp_r \proj{r}_{M_{F_A}}\otimes\proj{\chi_r}_{AW^-W^+F_CC}.
	\end{align}
	Because $\ket{\chi_r}$ is a pure state, the reduced states on regions $AW^-$ and $W^+F_CC$ must have the same spectra. The former $\sigma_r^{AW^-}=\Tr_{W^+F_CC}[\chi_r]$ is the same state as in \cref{eq:FAMeasureCQ}. The latter state $\sigma_r^{W^+F_CC}=\Tr_{AW^-}[\chi_r]$ we claim to be independent of $r$, by virtue of the fact that the measured region $F_A$ is sufficiently far from $W^+F_CC$.
	Formally, this can be shown by exchanging the order of $\Tr_{AW^-}$ and $\T^{F_A}$.
	If we first trace $AW^-$, we get $\Tr_{AW^-}[\Psi^{ABC}]=\rho^{F_A}\otimes\rho^{W^+F_CC}$ due to the fact that $\ket{\Psi^{ABC}}$ is a strictly local TNS [\cref{eq:LocalCorr}]. Since the states factorize, conditioning on the measurement result on $F_A$ has no effect on the state on $W^+F_CC$ and thus $\rho^{W^+F_CC}=\sigma_r^{W^+F_CC}$,  and in particular
	\begin{equation}
		S^{(2)}_W\coloneq S^{(2)}(\sigma_r^{AW^-}) = S^{(2)}(\rho^{W^+F_CC}).
		\label{eq:S2-equality}
	\end{equation}
	Taking the expressions from above, we have
	\begin{align}
		X(\mathcal{I}_W) =&\, \sigma^{F_CC}\otimes \sigma^{W^+} \otimes \sum_{r} p_r \ket{r}\bra{r}_{M_{F_A}} \otimes T_{r}, \nonumber\\  \text{where }T_{r} \coloneqq& \Big(\Theta^{W^-}\otimes \Theta^A\Big) [\sigma^{AW^-}_r],
		\label{eq:XWDomain}
	\end{align}
	with $\sigma^{AW^-}_r$ defined in \cref{eq:FAMeasureCQ}, and $\sigma^{F_C} = \mathcal{T}^{F_C}[\rho^{F_C}]$. We will now bound the norm $\|X(\mathcal{I}_W)\|_1$ as a step towards computing the error bound \eqref{eq:ErrorSumI}. Using the multiplicativity of the trace norm under tensor products, and the fact that $\|\sigma\|_1 = 1$ for any valid density matrix $\sigma$, we can reduce this quantity to
	\begin{align}
		\|X(\mathcal{I}_W)\|_1 = \sum_r p_r \|T_r\|_1.
	\end{align}
	Accordingly, we will fix an arbitrary $r$ and look to bound $\|T_r\|_1$ in the following.
	
	Since the map $\Theta^{W^-}$ appearing in Eq.~\eqref{eq:XWDomain} depends on the choice of local measurements on $B$, at this point we will need to specify the particular local POVM we use, which determines the measurement channels via Eq.~\eqref{eq:MeasurementPOVM}. For the specific case where the POVM on each site of $\Lambda$ forms a 2-design, we can employ tools introduced in Refs.~\cite{Colomer2023, Cheng2023} to analyse the quantity \eqref{eq:XWDomain}. This approach will be immediately applicable to the case of random holographic tensor networks, which feature 2-designs on each site by construction. For shallow-depth circuits with more general architectures, such as brickwork circuits, the arguments will need to be modified to account for the fact that 2-designs within each cell are not necessarily formed (recall that the sites we are working with are really cells which consist of $O(d_C^2)$ physical qudits). Let us defer this issue to Section \ref{sec:Application}, and work with 2-design measurements for now.

	Since the operators $T_{r}$ are not positive semi-definite, evaluating the norm $\|T\|_1 = \Tr[\sqrt{T^\dagger T}]$ involves a non-trivial square root of an operator, which is challenging to deal with analytically. A well-established technique for circumventing this issue involves the following characterization of the trace norm for any Hermitian (not necessarily positive) operator $L$ (Lemma 5.1.3 in Ref.~\cite{Renner2006})
	\begin{align}
		\|L\|_1 &\leq \|\sigma^{-1/4}L\sigma^{-1/4}\|_2 & \forall \sigma \succeq 0, \Tr[\sigma]=1.
		\label{eq:Schatten12}
	\end{align}
	(Equality can be achieved here taking $\sigma = \sqrt{L^\dagger L}/\|L\|_1$.) We apply the above to $T_r$, and since we are free to choose any density matrix $\sigma$ we can specialise to the case where $\sigma$ factorizes on each site of $W^-$ and $A'M_A$. Thus
	\begin{align}
		\|T_r\|_1 \leq \|(\tilde{\Theta}^{W^-}\otimes \tilde{\Theta}^A)[\sigma_r^{AW^-}]\|_2
		\label{eq:NormBoundSchatten2}
	\end{align}
	where the modified maps are $\tilde{\Theta}^{W^-} = \bigotimes_{i \in W^-} \tilde{\Theta}_i$, with
	\begin{align}
		\tilde{\Theta}_i[\,\cdot\,] &= \sigma_i^{-1/4}\Theta_i[\,\cdot\,]\sigma_i^{-1/4} \label{eq:ThetaTildeI}\\
		\tilde{\Theta}^A[\,\cdot\,] &= (\sigma^{M_AA'})^{-1/4}\Theta^A[\,\cdot\,](\sigma^{M_AA'})^{-1/4}
	\end{align}
	where $\sigma_i$, $\sigma^{M_AA'}$ are arbitrary valid states. The advantage of using Eq.~\eqref{eq:NormBoundSchatten2} is that the Schatten 2-norm can be expressed as $\|T\|_2^2 = \Tr[T^\dagger T]$, which is easier to compute.
	
	At this point, we need to consider the choice of local measurements on $B$. A key observation made in Refs.~\cite{Dupuis2021, Colomer2023, Cheng2023} is that when a unitary 2-design is applied just before measurement, the maps $\tilde{\Theta}_i$ are 2-norm contractive. In other words, we can find a choice of $\sigma_i$ such that the following inequality is satisfied
	\begin{align}
		\|\tilde{\Theta}_i[X]\|_2 &\leq \|X\|_2\;\; \forall X.
		\label{eq:ThetaContractiveLambda}
	\end{align}
	We note that the proof of \cref{eq:ThetaContractiveLambda}, which is given in Appendix \ref{app:RandomizngConstant} for completeness, is the only point in our argument that requires the circuit to be random.
	
	Finally, for the map $\tilde{\Theta}_A$ we can fix the state to be $\sigma^{M_A A'} = \pi^{A'}\otimes \omega^{M_A}$, where $\omega^{M_A} = \text{diag}(q_{r_A})$ is the classical distribution of unitaries $V_{r_A}$ appearing in the definition of the distillation channel \eqref{eq:MeasurementDistChannel}. Because these unitaries $V_{r_A}$ form a 2-design we can use the approach given in Ref.~\cite{Colomer2023} (see Appendix \ref{app:RandomizngConstant}) to obtain an analogous result
	\begin{align}
		\|\tilde{\Theta}_A[X]\|_2 &\leq \sqrt{d'}\|X\|_2 \;\; \forall X.
		\label{eq:ContractiveA}
	\end{align}
	These inequalities can be viewed as bounds on the induced norm $\|\Phi\|_{2\rightarrow 2} \coloneqq \sup_X \|\Phi[X]\|_2/\|X\|_2$ of the channels $\tilde{\Theta}_i$, $\tilde{\Theta}_A$. Since the $2 \rightarrow 2$ norm is multiplicative under tensor products, we can apply each of these inequalities sequentially for each site $i \in W^-$ and on $A$ to Eq.~\eqref{eq:NormBoundSchatten2}, to yield $\|T_r\|_1 \leq \sqrt{d'}\|\sigma^{AW^-}_r\|_2$. At this point, we use \cref{eq:S2-equality} to reach
	\begin{align}
		\|X(\mathcal{I}_W)\|_1 \leq \exp\left(\frac{1}{2}\Big[\log d' - S^{(2)}_W\Big]\right).
		\label{eq:XWBound}
	\end{align}
	
	Having bounded the norm of $X(\mathcal{I}_W)$ [Eq.~\eqref{eq:XWDef}], where $\mathcal{I}_W$ is the subset of all spin configurations that host a particular separating domain wall $W$, we are clearly close to the desired bound on our expression for $\overline{\epsilon}$ in Eq.~\eqref{eq:ErrorSumI}. It is tempting to write $\sum_\mathcal{I}(\cdots) = \sum_{W : \text{SDW}} \sum_{\mathcal{I}_W}(\cdots)$, where the sum over $W$ is over all separating domain walls. By the triangle inequality, this would yield a bound $\overline{\epsilon} \leq \sum_{W:\text{SDW}} \|X(\mathcal{I}_W)\|_1$, where the sum is over all separating domain walls $W$. Since domain walls take the form of self-avoiding walks on the dual lattice $\Lambda^*$, this would then yield the desired bound on $\bar{\epsilon}$ in terms of a self-avoiding walk partition function.
	
	In fact, this na{\"i}ve expression double counts spin configurations that contain more than one separating domain wall. Because the operators $X(\mathcal{I}_W)$ are not positive semi-definite this means such a bound is not necessarily valid. This issue is resolved in Appendix \ref{app:DoubleCounting}, where we prove that such configurations only lead to a subleading correction. As a result, we obtain the following.
	\begin{lemma}\label{lem:SAWQeBound}
		If $\ket{\Psi_\Lambda}$ is a strictly local state on a planar graph $\Lambda$ (\cref{def:SimpleTNS}), and we measure each site in region $B$ using a POVM that forms a 2-design, and apply the distillation channel $\mathcal{T}^{A \rightarrow A'M_A}$ from \cref{eq:MeasurementDistChannel}, then $\overline{\epsilon}$ defined in \cref{eq:EpsilonDef} satisfies
		\begin{align}
			\overline{\epsilon} \leq  \sqrt{d'} \mathcal{Z}_{\textup{SAW}} f(\mathcal{Z}_{\textup{SAW}})
			\label{eq:QeBoundFinal}
		\end{align}
		where $f(x) = (e^x - 1)/x$, and $\mathcal{Z}_{\textup{SAW}}$ is a partition function of self-avoiding walks $W \in \textup{SAW}(\Lambda^*, A, C)$ on the dual lattice $\Lambda^*$ which separate $A$ from $C$
		\begin{align}
			\mathcal{Z}_{\textup{SAW}} = \sum_{W \in \textup{SAW}(\Lambda^*, A, C)} \exp(-H[W]).
			\label{eq:SAWZDef}
		\end{align}
		The Boltzmann factor is given by
		\begin{align}
			H[W] = \frac{1}{2} S^{(2)}_W,
			\label{eq:BoltzmannS2}
		\end{align}
		where $S^{(2)}_W$, defined in \cref{eq:S2-equality}, is the R{\'e}nyi entropy across the cut $W$ in the pre-measurement state, whose length is $|W|$. 
	\end{lemma}
	\cref{eq:QeBoundFinal} includes a multiplicative factor $f(\mathcal{Z}_{\rm SAW})$ coming from configurations with multiple separating domain walls. In the regime of interest $0<\mathcal{Z}_{\rm SAW} < 1$, this correction is mild, namely $1<f(\mathcal{Z}_{\rm SAW}) < 2$. Self-avoiding walks is a well-studied statistical mechanics model in the field of polymer physics, and there exist rigorous results that we use later to obtain a bound in specific cases. \\
	
	Finally, we can combine \cref{thm:EntropyQeBound,lem:SAWQeBound}, which gives us a bound on MIE for any arbitrary integer choice of $d'$, namely
	\begin{align}
		\overline{E(A:C)} &\geq \left(1 - \frac{\sqrt{d'}g(F)}{2} \right)\log d' 
		- h_2\left(\frac{\sqrt{d'}g(F)}{2}\right)
	\end{align}
	where $F = -\log \mathcal{Z}_{\rm SAW}$ is the free energy of the statistical mechanics model defined above, and $g(x) \coloneqq \exp(e^{-x})-1$. We now look to optimize our choice of $d'$ to obtain the tightest possible bound. A simple yet near-optimal choice is $d' = \lceil F^{-2}e^{2F} \rceil$, which is exponentially large in $F$ while still ensuring that $\bar{\epsilon}$ decays with system size. If we make a further assertion that $F \geq 2$, then $\sqrt{d'}g(F) \leq \sqrt{F^{-2}e^{2F} + 1}(1 -\exp(-e^F)) \leq 2/F$, and we can then use the inequality $h_2(x) \leq 2x \log(1/x)$ for $x \leq 1/2$. Combining all the above we have
	\begin{theorem} \label{thm:EntBound}
		If $\ket{\Psi_\Lambda}$ obeys the strictly local property (\cref{def:SimpleTNS}) on a planar graph $\Lambda$ and each site in  region $B$ is measured with a POVM that forms a 2-design, with outcome $s_B$, then the average post-measurement entanglement entropy can be lower bounded as
		\begin{align}
			\sum_{s_B} p_{s_B} S(\rho^A_{s_B}) &\geq 2F_{\rm SAW} - 2\log(eF_{\rm SAW})
			\label{eq:EntropySAWBound}
		\end{align}
		provided $F_{\rm SAW} \geq 2$, where $F_{\rm SAW} = -\log \mathcal{Z}_{\rm SAW}$ is the free energy of a statistical mechanics model of self-avoiding walks on the dual lattice which separate the region $A$ from $C$ [defined in Eqs.~(\ref{eq:SAWZDef}, \ref{eq:BoltzmannS2})].
	\end{theorem}

	

	\section{Application to random shallow circuits\label{sec:Application}}
	
	Theorem \ref{thm:EntBound} gives us a general bound on measurement-induced entanglement in cases where the pre-measurement state is a strictly local isoTNS. Since states generated by constant-depth circuits can themselves be expressed as strictly local isoTNSs by blocking, we can now apply this result to specific families of random quantum circuits.
	
	\subsection{Random holographic tensor network states}
	
	We start with random holographic circuits, introduced in Section \ref{subsec:Holog}, which generate random holographic tensor network states distributed according to Eq.~\eqref{eq:HologDist}.  Since the circuit involves preparing entangled pairs on each bond (yielding a strictly local isoTNS), and then measuring each site of the lattice in a Haar-random basis (a particular instance of a 2-design POVM), Theorem \ref{thm:EntBound} can be applied immediately. In particular, because of the simple entanglement structure of the pre-measurement state, we can exactly evaluate the Boltzmann weights for each domain wall configuration $W$
	\begin{align}
		H[W] = \frac{S_e}{2}|W|
	\end{align}
	where $|W|$ is the length of the domain wall and $S_e = S^{(2)}(\omega_e)$ is the 2-R{\'e}nyi entanglement entropy of the bond state $\ket{\omega_e}$, which is often chosen to be maximally entangled with $S_e = \log \chi$, but more generally can be any bipartite entangled state. 
	
	With this expression for the Boltzmann weights, the partition function $\mathcal{Z}_{\rm SAW}$ then becomes equal to the canonical generating function for self-avoiding walks on the dual lattice $\mathcal{Z}_{\rm SAW}^\star[\beta] = \sum_W e^{-\beta |W|}$ \cite{Chayes1986}, with effective temperature $\beta = S_e/2$ and appropriate boundary conditions---here we assume the geometry shown in Fig.~\ref{fig:ABC} \footnote{The particular generating function corresponding to the geometry  shown in Fig.~\ref{fig:ABC} is the \textit{cylinder generating function} defined in Ref.~\cite{Chayes1986}.}. We can then make use of longstanding rigorous results on the statistical mechanics of self-avoiding walks: For values of $\beta$ strictly greater than the (lattice-dependent) critical temperature $\beta_{\rm crit}$, which is equal to $\log \mu_{\Lambda^*}$, with $\mu_{\Lambda^*}$ the so-called \textit{connective constant} of the lattice $\Lambda^*$ \cite{Hammersley1962}, the generating functions become strongly suppressed. 
	It can be shown that for $\beta > \beta_{\rm crit}$, there exists a function $m(\beta) > 0$ such that $\mathcal{Z}_{\rm SAW}^\star[\beta] \leq L_y e^{-m(\beta)L_x}$ \cite{Chayes1986, Duminil2012}. (The factor of $L_y$ is due to the number of locations where the domain wall can originate from.) In fact, because all configurations contributing to $\mathcal{Z}_{\rm SAW}^\star[\beta]$ have length at least $L_x$, we can bound $m(\beta) > \beta - \beta_{\rm crit}$. Since the logarithmic term in \eqref{eq:EntropySAWBound} is subleading for large $L_x$, we obtain the following explicit bound on MIE as function of $L_{x,y}$.
	\begin{corollary}
		For random holographic circuits with bond state $\ket{\omega_e}$ on a 2D lattice $\Lambda$ of dimensions $L_x \times L_y$, with $L_y = \textup{poly}(L_x)$,  the average measurement-induced entanglement between $A$ and $C$ in the geometry shown in Fig.~\ref{fig:Holog} satisfies
		\begin{align}
			\mathbbm{E}_{U \sim \U}\sum_{s_B} p_{s_B}S(\rho^A_{s_B}) \geq \kappa L_x
		\end{align}
		for $L_x$ exceeding some constant, and a constant $\kappa$ that can be taken to be arbitrarily close to
		\begin{align}
			\kappa \rightarrow S_e^{(2)} - 2\log \mu_{\Lambda^*},
			\label{eq:KappaEstimate}
		\end{align}
		where $\mu_{\Lambda^*}$ is the connective constant of the dual lattice $\Lambda^*$, and $S_e^{(2)}$ is the R{\'e}nyi-2 entropy of the bond state.  
		\label{thm:Holog}
	\end{corollary}
	This result establishes that if the entanglement of the pre-measurement state is above some threshold, then extensive long-ranged MIE is produced.
	For example, for the square lattice, there are known upper bounds on the connective constant $\log \mu_{\Lambda^*} \leq 0.97$ \cite{Ponitz2000}, yielding an upper bound on the threshold value of the R{\'e}nyi-2 bond entropy
	\begin{align}
		S_{e, {\rm crit}} &\leq  2.80 \text{ bits}. & \text{(square lattice)}
		\label{eq:SquareLatticeCritical}
	\end{align}
	The corresponding values for different lattices can be obtained from estimates of the corresponding connective constants. More broadly, this gives a concrete proof that a fixed (system-size independent) amount of entanglement on each bond suffices to generate highly entangled holographic states. In contrast, all previous rigorous estimates assumed a bond dimension growing in some way with system size.
	
	Provided one has the means to evaluate the appropriate generating function $\mathcal{Z}_{\rm SAW}[\beta]$, the above can be quickly generalized to different geometries of the regions $A$, $B$, $C$: The domain wall configurations included are always those that separate $A$ from $C$. Given that the transition of the self-avoiding walk partition function is a bulk property, we expect that long-ranged MIE will arise in the regime \eqref{eq:SquareLatticeCritical} for different geometries $A$, $B$, $C$, provided $A$ and $C$ make up a vanishing fraction of the system.

	The relationship we have derived between random holographic tensor networks and the statistical mechanics of self-avoiding walks can be thought of as a generalization of the Ryu-Takayanagi formula to regimes where bond dimension does not scale with system size \cite{Ryu2006}. 
	If the $q \rightarrow \infty$ limit is taken before the thermodynamic limit, then it is known that holographic entanglement is determined by the length of minimal cuts in the bulk that separate $A$ from $C$ \cite{Hayden2016} (this actually matches the asymptotically optimal achievable rate for entanglement of assistance \cite{Smolin2005,Horodecki2006}). Our result establishes a relationship (albeit an inequality) between holographic entanglement and the statistical mechanics of the same kind of cuts (domain walls), which now can fluctuate with an effective inverse temperature $\beta$ set by the entanglement of each bond state. In the limit of large bond dimension, which corresponds to $\beta \rightarrow \infty$, a Ryu-Takayanagi-like formula, where one considers only the highest-weight contribution to $\mathcal{Z}_{\rm SAW}$, is recovered.
	
	\cref{eq:SquareLatticeCritical} gives us a sufficient condition for long-ranged MIE to arise in random holographic circuits, and we can ask how closely this condition describes the true critical point. Numerical studies of random tensor networks in a square lattice geometry (which are equivalent to holographic circuits with maximally entangled bond states; see \cref{subsec:Holog}) indicate that the critical bond dimension is a little above 2, i.e.~$S_{e, {\rm crit}} \gtrsim 1 \text{ bit}$ \cite{Levy2021}. Thus, our bounds give an estimate of $S_{e, {\rm crit}}$ of the correct order of magnitude.
	
	\subsection{Random 4-local square lattice circuits \label{subsec:4local}}
	
	We now turn to random depth-$d_C$ circuits with the gate architecture shown in Fig.~\ref{fig:2dArch}(b), where 4-site gates act around each plaquette of the square lattice. States generated from $d_C$ layers of these circuits can be written as strictly local isoTNSs by blocking, as we showed in Section \ref{subsec:TNS}. However, because of this blocking, we cannot represent the measurement step in terms of a 2-design POVM within each cell, even if the local gates themselves each form 2-designs on smaller regions. We therefore need to modify the arguments that led to Theorem \ref{thm:EntBound} slightly. To simplify the analysis, we will focus on the case $d_C = 2$ here, and consider measurement-induced entanglement as a function of the local Hilbert space dimension $q$. Intuitively, MIE should only increase as a function of $d_C$ in the shallow-depth regime, and thus we expect any bounds we find for $d_C = 2$ should also apply for $d_C \geq 2$ (though this argument is not yet rigorous). We obtain the following result.

	\begin{proposition}
		\label{thm:4local}
		Starting from a state generated by 2 layers of Haar-random 4-local gates in the architecture shown in Fig.~\ref{fig:2dArch}(b), after measuring sites $B$ in the geometry shown in Fig.~\ref{fig:ABC}, the average post-measurement entanglement between $A$ and $C$ scales as
		\begin{align}
			\mathbbm{E}_{U \sim \U} \sum_{s_B} p_{s_B}S(\rho^A_{s_B}) \geq \kappa L_x
			\label{eq:Bound4Local}
		\end{align}
		for some constant $\kappa > 0$ and $L_x$ exceeding some constant value for local Hilbert space dimension $q \geq 88$.
	\end{proposition}
	The proof of this statement is given in Appendix \ref{app:4Local}. In brief, we can use Eq.~\eqref{eq:XWDomain} as a starting point, since no assumptions about the POVMs had been made at that point (only that the state obeyed the strict locality condition), and look in more detail at the intra-cell structure of the measurement maps $\Theta_i$. We arrive at an analogous mapping to a self-avoiding walk partition function $\mathcal{Z}_{\rm SAW}$, with slightly modified parameters, and so we can use the same rigorous statistical mechanics results as before to get the bound \eqref{eq:Bound4Local}.
	
	The specific inequalities we have employed to characterise the intra-cell structure of $\Theta_i$ are fairly crude. This is why we obtain a relatively conservative bound on $q$, whereas we would expect the true critical value to be somewhat similar to the numerically observed value for the holographic architecture $q_{\rm crit} \gtrsim 2$ \cite{Levy2021}. Even so, the key insight from \cref{thm:4local} is that we can infer the existence of extensive long-ranged post-measurement entanglement in circuits with constant Hilbert space dimension and constant depth, which is a significant improvement over previous bounds which required $q$ to grow with system size.

	\subsection{Random 2D brickwork circuits}
	
	Finally, we consider random brickwork circuits in the architecture shown in Fig.~\ref{fig:2dArch}(a). We will fix a depth $d_C = 4$ here, which will allow us to use similar techniques to the previous section, since a depth-4 brickwork circuit can be compiled into a depth-2 circuit of 4-local gates as shown in Fig.~\ref{fig:2dArch}(b). Again, we intuitively expect that if random depth-4 brickwork circuits host long-ranged MIE, so should random circuits of all depths $d_C \geq 4$.
	\begin{proposition}\label{thm:Brickwork}
		Starting from a state generated by 4 layers of Haar-random 2-local gates in the brickwork architecture shown in Fig.~\ref{fig:2dArch}(a), after measuring sites $B$ in the geometry shown in Fig.~\ref{fig:ABC}, the average post-measurement entanglement between $A$ and $C$ scales as
		\begin{align}
			\mathbbm{E}_{U \sim \U} \sum_{s_B} p_{s_B}S(\rho^A_{s_B}) \geq \kappa L_x
		\end{align}
		for some constant $\kappa > 0$ and $L_x$  exceeding some constant value for local Hilbert space dimension $q \geq 134$.
	\end{proposition}
	The proof of the above is analogous to that of \cref{thm:4local}, and can be found in Appendix \ref{app:4Local}.

	\subsection{Simulating sampling and contracting random tensor networks\label{subsec:Contract}}
	
	As mentioned in the introduction, our results on measurement-induced entanglement can be translated into statements about the hardness of sampling from random constant-depth circuits, as well as the related task of contracting random tensor networks. Here we show that standard algorithms for these tasks based on MPSs will fail in the regimes where our lower bound on MIE is nontrivial. For concreteness, we will specifically focus on the task of sampling from holographic circuits on the square lattice, though our results readily generalize to other architectures and geometries. By the correspondence of Ref.~\cite{Hayden2016}, this is directly related to the problem of contracting tensor networks that are randomly distributed according to Eq.~\eqref{eq:HologDist}.

\begin{figure}
		\includegraphics{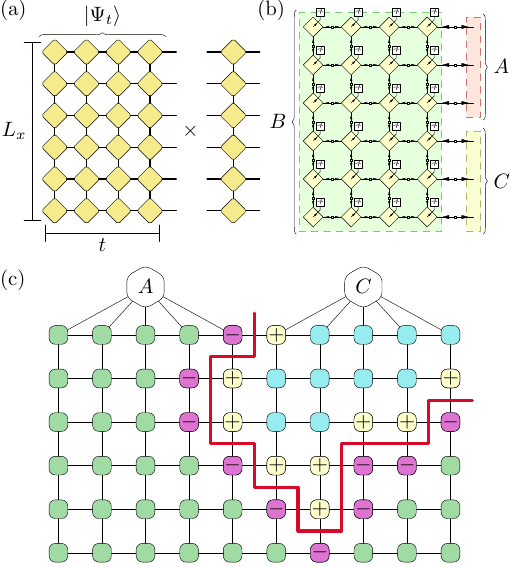}
		\caption{(a) In the bMPS algorithm, one contracts a tensor network column by column. At each step $t$, one stores a MPS representation of a $L_x$-leg tensor given by contracting the first $t$ column, which can be thought of as a $L_x$-site state $\ket{\Psi_t}$. This state advances by contracting with the next column until the whole network is contracted. (b) The analogous algorithm for sampling problems involves measuring sites column by column, and storing a post-measurement state on the virtual degrees of freedom. We look to lower bound the half-chain entanglement of this state using the geometry $ABC$ shown here. (c) In this new geometry, the relevant domain walls are those that originate from the boundary of $A$ and $C$ and terminate on one of the three other edges of the system. Sites are coloured according to the convention used in Fig.~\ref{fig:DomainWall}.}
		\label{fig:contract}
	\end{figure}

	The boundary MPS (bMPS) method \cite{Verstraete2005, Jordan2008} and the related `sideways evolving block decimation' (SEBD) algoirthm for circuit sampling \cite{Napp2022} each involve sweeping through a 2D system in a chosen spatial direction. For tensor network contraction, the network is contracted by applying a transfer matrix column-by-column. At the $t^\mathrm{th}$ step, one stores an MPS representation of the $L_x$-leg tensor formed from contracting the first $t$ columns and leaving $L_x$ virtual legs free, as depicted in Fig.~\ref{fig:contract}(a). One then updates this tensor by contracting with the $(t+1)$th column of tensors (the transfer matrix), resulting in a new rank-$L_x$ tensor with larger MPS bond dimension. If this bond dimension cannot be truncated to some polynomial-in-$L_x$ value while keeping the error bounded, then the algorithm aborts. Similarly, when using SEBD to simulate circuit sampling, at the $t$th step one stores an MPS representation of the conditional state of the $(t+1)^{\textrm{th}}$ column of sites, after the first $t$ columns have been measured  [Fig.~\ref{fig:contract}(b)]. Because this conditional state is by assumption an MPS with polynomial bond dimension, one can efficiently sample the measurement outcomes for the next column of sites from the appropriate conditional distribution \cite{Napp2022}, and in turn compute the conditional state for the next column, which should be truncated to keep the bond dimension bounded.  We will refer to the states that are stored at intermediate points in the corresponding algorithm as virtual states.

	To store the virtual state efficiently (polynomial bond dimension) as an MPS with bounded error, this state must have low entanglement, in a sense made precise in Ref.~\cite{Schuch2008}. Depending on the task in question, this virtual state is either a post-measurement state (sampling) or a holographic tensor network state (tensor contraction). Theorem \ref{thm:EntBound} gives us the means to lower bound the entanglement of post-measurement states, and also by the correspondence established in Ref.~\cite{Hayden2016} (Section \ref{subsec:Holog}), we can also make statements about random holographic tensor networks that are distributed according to Eq.~\eqref{eq:HologDist}. 
	
	Specifically, we will look at the half-chain entanglement entropy of the virtual state. This requires us to consider a different geometry for $A$ and $C$, as depicted in Fig.~\ref{fig:contract}(b). The calculation described previously still holds, yielding a bound in terms of the free energy of a model of self-avoiding walks $W$. For holographic circuits, the Boltzmann weights are $\exp(- \frac{1}{2}|W|S_e^{(2)})$, but now the boundary conditions specify that $W$ should originate at the interface between $A$ and $C$, and terminate at one of the fully contracted edges of the system [Fig.~\ref{fig:contract}(c)]. We can again relate this to canonical generating functions for self-avoiding walks with effective temperature $\beta = \frac{1}{2} S_e^{(2)}$, and thus we expect to see high entanglement in the ordered phase $\beta > \beta_{\rm crit} = \log \mu_{\rm \Lambda^*}$. For the square lattice considered here, this means $S_e^{(2)} \geq 2.80 \text{ bits}$. When considering random tensor networks with maximally entangled bond states, we have a corresponding condition on the bond dimension $\chi \geq 7$.
	
	To get a specific upper bound on the relevant partition function in the ordered regime, we use the cylinder generating function defined in Ref.~\cite{Chayes1986} for the square lattice, which counts the total weight of all self-avoiding walks that originate from the origin and terminate at some plane a distance $\ell$ away, without leaving the region $0 \leq x \leq \ell$. The weighted sum over such walks can be upper bounded by $e^{-m(\beta)\ell}$, where $m(\beta)$ is the effective mass of the model, which is positive for $\beta > \beta_{\rm crit}$. In the current geometry, all walks terminate a distance at least $\xi = \text{min}(t, L_x/2)$ away from their origin, and so using Lemma \ref{lem:SAWQeBound} we obtain an average distillation error $\overline{\epsilon} \leq \sqrt{d'}e^{-m(\beta) \xi}$. Using our concentration bound \eqref{eq:EntropyConcO1}, and choosing $d' = \exp(\kappa \xi)$ for some $0 < \kappa < m(\beta)/2$, we see that the boundary state has an entanglement entropy of order $\Omega(\xi)$ with probability exponentially close to 1. Since a faithful representation of such a state as an MPS with polynomial bond dimension cannot exist \cite{Schuch2008}, we obtain the following result.
	\begin{corollary}[Formal version of \cref{thm:ContractInf}]
		\label{thm:Contract}
		Using the SEBD algorithm to sample from a random holographic circuit on a square $\sqrt{N} \times \sqrt{N}$ lattice where the bond state $\ket{\omega_e}$ has R{\'e}nyi-2 entanglement entropy $S^{(2)}_{e} > 2.8 \textup{ bits}$, the probability of the algorithm aborting is $1 - e^{-\Omega(\sqrt{N})}$. The same holds for using bMPS to contract random tensor networks with bond dimension $\chi \geq 7$, where the distribution of tensor components is given by Eq.~\eqref{eq:HologDist}.
	\end{corollary}
	More generally, if we look at different circuit architectures, then in regimes where the relevant self-avoiding walk model is ordered, by Lemma \ref{lem:SAWQeBound} we will find that $\overline{\epsilon}$ is exponentially small in the linear system size, and in turn an analogous result will apply.

	The above can be compared to related results on the hardness of certain tensor network contraction problems in 2D. Worst-case hardness results are known for contracting TNSs, both in general \cite{Schuch2007} and for restricted classes such as isometric and injective TNSs \cite{Anshu2024,Malz2024}. In  Ref.~\cite{Haferkamp2020}, it was shown that exactly evaluating normalized expectation values $\braket{\Psi|A|\Psi}/\braket{\Psi|\Psi}$ is average-case hard, while the problem becomes classically tractable when a sufficient amount of bias is added to the distribution of tensor components \cite{Chen2024}.  These complexity-theoretic results are more powerful in that they are not specific to a particular choice of classical algorithm, whereas here we obtain a statement about average-case, relative-error problems in relation to the bMPS algorithm. Nevertheless, given that the bMPS method is widely employed in practice, the above result is of practical significance.
	
	The exponential dependence on $L_x$ puts severe quantitative limits on the system sizes one can simulate with these methods. Consider the case of holographic circuits on the square lattice with maximally entangled bond states of dimension $q$. With high probability, the state being represented by the bMPS has a half-chain entanglement close to $m\kappa L_x/2$, where by Eq.~\eqref{eq:KappaEstimate}, we can take $\kappa$ to be $\frac{1}{2}\log q  - \log \mu_{\Lambda^*}$. Any approximation of such a state with error $\delta$ must have bond dimension $\log \chi_{\rm bMPS} \geq \frac{1}{2}(m(\beta)-\frac{1}{2}\delta \log q)L_x$ \cite{Schuch2008}. Taking an example of $q = 16$, and demanding an approximation error of $\delta \leq 5\%$, this gives $\chi_{\rm bMPS} \geq e^{0.2 L_x}$. The number of coefficients in such an MPS is $L_xq\chi_{\rm bMPS}^2$; thus, with $1\text{ TB}$ of random access memory and working at double floating point precision, the maximum system size one can reach is $L_{x, {\rm max}} \leq 48$. 
	Given that our rigorous results are likely to give conservative lower bounds on the entanglement of the bMPS, we expect that the maximum system size will actually be smaller by an appreciable factor.
	
	\subsection{Quantum advantage in random constant-depth circuits}
	
	As a final application, we prove that random constant-depth circuits exhibit a certain kind of quantum advantage, which was first demonstrated for specific instances of circuits in Ref.~\cite{Bravyi2018}. Specifically, we show that random 2D constant depth circuits cannot be simulated by any sub-logarithmic depth classical circuit, even with all-to-all connectivity. To demonstrate this, we make use of a rigorous relationship between MIE and classical simulability that was recently established in Ref.~\cite{Watts2024}. Specifically, the authors of that work showed that if long-ranged tripartite entangled states (specifically Greenberger-Horne-Zeilinger states) are generated upon measuring a given stabilizer state, then any classical simulation of sampling from that state must have at least logarithmic depth.
	
	Our bounds so far are framed in terms of bipartite (as opposed to tripartite) entanglement, and thus we cannot directly use the corollaries described in the previous sections in their current forms to argue for such a quantum advantage. Nevertheless, we are able to adapt our arguments to show that the tripartite MIE condition identified in Ref.~\cite{Watts2024} is indeed satisfied for random Clifford circuits of sufficient depth. For simplicity, we focus on the square lattice holographic circuit with maximally entangled states of dimension $q = 2^m$ (for some integer $m$) on each bond, and random four-qudit Clifford unitaries on each site. The detailed arguments are given in Appendix \ref{app:Advantage}, and we find the following result.
	\begin{corollary}[Formal version of \cref{thm:AdvantageInformal}]
		\label{thm:AdvantageFormal}
		Consider the holographic circuit defined on $q = 2^m$-dimensional qudits, with four-qudit gates $U_i$ independently sampled from the Clifford group at each site $i$. For some constant $\delta > 0$, and for any $m \geq 6$, any classical probabilistic circuit that takes $\{U_i\}$ as input and outputs a distribution that is on average $\delta$-close in total variational distance to the output of the quantum circuit must have depth $\Omega(\log N)$.
	\end{corollary}
	In Ref.~\cite{Watts2024}, it was shown that this type of condition holds for depth-2 4-local circuits with qudits of dimension $q = \Omega(N)$, i.e.~$m = \Omega(\log N)$. In contrast, our result holds for constant depth and constant $q$. In principle, we can also combine the arguments that led to \cref{thm:4local} with those that led to \cref{thm:AdvantageFormal} to get a similar result for  depth-2 4-local circuits for $q = O(1)$ as well.

	\section{Discussion \label{sec:Discussion}}
	
	We have now presented our main technical result, Theorem \ref{thm:EntBound}, which gives us a lower bound on post-measurement entanglement, and described a number of important corollaries as well as closely related propositions that connect to both MIE and the hardness of certain computational problems related to sampling and contracting tensor networks. Here we provide some additional remarks to help contextualize our findings.
	
	\subsection{Monitored quantum dynamics from sampling shallow circuits and spacetime duality}
	
	\begin{figure*}
		\includegraphics[width=512pt]{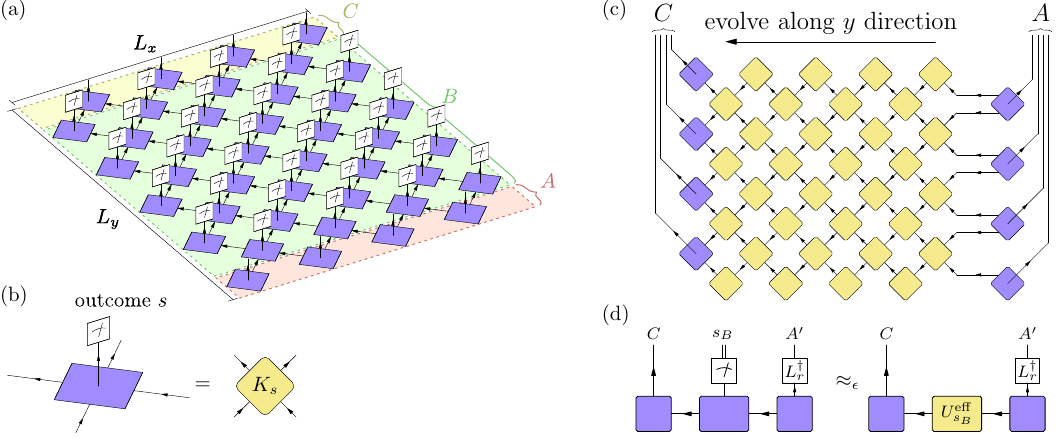}
		\caption{(a) Starting from an isoTNS, we measure degrees of freedom in $B$. (b) Applying a measurement to a physical degree of freedom converts the local isometric tensor to an unravelled quantum channel, described by properly normalized Kraus operators $K_s$ which act on virtual degrees of freedom. (c) Concatenating the virtual legs together, we obtain a (1+1)D monitored quantum dynamics with a brickwork architecture, with the time-like dimension (here $y$) determined by the direction of the isometries. (d) In the distillation protocol, a projection $L_r^\dagger$ is applied to $A$ to get a state on a smaller Hilbert space $A'$. Within this subspace, the sideways-evolving dynamics can be approximated by an ensemble of effective unitaries $U^{\rm eff}_{s_B}$ which depend on the random outcomes $s_B$.}
		\label{fig:IsomMapping}
	\end{figure*}
	
	We have already seen that states prepared by 2D finite-depth circuits can be represented as isometric tensor network states with constant bond dimension. Using this construction, we can establish a relationship between sampling from shallow-depth circuits with (1+1)D monitored quantum dynamics, which allows us to make connection with the large body of work on measurement-induced phase transitions in these settings. Connections of this type have been made before in particular models \cite{Jian2022,Napp2022,Liu2022,Liu2023}, but by using isoTNSs we can make the association highly concrete and general.
	
	Figure \ref{fig:IsomMapping}(a) illustrates the setup we consider, where one performs projective measurements on an isoTNS with the regions $ABC$ chosen as in Fig.~\ref{fig:ABC}. The significance of the isometric property of the local tensors is that isometries represent physical quantum operations. In this case, each tensor constitutes a map from two virtual legs $i_1i_2$ to the physical leg $Q$ plus the remaining two virtual legs $o_1o_2$, and in principle this map could be implemented using unitaries and ancillas. If we append a measurement to the physical leg $Q$, which itself is also a physical operation, then we obtain a \textit{unravelled quantum channel} from $i_1i_2$ to $o_1o_2$ (see Ref.~\cite{Gullans2020}). These unravelled channels, which are specified by a set of Kraus operators $\{K_s\}$, describe stochastic quantum processes where the state $\rho$ evolves to $\rho_s' = K_s \rho K_s^\dagger/p(s|\rho)$ with probability $p(s|\rho) = \Tr[K_s^\dagger K_s \rho]$. The Kraus operators for a given site $i$ can be related to the local tensor by projecting out the physical leg $K_{s_i} = (\bra{s_i}_{Q_i} \otimes I_{o_1o_2})V^{i_1i_2\rightarrow Po_1o_2}$ [Fig.~\ref{fig:IsomMapping}(b)]. Using the fact that $V^{i_1i_2\rightarrow Po_1o_2}$ is an isometry, one can verify that the unravelled map
	\begin{align}
		\mathcal{N}_u^{i_1i_2 \rightarrow M_i o_1o_2}[\rho^{i_1i_2}] = \sum_{s_i} \proj{s_i}_{M_i} \otimes K_s \rho^{i_1i_2} K_s^\dagger,
		\label{eq:UnravelledChannel}
	\end{align}
	which outputs both the quantum state on $o_1o_2$ and the measurement outcome $s_i$ on a classical register $M_i$, is a valid channel, i.e.~a CPTP map.
	
	When the virtual legs are concatenated, one obtains a brickwork circuit of unravelled channels [Fig.~\ref{fig:IsomMapping}(c)], described by a larger family of Kraus operators $K_{s_B}$ formed from products of the constituent operators $K_s$. These are the same objects which describe monitored quantum dynamics of the kind studied in hybrid unitary-projective circuits \cite{Li2018, Chan2018, Choi2019, Skinner2019, Garratt2024,McGinley2024}. The latter are known to exhibit `measurement-induced phase transitions' (MIPTs) between low- and high-entanglement states as a function of the degree of non-unitarity in the circuit, usually controlled by a rate of projective measurements. Thus, from this mapping we can relate the existence of a phase transition in shallow 2D circuits with measurements to the existence of MIPTs in (1+1)D circuits. This would suggest that these two phase transitions are of the same universality class. Moreover, the correspondence described above indicates that our methods could be used to obtain rigorous results on MIPTs, which heretofore have only been obtained in settings with sufficiently simple structures, e.g.~all-to-all connectivity \cite{Fava2023,Giachetti2023,Bulchandani2024, Deluca2024}, or in cases where $q$ grows with system size \cite{Bao2020}. 
	
	Beyond the settings we consider here, relationships between different forms of dynamics that involve re-interpreting a spatial direction as a temporal one are referred to as `spacetime dualities'. Correspondences of this kind have helped establish a number of important insights into many-body quantum dynamics \cite{Ippoliti2021,Lu2021,Ippoliti2022}, and led to the discovery of new exactly solvable one-dimensional circuit models \cite{Bertini2018,Bertini2019,Piroli2020a,Claeys2021}. For systems in two or higher spatial dimensions, we anticipate that isoTNSs can provide a useful formalism to better understand these spacetime dualities, and potentially form the basis of new exact solutions to quantum many-body problems. 
	
	\subsection{Comparison to measurement-based quantum computation}
	
	The cluster state---the canonical resource for performing measurement-based quantum computation (MBQC) \cite{Raussendorf2001}---is a specific example of a state that can be prepared in constant depth and for which we can guarantee long-ranged MIE (provided we measure in the appropriate basis): By construction, this state has the property that upon measuring qubits in bases within the $x$-$y$ plane, a random unitary circuit is executed in the sideways direction. The specific unitary implemented in this sideways circuit depends on the measurement outcomes $s_B$, and thus the post-measurement state received by $A$ and $C$ after this process is given by $\ket{\phi^{AC}_{s_B}} = (I_A\otimes U^{\rm MBQC}_{s_B})\ket{\Phi^+_{AC}}$, where $U^{\rm MBQC}_{s_B}$ describes the unitary circuit conditioned on the outcomes, and $\ket{\Phi^+_{AC}}$ is some maximally entangled state. We can relate the unitarity of this sideways circuit to the fact that such states have maximal measurement-induced entanglement (this is a consequence of the Schr{\"o}dinger-HJW theorem \cite{Hughston1993}).
	
	When considering generic finite-depth states, the effective sideways circuit is no longer unitary; rather, it is described by a network of non-unitary Kraus operators $K_{s_B}$, as explained in the previous section and in Fig.~\ref{fig:IsomMapping}. Using the same arguments as in Ref.~\cite{Gullans2020}, we can interpret our results as a lower bound on the quantum capacity of this unravelled channel acting within the virtual space of the isoTNS. Specifically, our bounds on entanglement translate to a lower bound on the coherent information of the unravelled channel $\mathcal{N}^{A \rightarrow M_B C}_u[\,\cdot\,] = \sum_{s_B} \proj{s_B}_{M_B} \otimes K_{s_B}\,\cdot\, K_{s_B}^\dagger$ [cf.~Eq.~\eqref{eq:UnravelledChannel}], which in turn is known to be a lower bound on its quantum capacity. In the context of (1+1)D unitary-projective dynamics, the fact that the capacity remains nonzero after concatenating many layers of non-unitary evolution has been referred to as dynamically self-correcting quantum memory \cite{Choi2019,Gullans2020,Fan2021,Fidkowski2021}.
	
	Our distillation argument (Section \ref{subsec:Distil}) can also be interpreted as a quantum communication task through the channel $\mathcal{N}^{A \rightarrow M_B C}_u$. When we act with the distillation channel on $A$ [Eqs.~(\ref{eq:ChannelRandomProj}, \ref{eq:ChannelRandomProjStoch})], we project the state onto some subspace defined by $L_rL_r^\dagger$, which is a random subspace of dimension $d'$. Then, once measurements on $B$ are performed, we are left with a state on $A'C$ that is $\epsilon$-close to maximally entangled. By standard arguments from the field of approximate error correction \cite{Schumacher2001}, this implies that within the subspace defined by $L_rL_r^\dagger$, the dynamics of this sideways circuit is effectively unitary, up to corrections that are correspondingly small in $\epsilon$. 
	In other words, we have $K_{s_B}L_r^\dagger \approx U^{\rm eff}_{s_B, r}L_r^\dagger$, where $U^{\rm eff}_{s_B, r}$ is determined by the random variables $s_B$ and $r$  [Fig.~\ref{fig:IsomMapping}(d)]. The random projection we apply can thus be thought of as an encoding of $\log d'$ bits of quantum information in a subspace that is protected against the non-unitary effects of $\mathcal{N}^{A \rightarrow M_B C}_u$. Such random encodings are known to be optimal when sending quantum information through many identical copies of a given quantum channel \cite{Hayden2008}, and we speculate that this is also true in this many-body (but non-asymptotic) setting in the thermodynamic limit.
	
	Comparing our generic shallow circuits to the MBQC setting, we see that even away from the cluster state limit, the effect of measurements can be interpreted as a unitary evolution in the sideways direction. The difference is that this evolution occurs within some protected subspace, and that the effective unitaries $U^{\rm eff}_{s_B, r}$ are depend on the measurement outcomes in a more complicated way. This perspective allows one to make connection to the phenomenon of deep thermalization, namely the formation of complex-projective $k$-designs in the projected ensemble \cite{Ho2022,Cotler2021,Claeys2022}. For cluster states and other states with flat entanglement spectra, such as those generated by dual-unitary circuits \cite{Ho2022}, the random sideways-evolving unitary circuits generated by the measurement provides a mechanism for the generation of maximally random states \cite{Brown2008,Mezher2018, Ho2022}. Here, it seems natural to expect that the states in the projected ensemble are maximally random within the relevant error-corrected subspace. This could yield an alternative picture of deep thermalization that does not explicitly require states to cover the whole Hilbert space uniformly, as occurs e.g.~in finite-temperature state ensembles \cite{Mark2024,Liu2024}.
	
	We remark that the picture described here very naturally applies to the problem of measuring cluster states in bases out of the $x$-$y$ plane, where entanglement transitions have been observed numerically \cite{Liu2022,Liu2023}. Additionally, the sideways-evolving circuit picture provides a new way to analyse strategies for distributing entanglement in quantum networks \cite{Acin2007, Perseguers2010}, where similar kinds of short- to long-ranged entanglement transitions can occur.  

	\subsection{Connections to hardness of sampling}
	
	As mentioned in the introduction, there is a close relationship between measurement-induced entanglement in a given setting and the hardness of simulating the output distribution of the underlying state \cite{Napp2022, Watts2024, Lin2023}. Indeed, the known worst-case examples of finite-depth states that are provably hard to sample from \cite{Terhal2004,BermejoVega2018} (assuming certain widely-believed conjectures in complexity theory) use resource states for MBQC, which as we saw above can exhibit maximal MIE. Here we discuss how our results can be placed within the context of quantum advantage in shallow-depth circuits.
	
	In Ref.~\cite{Napp2022}, classical algorithms for simulating the output distribution of finite-depth circuits were put forward, the success of which relies on the post-measurement states having weak entanglement. The sideways-evolving block decimation (SEBD) algorithm is akin to the bMPS method for contracting tensor networks discussed in Section \ref{subsec:Contract}, and requires the boundary state after measuring $t$ columns of sites to be representable by an MPS with polynomial bond dimension. We have found sufficient conditions for such boundary states to have extensive bipartite entanglement entropy, and using the results of Ref.~\cite{Schuch2008}, this is sufficient to show that no approximate MPS representation can be found. Thus, the SEBD algorithm will fail if the conditions described in the various corollaries of Section \ref{sec:Application} are met.
	
	Evidently, the existence of long-ranged MIE implies that algorithms exploiting the short-ranged entanglement structure of post-measurement states will fail in constant-depth circuits---this is made formal for the specific case of bMPS-based algorithms in \cref{thm:Contract}. This complements rigorous results in Ref.~\cite{Napp2022} which give sufficient conditions for shallow circuits with certain architectures to be easy to simulate classically. Of course, MIE alone is not sufficient to establish hardness, since Clifford circuits with Pauli measurements can exhibit extensive MIE whilst being efficiently simulable by the Gottesman-Knill theorem \cite{Gottesman1998}. Intuitively, one would expect that for typical instances of Haar random circuits, the post-measurement states will be both highly entangled and very far from stabilizer states, making methods that exploit this kind of structure inapplicable. Since tensor network and stabilizer methods are the primary techniques for simulating many-body quantum circuits, our results suggest that all currently known classical algorithms for simulating sampling will fail for constant-depth circuits in the long-ranged MIE phase.
	
	Unsurprisingly, a complexity-theoretic proof that sampling from random constant-depth circuits cannot be simulated by polynomial classical circuits remains elusive. (Indeed, such a result would imply $\textsf{SampBQP} \neq \textsf{SampBPP}$ \footnote{Here, $\textsf{SampBPP}$ and $\textsf{SampBQP}$ are the classes of sampling problems that can be solved in polynomial time with bounded error by probabilistic classical circuits and quantum circuits, respectively.}, and in turn several longstanding conjectures in computer science \cite{Lund2017}). Nevertheless, \cref{thm:AdvantageFormal} establishes an unconditional separation between constant-depth classical circuits and random constant-depth quantum circuits, which is a significant step forward. Together, our algorithm-specific statements combined with this weaker but highly general theorem provide strong evidence that typical instances of constant-depth circuits are indeed hard to sample from.

	\subsection{Interpretation of statistical mechanics mapping \label{subsec:StatMechInterp}}
	
	The emergence of the self-avoiding walk partition function in our calculations (Lemma \ref{thm:EntropyQeBound}) is reminiscent of other statistical mechanics mappings seen in various analyses of random quantum circuits with measurements. For the most part, to alleviate the issue of the nonlinear relationship between pre- and post-measurement quantum states, the mappings previously derived involve calculating some kind of quasientropy, which we reviewed in Section \ref{subsec:PrevQuasiEnt}. With the exception of systems with infinite local Hilbert space dimension $q \rightarrow \infty$, there is no known rigorous relationship between quasi-entropy (for fixed replica index $\alpha \geq 2$) and the correctly averaged MIE. Nevertheless, there is a qualitative similarity between the Ising model that describes quasientropy [Eq.~\eqref{eq:QuasiRatioIsing}] and the self-avoiding walk model derived here [Eq.~\eqref{eq:SAWZDef}]. In the latter, the degrees of freedom can be thought of as domain walls that separate $A$ from $C$, which resembles the domain walls that are the relevant degrees of freedom in the Ising model representation of quasientropy \eqref{eq:QuasiRatioIsing} in the case of $\alpha=2$ replicas. Moreover, in both mappings, the free energy cost of these domain walls is related to the entanglement of the pre-measurement state across this cut (albeit with quantitatively different Boltzmann factors).
	
	As well as the difference in rigour between the two approaches, from a purely statistical mechanics perspective, we note that in the Ising model partition function, one must include all spin configurations including those that host closed domain wall loops, whereas in our model we just keep track of a single separating domain wall. Since such closed loops are not sensitive to the boundary conditions on $A$ and $C$, these contributions effectively cancel each other out in the numerator and denominator of Eq.~\eqref{eq:QuasiRatioIsing} (although they could in principle renormalize the surface tension of larger domain walls). This leads to qualitatively similar behaviour to the self-avoiding walk model, namely a phase transition into an ordered phase where the quantities in question are exponentially suppressed at some constant critical temperature. The fact that our rigorous results are consistent with the previously proposed `teleportation phase transition' phenomenology, proposed on the basis of quasientropy calculations in Ref.~\cite{Bao2024}, can be seen as a reflection of the similarity of these models. In fact, in the closely related setting of (1+1)D monitored quantum circuits, suggestive arguments have been put forward indicating that upon taking the proper replica limit $\alpha \rightarrow 1$, the appropriate theory for describing quasientropy $Q^{(\alpha)}$ is a directed polymer in a random environment \cite{Li2023}---a statistical mechanics model with some qualitative similarities to the self-avoiding walk model we obtain here.
	
	We emphasise that the domain walls appearing in our analysis are not associated with any replica technique, but rather originate from the multiparty splitting formula \eqref{eq:ThetaDef}, which we adopted from Refs.~\cite{Colomer2023,Cheng2023}. Indeed, this decomposition can be used  even in settings without any random gates---in our proof, we only invoked randomness at the very end to prove the contractivity property \eqref{eq:ThetaContractiveLambda}. As explained before, this formula decomposes operators according to those that are either traceless or proportional to the identity on each given site. This bears some similarity to the entanglement membrane formalism \cite{Zhou2020}, which can also be defined without requiring random gates, although this picture explicitly invokes replicas. Whether such ideas can be unified with our rigorous techniques remains an interesting problem for future work.

	\section{Conclusion and outlook \label{sec:Conclusion}}
	
	We have rigorously demonstrated that a macroscopic amount of long-ranged entanglement is generated upon measuring states generated by random two-dimensional quantum circuits of constant depth and qudit dimension. To derive such a result, we have established a new theoretical technique for lower bounding post-measurement entanglement, which allows us to bypass the replica method. Our findings were shown to have implications for the classical hardness of sampling from such circuits, as well as the problem of contracting random tensor networks. In particular, we proved that random constant-depth quantum circuits cannot be simulated by any classical circuit of sublogarithmic depth, thereby generalizing the result of Ref.~\cite{Bravyi2018} to random circuits.
	
	Our results raise several important open questions. One natural direction would be to consider the effects of experimentally relevant perturbations, such as noise and measurement readout errors. While a constant rate of noise without any change of circuit architecture is likely to lead to result in a distribution that is easy to sample from classically (as is the case in deep random circuit sampling \cite{Aharonov2023}), hardness results for noisy circuit sampling in the constant-depth regime have been put forward in three-dimensional systems \cite{bravyi2020,Caha2023}, and also in certain adaptive settings \cite{Perseguers2008}. One can also ask the separate question of whether post-measurement states in these settings with noise remain long-ranged entangled, which could be addressed by generalizing our distillation-based arguments to mixed states, e.g.~using a decoupling approach \cite{Dupuis2014}. In addition, it would be interesting to consider rigorous approaches to upper bounding MIE, in order to prove the stability of the conjectured low-MIE, classically simulable phase \cite{Napp2022}.

	The states we have considered in this work are by definition short-ranged entangled. Using our isoTNS representation of these states, we showed how the sampling problem is equivalent to a non-unitary monitored quantum dynamics evolving in the spatial direction. More generally, isometric tensor networks can be used to represent a more general class of wavefunctions, including those with topological order \cite{Soejima2020}. It would be interesting to understand how the structure of such long-ranged entangled states imprints itself on the effective monitored dynamics, and in turn, how the post-measurement states are themselves affected, particularly given recent results linking topological phases with particular patterns of MIE \cite{Cheng2024}.
	
	More generally, there are a wide range of different settings in which such transitions in MIE can occur, and a more detailed understanding of the physics of these phenomena would shed new light on the role of entanglement in many-body quantum dynamics. There is scope for more extensive numerical studies of these transitions, e.g.~to better pinpoint the location and  nature of the critical point in different circuit architectures, which would also help quantify the tightness of our bounds. Similarly, we remark on the possibility of experimental studies of these phenomena, building on recent realisations of shallow circuit sampling in trapped ion quantum computers \cite{Ringbauer2025}. While detecting and verifying MIE can be challenging due to the randomness of the post-measurement state, a number of recent strategies have been put forward which could be used to directly observe these transitions \cite{Garratt2024, McGinley2024}.
	
	Finally, the techniques we have introduced in this work have the potential to be used in many other settings where one wishes to quantify entanglement in many-body systems. In several such settings, including decoding thresholds in error correcting codes \cite{Fan2024,Sala2024}, and phase transitions in measurement-induced dynamics \cite{Bao2020}, the replica technique has been the primary theoretical tool, but the quantities that one can calculate in this formalism are not concretely connected to entanglement without being able to take the replica limit, as we saw for quasientropy. By using the distillation-based argument described in this work, one could obtain rigorous results for this kind of physics, thereby shedding more light on the nature of entanglement dynamics in complex many-body quantum systems.

	\acknowledgements{MM acknowledges support from Trinity College, Cambridge.
		WWH is supported by the Singapore NRF Fellowship, NRF-NRFF15-2023-0008, and through the National Quantum Office, hosted in A*STAR, under its Centre for Quantum Technologies Funding Initiative (S24Q2d0009).  
		DM acknowledges support from the Novo Nordisk Fonden under grant number NNF22OC0071934 and NNF20OC0059939. Discussions between MM and WWH on the subject of this work were initiated in part at the Aspen Center for Physics, which is supported by National Science Foundation grant PHY-1607611.
	}

	\appendix
	
	\section{Isometric tensor network representation of finite-depth states \label{app:isoTNS}}
	
	In this appendix, we describe an explicit construction to get a representation of any state prepared by $d_C$ layers of 4-local gates in the architecture shown in Fig.~\ref{fig:2dArch}(b) in terms of a strictly local isoTNS on a triangular lattice, where each physical leg corresponds to a cell of $O(d_C^2)$ qudits. This result immediately applies to 2D brickwork circuits of depth $2d_C$, since these can be compiled into 4-local circuits. While it has already been shown that states obtained from  circuits of depth $d_C$ can be expressed as isoTNS with bond dimension $\exp(d_C)$~\cite{Soejima2020}, here require a slightly different construction that involves blocking, such that we obtain isoTNS with strictly local correlations.

	\begin{figure*}
		\includegraphics[width=512pt]{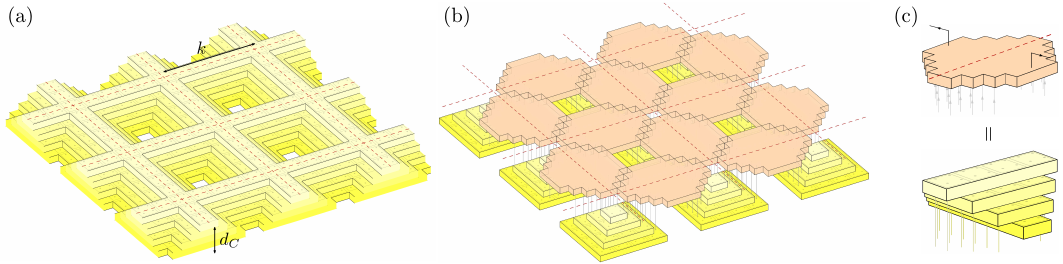}
		\caption{Decomposing a finite-depth 2D circuit into separate steps. (a) We define square cells of size $k\times k$, with $k \geq 2d_C$ (red dashed lines). We remove all gates that only yield unitary rotations within a cell to obtain a state that is equivalent to $\ket{\Psi_\Lambda}$ up to intra-cell unitaries. (b) Along the edges of each cell, we pick out a particular collection of gates, denoted with orange tensors in panel (c). This leaves non-overlapping square-shaped patches of gates in a pyramid arrangement. Note that the inputs to the group of gates in (c) are split into two subsets corresponding to different pyramids, while the outputs are split into two in a different way, namely according to the cells each site belongs to (red dashed line).}
		\label{fig:2DPyramid}
	\end{figure*}
	
	The general steps we take to arrive at such a representation are depicted in Fig.~\ref{fig:2DPyramid}. They work for any local circuit on any lattice, but here we assume a square lattice for presentation reasons. We partition the lattice into square cells of size $k \times k$, where the cell size $k$ is even and satisfies $k \geq 2d_C$.  First, we separate out those gates whose light cone lies entirely within a cell, as they correspond only to a basis change in the final state. These gates can be removed with tensor product of local unitaries, which we call $U_L$. Removing these gates yields the grid network of gates illustrated in Fig.~\ref{fig:2DPyramid}(a).
	In the next step, we similarly collect the gates along the ridges of the remaining grid, illustrated as orange blocks of gates in Fig.~\ref{fig:2DPyramid}(c), which are chosen such that they do not overlap and can be implemented by a product of local unitaries, which we call $U_R$. Removing also these gates yields a state that factorizes on square patches of linear size $2d_C \leq k$ that lie on the corners of the initially chosen cells. This remaining state can be disentangled by a final local unitary $U_P$ that similarly factorizes into a tensor product.
	Thus we see that $\ket{\Psi_\Lambda}=U_L\dagg U_R\dagg U_P\dagg\ket{0}^{\otimes |\Lambda|}$, where each of the unitaries is a tensor product of strictly local unitaries.
	
\begin{figure}
		\includegraphics[scale=1]{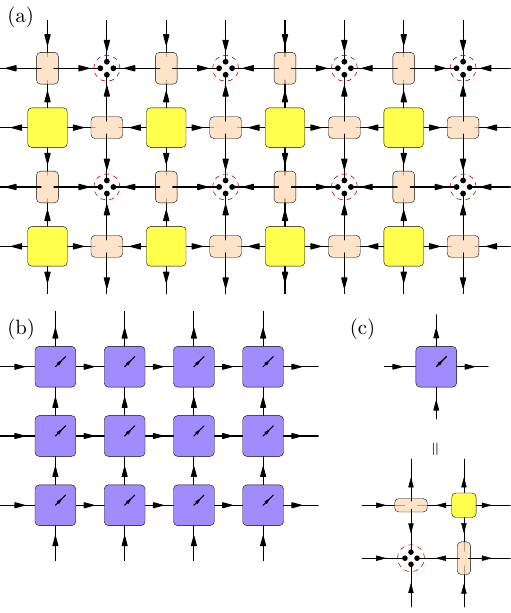}
		\caption{(a) After grouping legs together appropriately, the collections of gates depicted in Fig.~\ref{fig:2DPyramid} can be connected up into the network shown. Black dots representing physical (blocked) qudits, which are arranged into cells, demarcated with red dashed lines. The intra-cell unitaries are left implicit here. (b) If we consider all qudits within a cell as one collective degree of freedom, we can write $\ket{\Psi_{\rm 2D}}$ as a tensor network state, where each five-legged tensor (c, blue squares) is an isometry from two virtual legs to the physical leg (out of the page) plus the other two virtual legs. This defines an isoTNS state \cite{Zaletel2020}, with an orthogonality center in a particular corner of the lattice.}
		\label{fig:2DTNS}
	\end{figure}

	By grouping the inputs and outputs of the unitaries in each step into larger collective qudits, we end up with a blocked network depicted in Fig.~\ref{fig:2DTNS}(a), with arrows indicating the directions of isometries. In \cref{fig:2DTNS}(a), the yellow squares describe the state of the patches given by $U_P\dagg\ket0^{\otimes|\Lambda|}$. The orange rectangles are applied afterwards and are the gates on the ridges that we collected into $U_R\dagg$. Their output qudits are separated and grouped with neighbouring qudits into the cells that represent our original partition. Note that in \cref{fig:2DTNS}(a) we have not included the on-site unitaries from $U_L\dagg$ above.

	We can use the representation in Fig.~\ref{fig:2DTNS}(a) to see that after blocking, $\ket{\Psi_{\rm \Lambda}}$ can be viewed as a TNS where each tensor has the form of Fig.~\ref{fig:2DTNS}(b). By looking at the arrows in this diagram, one can see that the tensors in question are isometries from two virtual legs to the remaining three. This is the defining feature of an isoTNS on the 2D square lattice. Additionally, by counting legs we can obtain an upper bound on the bond dimension $\chi \leq q^{O(d_C^2)}$, which is independent of system size $N$. Therefore, in the shallow-depth regime $d_C = O(1)$, this is an efficient representation of $\ket{\Psi_{\rm \Lambda}}$. The special case where the yellow squares in Fig.~\ref{fig:2DTNS} have bond dimension 1 describes the output of a holographic circuit (Section \ref{subsec:Holog}), with bond states $\ket{\omega_e}$ determined by the orange rectangular tensors. This demonstrates that these circuits yield states that are manifestly strictly local isoTNSs, even for bond states that are not maximally entangled.
	
	Note that for the particular choice of grouping chosen in Fig.~\ref{fig:2DTNS}(b), the virtual legs have arrows pointing to the North and East. In the language of Ref.~\cite{Zaletel2020}, this corresponds to an isoTNS with an `orthogonality center' in the South West corner, namely a point in the lattice for which arrows are always directed away from this site. By symmetry, we can group the tensors differently to arrive at similar isoTNS representations with orthogonality centers at any corner.
	
	We have now obtained a representation of $\ket{\Psi_\Lambda}$ in terms of a square-lattice isoTNS, but this is not yet a strictly local isoTNS. This is because two cells that share a corner can be correlated with one another, even though they do not share a bond in the square lattice. Therefore, to obtain a strictly local isoTNS, we take pairs of adjacent $k\times k$ cells and block them into larger cells containing $2k \times k$ physical qudits. We choose the pairs such that they form a brickwork layout, as illustrated in Fig.~\ref{fig:Tessellation}(a). The network of such cells forms a triangular lattice, with each cell having six neighbours as shown in Fig.~\ref{fig:Tessellation}(b). Since we choose $k \geq 2d_C$, it is straightforward to show that any two rectangular cells that are not connected by an edge in the triangular lattice will be fully uncorrelated, and in turn we obtain Eq.~\eqref{eq:LocalCorr}.

	\section{Ising partition function bound for holographic model \label{app:HolographicBoundVac}}
	
	In this appendix, we derive a bound on entanglement in holographic states by directly applying the formulae derived in Ref.~\cite{Colomer2023}. The bound one obtains using this method turns out to be vacuous in the thermodynamic limit, which is why we need to adapt these techniques for the scenarios considered in this work (Section \ref{subsec:DWs}).
	
	The approach taken in Ref.~\cite{Colomer2023} starts with Eq.~\eqref{eq:ErrorSumI}, and applies the triangle inequality to the sum to obtain
	\begin{align}
		\overline{\epsilon} \leq \sum_{I \in \mathcal{I}} \|\Theta_I[\rho^I]\|_1
	\end{align}
	Each term in the above sum can be bounded by following the same steps as presented in the main text and using Eq.~\eqref{eq:ThetaContractiveLambda}, yielding $\|\Theta_I[\rho^I]\|_1 \leq \sqrt{d'} \|\rho^I\|_2$. Again we can represent the sum over $I$ in terms of Ising spin variables $\sigma_i$ with fixed boundary conditions on $A$ and $C$, at which point we obtain
	\begin{align}
		\overline{\epsilon} \leq \sqrt{d'} \sum_{\{\sigma_i\}} \delta_{\sigma_A, -1}\delta_{\sigma_C,+1} e^{-\mathcal{H}[\{\sigma_i\}]/2}
		\label{eq:AppQeBound}
	\end{align}
	where $\mathcal{H}[\{\sigma_i\}]$ is equal to the 2-R{\'e}nyi entropy of the corresponding region $I$ in the pre-measurement state, as we saw in Eq.~\eqref{eq:FreeEnergy}. In the holographic circuit model, $H[\{\sigma_i\}]$ is equal to the entropy of the bond state $S_e^{(2)}$ times the total number of bonds on which the Ising spins anti-align. The right hand side of \eqref{eq:AppQeBound} then becomes the partition function for the Ising model with opposite boundary conditions on $A$ and $C$, where the inverse temperature is $\beta = S_e^{(2)}/2$. We can compare this to $\mathcal{Z}_{+-}$ [Eq.~\eqref{eq:QuasiNumeratorPartition}], which we encountered in the calculation of quasientropy: the difference is just a factor of 2 in the effective temperature. By the Cauchy-Schwartz inequality $(\sum_{\{\sigma_i\}} e^{-H[\{\sigma_i\}]/2})^2 \geq \sum_{\{\sigma_i\}} e^{-H[\{\sigma_i\}]}$, we see that the right hand side of \eqref{eq:AppQeBound} is at least as big as $\sqrt{d'\mathcal{Z}_{+-}}$, as claimed in the main text. 
	
	In the thermodynamic limit $L_{x,y} \rightarrow \infty$ at constant bond entropy $S_e^{(2)} = O(1)$, we can identify two separate contributions to (minus) the free energy $\log \mathcal{Z}_{+-}$: a boundary term $-\zeta L_x$ due to the opposite boundary conditions on $A$ and $C$, plus a bulk term $+\alpha L_x L_y$, with $\alpha = \int_0^T \dif T' f(T') > 0$, where $f(T)$ is the free energy density of the Ising model at temperature $T$ \cite{Abraham1973}. Because both $\zeta$ and $\alpha$ are constants, the latter `bulk' term dominates. Overall, the bound on $\overline{\epsilon}$ that we get ends up being of the form
	\begin{align}
		\overline{\epsilon} \leq \exp(\Theta(N)).
	\end{align}
	Evidently, we do not get any useful information from this bound.
	
	If on the other hand we let the bond state entropy scale with $L_{x,y}$, then we can obtain a meaningful bound. At low temperatures, $\alpha$ behaves as $\exp(-\Omega(1/T))$, and hence if we take the bond entropy to scale as $S_e^{(2)} = \Omega(\log N)$, the bulk contribution to the free energy ends up being suppressed in the thermodynamic limit. Note that this regime requires a local Hilbert space dimension $q = \Omega(\text{Poly} N)$. The only surviving contribution to the Ising free energy is then the boundary term, and this gives us a nontrivial bound $\overline{\epsilon} \leq e^{-\omega(L_x)}$. In fact, in this scaling regime where the inverse temperature (bond entropy) is growing sufficiently quickly with system size, the partition function itself can actually be well approximated by taking only the lowest energy contributions to $\mathcal{Z}_{-+}$. These are minimal-length domain walls, so the expression for the post-measurement entanglement entropy that one ends up with resembles the Ryu-Takayanagi formula \cite{Ryu2006}. This is consistent with the discussion of Ref.~\cite{Hayden2016}, where intuitive arguments are put forward to justify these approximations in the case where $q$ grows with system size.
	
	Note that when calculating quasientropy, the expression one obtains [Eq.~\eqref{eq:QuasiNumeratorPartition}] is a ratio of partition functions with different boundary conditions $\mathcal{Z}_{-+}/\mathcal{Z}_{++}$. There, the bulk contribution to the free energy cancels between the numerator and denominator. In contrast, the bounds we obtain here feature no such denominator. The two agree in the limit where $S_e^{(2)}$ scales with system size, because in this case $\mathcal{Z}_{++}$ is dominated by the confiuration where $\sigma_i = +1$ for all $i$, which gives   $\mathcal{Z}_{++} \approx 1$.

	\section{Contractivity of measurement channels \label{app:RandomizngConstant}}
	
	In this appendix, we prove contractivity properties of the maps $\tilde{\Theta}_i$ [Eq.~\eqref{eq:ThetaTildeI}] and $\tilde{\Theta}_A$ [Eq.~\eqref{eq:ContractiveA}], following the derivations given in Refs.~\cite{Horodecki2006, Colomer2023}. For convenience, we recall the definition of the distillation channel applied to $A$
	\begin{align}
		\mathcal{T}^{A \rightarrow A'M_A}[\rho^A] = \frac{d_A}{d'} \sum_{r_A}  q_{r_A} L_{r_A}^\dagger\rho^A L_{r_A}  \otimes \proj{{r_A}}_{M_A},
		\label{eq:ChannelRandomProjAppendix}
	\end{align}
	in terms of which $\Theta_A \coloneqq (\text{id}^{A'}-\mathcal{D}^{A'})\circ \mathcal{T}^{A \rightarrow A'M_A}$. Here, $L_{r_A}^\dagger = L_0^\dagger V_{r_A}$, where $L_0$ is a fixed isometry from $A'$ to $A$, and $\{(q_{r_A}, V_{r_A})\}$ is a unitary 2-design.  Because the measurement channels on $B$ are a special case of the above channel for $d' = 1$, we will prove the more general bound \eqref{eq:ContractiveA} here.
	
	For an arbitrary operator $X$ on $A$, we want to find an upper bound for
	\begin{align}
		&\min_{\sigma^{A'M_A}} \|\tilde{\Theta}_A[X]\|_2^2 \nonumber\\ = &\min_{\sigma^{A'M_A}}\Tr\left[(\sigma^{A'M_A})^{-1/2}\Theta_i[X](\sigma^{A'M_A})^{-1/2}\Theta_i[X]\right]
	\end{align}
	where the minimum is over all density matrices on $A'M_A$. Finding this minimum is in general challenging; however, we can get the required upper bound by using any fixed choice of $\sigma_{A'M_A}$. In particular, we choose
	\begin{align}
		\sigma^{A'M_A} = \pi^{A'} \otimes \text{diag}(q_r)^{M_A}.
	\end{align}
	We then obtain
	\begin{align}
		\min_{\sigma^{A'M_A}} \|\tilde{\Theta}_A[X]\|_2^2 &\leq \frac{d_A^2}{d'}\sum_{r_A} q_{r_A}\bigg( \Tr[(L_0^\dagger V_{r_A}X V_{r_A}L_0)^2] \nonumber\\ &- \frac{1}{d'}\Tr[L_0^\dagger V_{r_A}X V_{r_A}L_0]^2\bigg),
	\end{align}
	where $\Pi_0 \coloneqq L_0L_0^\dagger$ is a projector of rank $d'$. Noticing that the above involves only second moments of the ensemble of unitaries $\{(q_{r_A}, V_{r_A})\}$, we can us the 2-design property to replace the average  with an integral with respect to the Haar measure $\dif\mu_H(U_A)$ over $U_A \in \mathrm{U}(d_A)$
	\begin{align}
		&\min_{\sigma^{A'M_A}} \|\tilde{\Theta}_A[X]\|_2^2 \nonumber\\&\leq \frac{d_A^2}{d'}\int \dif \mu_H(U_A) \Tr[(\mathcal{F}_{A'} - \mathcal{I}_{A'}/d')\cdot(L_0^\dagger U_A X U_A^\dagger L_0 )^{\otimes 2}],
		\label{eq:ContractiveHaarIntegral}
	\end{align}
	where we have used the swap operator $\mathcal{F}_{A'}$ to express the first term using the identity $\Tr[(Y_{A'})^2] = \Tr[\mathcal{F}_{A'} Y_{A'}^{\otimes 2}]$, and $\mathcal{I}_{A'} \coloneqq I_{A'}\otimes I_{A'}$. Now all that remains is to evaluate this integral with the help of Weingarten calculus \cite{Collins2006}. Noting that $\Tr[(\mathcal{F}_{A'} - \mathcal{I}_{A'}/d')\cdot (L_0L_0^{\dagger})^{\otimes 2}] = 0$, we can use the following expression which holds for traceless $Y$
	\begin{align}
		&\int \dif \mu_H(U_A) \Tr[Y U^{\otimes 2}_A X(U^\dagger_A)^{\otimes 2}] \nonumber\\ = &\frac{1}{d_A^2-1}\Bigg[ \Tr[Y^2]\Tr\Big[\big((\text{id}^A-\mathcal{D}^A)[X]\big)^2\Big] \Bigg] & \text{if }\Tr[Y] = 0
	\end{align}
	Identifying $Y = L_0^{\otimes 2}(\mathcal{F}_{A'}-\mathcal{I}_{A'}/d') (L_0^\dagger)^{\otimes 2}$, we obtain
	\begin{align}
		\min_{\sigma^{A'M_A}} \|\tilde{\Theta}_A[X]\|_2^2 &\leq \frac{d_A^2}{d'}\frac{(d')^2-d'/d_A}{d_A^2 - 1}\|(\text{id}^A - \mathcal{D}^A)[X]\|_2^2 \nonumber\\ &\leq d'\|X\|_2^2,
	\end{align}
	which gives us Eq.~\eqref{eq:ContractiveA}. Since the measurement channel for each qudit in $B$ is equivalent to the distillation channel \eqref{eq:ChannelRandomProjAppendix} for the case of fully destructive measurements $d' = 1$, we immediately obtain Eq.~\eqref{eq:ThetaContractiveLambda} as a corollary.

	Note that due to the prefactor of $d_A^2/d'$ in Eq.~\eqref{eq:ContractiveHaarIntegral}, if we generalize to measurement operations for which the POVM ensemble forms a $\epsilon$-approximate state design, we can estimate the size of the correction correction to the contractivity constant $\lambda_i$ as being $\epsilon d_A^2/d' \geq \epsilon d_A $. This may be somewhat larger than one might na{\"i}vely expect in cases where $\epsilon$ is taken to have a constant magnitude---an issue that was noted in the context of decoupling problems in Ref.~\cite{Szehr2013}. Unfortunately, this means we cannot directly apply powerful results on the formation of approximate designs in shallow quantum circuits \cite{Schuster2024, Laracuente2024}, which is why we require a more specialised analysis the spatial structure of entanglement in these circuits (Appendix \ref{app:4Local}).

	\section{Summing over domain walls \label{app:DoubleCounting}}
	
	In this appendix, we describe how the sum over $I\in \mathcal{I}$ in Eq.~\eqref{eq:ErrorSumI} can be expressed in terms of a sum over separating domain walls. As explained in the main text, a separating domain wall $W$ is a division of sites with $A$ contained on one side and $C$ on the other, with spins immediately on the $A$-side of $W$ set to $\sigma_i = -1$, and those immediately on the $C$-side set to $\sigma_i = +1$ (see Fig.~\ref{fig:DomainWall}). Note that we are being specific about the orientation of the domain wall here, i.e.~we do not explicitly count domain walls where $\sigma_i = +1$ on the $A$-side and $\sigma_i = -1$ on the $C$-side. We write $\mathcal{I}_W \subset \mathcal{I}$ for the subset of configurations that feature a particular $W$. 
	
	Because the majority of spins in the bulk are not constrained, $\mathcal{I}_W$ will contain some configurations that host extra spanning domain walls in addition to $W$. Thus, if we take the joint sum $\sum_{W:\text{SDW}}\sum_{I \in \mathcal{I}_W} X_I$, we will actually be counting configurations with more than one spanning domain wall multiple times. Since each operator $X_I$ is not necessarily positive semi-definite, we cannot yet bound $\overline{\epsilon}$ with $\sum_{W:\text{SDW}} \|X(\mathcal{I}_W)\|_1$. We therefore consider these problematic configurations more explicitly.
	
	Since we are working on the triangular lattice, domain walls do not intersect, so for any spin configuration we can unambiguously identify an exhaustive set of $n$ spanning domain walls for some $1 \leq n < L_y/2$. Moreover, we can label these walls $W_1, \ldots, W_n$ such that $W_i$ is on the $A$-side of $W_j$ for all $i < j$. We treat each $n$ sector separately.
	
	In the sum $\sum_{W} X(\mathcal{I}_W)$, configurations with $n=2$ domain walls are counted twice. We can therefore define $\mathcal{I}[W_1, W_2]$ as the subset of configurations with separating domain walls at $W_1$ and $W_2$ (possibly among others), and subtract off the sum $\sum_{W_1, W_2}X(\mathcal{I}[W_1, W_2])$ to compensate. The configurations with $n = 1$ and $n = 2$ are now properly accounted for, but those with $n = 3$ are now counted ${3 \choose 1} - {3 \choose 2} = 0$ times. We can therefore add an extra term $+\sum_{W_1, W_2, W_3}X(\mathcal{I}[W_1, W_2, W_3])$ to deal with the $n = 3$ sector, leaving those configurations with $n = 4$ counted ${4 \choose 1} - {4 \choose 2} + {4 \choose 3} = 2$ times. Iterating this reasoning, we find that to count each configuration exactly once, we need to consider
	\begin{align}
		\sum_{I \in \mathcal{I}} X_I = \sum_{n=1}^{L_y/2} \sum_{\substack{W_1, \ldots, W_n \\ \text{ordered SDWs}}} (-1)^{n+1} X(\mathcal{I}[W_1, \ldots, W_n]),
	\end{align}
	where the second sum on the right hand side is over all choices of $W_i$ such that the ordering described above is obeyed. We can thus infer the bound $\overline{Q}_e \leq \sum_n \sum_{W_1, \ldots, W_n} \|X(\mathcal{I}[W_1, \ldots, W_n])\|_1$.
	
	Now, to compute $\|X(\mathcal{I}[W_1, \ldots, W_n])\|_1$, we note that for each of the $n$ spanning domain walls there will be a corresponding region of $+$ spins of the same structure as the region $W^+$ depicted in Fig.~\ref{fig:DomainWall}. Since these sites are traced out and the underlying state is a strictly local TNS (Definition \ref{def:SimpleTNS}), the reduced state of the remaining degrees of freedom will factorize into $(n+1)$ disconnected regions which are all mutually uncorrelated with one another, as we saw in Eq.~\eqref{eq:DWStateFact}. The same arguments as given for the $n = 1$ case in Section \ref{subsec:DWs} can then be applied to get
	\begin{align}
		&\|X(\mathcal{I}[W_1, \ldots, W_n])\|_1 \nonumber\\ \leq& \exp\left( \frac{1}{2}\left[\log d' - \sum_{j=1}^n S^{(2)}_{W_j}\right] \right)
		\label{eq:XWBoundMulti}
	\end{align}
	Now, in the sum over $\{W_j\}_{j=1}^n$, the SDWs do not intersect with one another, and are ordered. The sum is therefore equivalent to a sum over unordered non-overlapping $W_j$, times a factor of $1/(n!)$. Since the norms \eqref{eq:XWBoundMulti} are non-negative, we can get an upper bound by ignoring the non-intersecting condition at this point, and so each of the $n$ domain walls can be considered independent. Thus we have
	\begin{align}
		\left\|\sum_{I\in \mathcal{I}} X_I\right\|_1 \leq& \sqrt{d'} \sum_{n=1}^{L_y/2} \frac{1}{n!} \sum_{W_1} \cdots \sum_{W_n} e^{-\sum_{j=1}^n S^{(2)}_{W_j}/2}  \nonumber\\
		\leq& \sqrt{d'} \sum_{n=1}^{L_y/2} \frac{1}{n!} \left(\Z_{\rm SAW}\right)^n  \nonumber\\
		\leq& \sqrt{d'}(e^{\mathcal{Z}_{\rm SAW}} - 1)
	\end{align}
	where $\mathcal{Z}_{\rm SAW} = \sum_{W} e^{-S^{(2)}_W/2}$ is the partition function for a single self-avoiding walk. This proves the expression quoted in Lemma \ref{lem:SAWQeBound}.
	
	As a technical note, if the original lattice $\Lambda$ is a planar graph that is not a triangulation, then we have the added complication that there can be faces where domain walls intersect. As explained previously, we can always add edges to $\Lambda$ to make a triangulation $\Lambda'$, while ensuring that the strictly local condition \cref{def:SimpleTNS} still holds with respect to $\Lambda'$. This will yield a bound on $\overline{\epsilon}$ in terms of self-avoiding walks on the dual lattice of $(\Lambda')^*$, which may be distinct from $\Lambda^*$. For each self-avoiding walk on $(\Lambda')^*$, there is a corresponding self-avoiding walk on $\Lambda^*$, and because the Boltzmann weights only depend on the R{\'e}nyi entanglement entropy across $W$ (which does not change when extra fictitious bonds are added), we can substitute the partition function for the original dual lattice $\Lambda^*$.
	
	\section{Detailed bound for random 4-local circuits \label{app:4Local}}
	
	In this appendix, we present details of our proofs of \cref{thm:4local,thm:Brickwork}, namely the bounds on MIE for depth-2, 4-local Haar random circuits, and depth-4 brickwork circuits. Because these circuits are made up of random gates that do not span an entire cell, the majority of these arguments involve characterising the entanglement structure of the states in question at the intra-cell level as necessary.
	
	\subsection{4-local circuits}
	
	As explained in Section \ref{subsec:TNS}, to ensure that we have a strictly local TNS representation of our state, we block sites into rectangular cells that tessellate in the way shown in Fig.~\ref{fig:Tessellation}(a).  For 4-local circuits at depth $d_C = 2$, the issue we have is that our assumption that each cell is measured in a uniformly random basis no longer holds, since the cells are strictly larger than the size of each random gates. Thus, Eq.~\eqref{eq:ThetaContractiveLambda} cannot be immediately applied.
	
	To get around this issue, we will consider the circuit structure within each cell more carefully. In the following, we choose the cells to have size $8 \times 4$. We can also ensure that the second of the two layers of unitaries in the circuit contains gates that act within single cells, and so these can be absorbed into the measurement channel for each cell using Eq.~\eqref{eq:POVMRandomUnitary}. Having done so, we can then think of the setup as a series of measurements on the state that results from the first layer of unitaries. This depth-1 state consists of many independent random 4-qudit states, which has a simple enough entanglement structure for us to explicitly compute the necessary bounds.

	Evidently, the POVM describing the effective measurement process within each cell does not form a 2-design, but rather a tensor product of eight separate 2-designs, corresponding to each of the eight four-site gates applied immediately before measurement. While Eq.~\eqref{eq:XWDomain} was obtained by considering coarse-grained cells, we will now have to `fine-grain' again and separate out each quartet of four qudits that share the same gate in the final layer, of which there are 8 per coarse-grained cell. We label each quartet within a cell $i$ by an index $a = 1, \ldots, 8$, so we can write $\mathcal{T}^{Q_i \rightarrow M_i} = \bigotimes_{a=1}^8 \mathcal{T}_{i,a}$, where $\mathcal{T}_{i,a}$ are the measurements on a single quartet $a$.  We can use the same approach as Eq.~\eqref{eq:TDecompSum} to split $\Theta_i$ into a sum of $2^8 - 1 = 255$ terms
	\begin{align}
		\Theta_i = \sum_{\substack{\{\tau_a\}\\ \vec{\tau} \neq (+1, \ldots, +1)}} \left(\bigotimes_{a : \tau_a = -1} \Theta_{i,a}\right)\otimes \left( \bigotimes_{a : \tau_a = +1}\hat{\mathcal{D}}_{i,a}\right),
		\label{eq:Theta4LocalDecomp}
	\end{align}
	where the sum in the above excludes the term where all $\tau_a = +1$. The fine-grained operator $\Theta_{i,a} = \mathcal{T}_{i,a} - \hat{D}_{i,a}$---defined in the same way as \eqref{eq:ThetaDef}---is determined by the measurement channel $\mathcal{T}_{i,a}$ on a single quartet. Because random gates are applied to each quartet immediately beforehand, $\mathcal{T}_{i,a}$ does describe a 2-design measurement on these four qudits. We can therefore make use of the fact that the modified map $\tilde{\Theta}_{i,a}[\,\cdot\,] = \sigma_{i,a}^{-1/4}\Theta_{i,a}[\,\cdot\,]\sigma_{i,a}^{-1/4}$ [for some arbitrary state $\sigma_{i,a}$, by analogy to Eq.~\eqref{eq:ThetaTildeI}] is 2-norm contractive [Eq.~\eqref{eq:ThetaContractiveLambda}].
	
	Substituting Eq.~\eqref{eq:Theta4LocalDecomp} into our expression for $T_r$ \eqref{eq:XWDomain}, we obtain a sum of $255^{|W^-|}$ operators. Taking the norm of this sum and applying the triangle inequality, we obtain an upper bound
	\begin{align}
		\|T_r\|_1 \leq \sum_{\{\tau_{i,a}\}} \left\| \left(\Theta^A \otimes \bigotimes_{i \in W^-} \bigotimes_{a : \tau_{i,a} = -1} \Theta_{i,a}\right)[\sigma^{AW_{\tau}}_r]\right\|_1,
	\end{align}
	where $W_\tau \subseteq W^-$ is the set of sites that are not traced out, and $\sigma^{AW_\tau}_r$ is the reduced density matrix of the conditional state $\sigma^{AW^-}_r$ in the corresponding region. Using Eq.~\eqref{eq:Schatten12} and the contractivity properties of the maps $\tilde{\Theta}_{i,a}$, we get the bound
	\begin{align}
		\|T_r\|_1 \leq \sqrt{d'}\sum_{\{\tau_{i,a}\}} \|\sigma^{AW_\tau}_r\|_2
		\label{eq:TrBoundFineGrained}
	\end{align}
	One can think of this expression as a fine-grained version of Eq.~\eqref{eq:XWBound}. All that remains now is to obtain upper bounds on each term in the above sum by examining the entanglement structure of the states $\sigma^{AW^-}_r$. 
	
\begin{figure}
		\centering
		\includegraphics[width=\columnwidth]{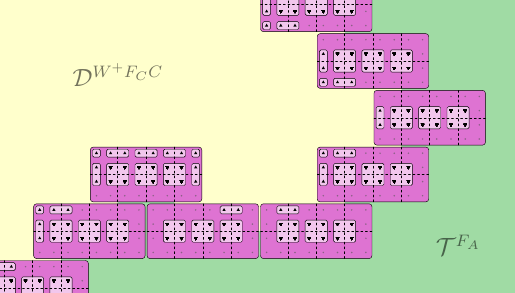}
		\caption{Substructure of the region $W^-$. Each $8 \times 4$ cell $i \in W^-$ (which in Fig.~\ref{fig:DomainWall} was drawn as a single circle) is now drawn as a dark purple rectangle. Within each rectangle, sites are divided into eight quartets (separated by dashed lines), according to which qudits are acted on by the same gate in the second layer of unitaries. The variables $\tau_{i,a} = \pm 1$ determine whether each quartet is acted on by $\Theta_{i,a}$ or $\mathcal{D}_{i,a}$ in Eq.~\eqref{eq:Theta4LocalDecomp}. We also make use of a complementary partitioning into regions $W^{1,2,3,4}$ according to the gate positions in the first layer of unitaries. Here, qudits in $W^1$ or $W^4$ are marked as faint grey dots, and those in $W^2$ ($W^3$) are shown as downward- (upward-)pointing triangles, which are correlated with each other according to the light purple blocks. As an example, the rightmost cell in this figure has just two qudits in $W^3$, which are jointly in a mixed state due to their entanglement with other qudits in $W^+$ (which are traced out). Regardless of the shape of the domain wall $W$, every cell will feature at least one such pair of qudits.}
		\label{fig:WRegions}
	\end{figure}

	Recall that $\sigma^{AW^-}_r$ is the state that results from applying the first layer of unitaries, measuring on $F_A$ with outcome $r$, and tracing out $W^+F_CC$.
	Thanks to the product structure of the pre-measurement state, we can separate out $W^-$ into four subregions $W^{1,2,3,4}$ which are mutually uncorrelated, as depicted in Fig.~\ref{fig:WRegions}. More specifically, we will argue that one can write $\sigma^{AW^-}_r = \sigma^{AW^1}_r \otimes \sigma^{W^2} \otimes \sigma^{W^3} \otimes \sigma^{W^4}$, where only the first factor is $r$-dependent. Then, for each choice of $\{\tau_{i,a}\}$, the corresponding state $\sigma^{AW_\tau}_r$, for which some sites are traced out, will also respect the same product structure, i.e.
	\begin{align}
		\sigma^{AW_\tau}_r = \sigma^{AW^1_\tau}_r \otimes \sigma^{W^2_\tau} \otimes \sigma^{W^3_\tau} \otimes \sigma^{W^4_\tau},
	\end{align}
	where we define $W^{b}_\tau \coloneqq W^b \cap W_\tau$, for $b = 1, \ldots, 4$.

	Our decomposition of $W^-$ into four non-overlapping subregions goes as follows: First, there are those sites $W^1$ that were entangled with qudits in $F_A$ before the measurement channel $\mathcal{T}^{F_A}$ was applied. After this measurement operation on $F_A$, we obtain a $r$-dependent state $\sigma^{AW^1}_r$, which is always a valid density matrix, and so $\|\sigma^{AW^1}_r\|_2 \leq 1$. Second, there are those sites $W^2$ that have been acted on by unitaries in the first layer of gates that are entirely within a cell (there are always 12 such sites per cell of $W^-$). Thirdly, $W^3$ is made up of sites that are acted on by gates that connect a single cell of $W^-$ with some qudits in $W^+$. Finally, $W^4$ is made up of all the remaining qudits in $W^-$, namely those that are acted on by gates that span more than one cell of $W^-$, but do not act on $F_A$.
	
	Using the decomposition described above, we can get a fairly crude upper bound on the sum in Eq.~\eqref{eq:TrBoundFineGrained} as follows. We ignore the structure of $\sigma^{AW^1_\tau}_r$ and $\sigma^{W^4_\tau}$, and simply use the fact that their 2-norms are upper bounded by unity. Then, since $\sigma^{W^2_\tau}$ and $\sigma^{W^3_\tau}$ possess no correlations between different cells of $W^-$, we can factorize the sum as
	\begin{align}
		\|T_r\|_1 \leq \sqrt{d'}\prod_{i \in W^-}\left(\sum_{\{\tau_{a}\}} \|\sigma^{W^2_{i,\tau}} \|_2 \cdot \|\sigma^{W^3_{i,\tau}} \|_2 \right)
		\label{eq:W23Factorize}
	\end{align}
	where $W^{2,3}_{i,\tau}$ are the qudits in cell $i$ that also belong to $W^{2,3}_\tau$. Note that for every cell $i$, the region $W^{3}_i$ features at least a pair of qudits that are entangled with another pair of qudits in $W^+$ (which we have traced out). Thus, the corresponding state $\sigma^{W^3_i}$ is generically mixed. For choices of $\tau_{i,a}$ where these sites are not traced out, this will contribute a multiplicative factor strictly less than unity to the norm. On the other hand, if any of the sites within a given cell \textit{are} traced out, then $\sigma^{W^2_\tau}$ will necessarily be mixed, due to the entanglement generated by intra-cell unitaries applied during the first layer of gates.
	
	Thus, for every cell $i$, and for any of the 255 choices of $\tau_{i,a}$, the 2-norm is suppressed by some factor, associated with the mixedness of the states in $W^2_i$ and/or $W^3_i$. The states in question are all tensor products of 1-, 2-, and 3-qudit states $\sigma_{1,2,3}$ generated by tracing out qudits from a Haar-random 4-qudit state. Using Haar integration techniques, along with Jensen's inequality $\mathbbm{E}_X \|X\|_2 \leq \sqrt{\mathbbm{E}_X \|X\|_2^2}$, we can bound the norm of such states (averaged over the unitary gates in the random circuit). Specifically,
	\begin{align}
		\mathbbm{E}_{\rm gates} \|\sigma_{1,3}\|_2 &\leq c_1, & \mathbbm{E}_{\rm gates} \|\sigma_{2}\|_2 &\leq c_2,
		\label{eq:PurityBounds4local}
	\end{align}
	where
	\begin{align}
		c_1^2 &= \frac{q(q^2+1)}{q^4 + 1} & c_2^2 &= \frac{2q^2}{q^4 + 1}.
	\end{align}
	One can very quickly see that each of the 255 terms that form each factor in Eq.~\eqref{eq:W23Factorize} is upper bounded by $\max(c_1^2, c_2) = c_2$, and thus gives us a straightforward bound $\|T_r\|_1 \leq \sqrt{d'} (255c_2)^{|W^-|}$. This then gives us a bound on MIE in terms of the self-avoiding walk generating function with effective temperature $\beta = -\log(255 c_2)$. Since $c_2$ itself decays as $\sim \sqrt{2}/q$ as we increase $q$, there will be some constant value of $q$ above which $\beta$ becomes larger than the critical value $\beta_c = \log \mu_{\Lambda^*}$, and this is sufficient to prove long-ranged MIE using the same arguments as in \cref{thm:Holog}.
	
	We can actually be slightly more careful by going through all 255 configurations per site separately and bounding each one by looking at the spatial structure of the pre-measurement state. A straightforward computation gives us a bound $\|T_r\|_1 \leq \sqrt{d'} g(q)^{|W^-|}$, for some complicated function $g(q)$ which we have determined numerically. Its form is not particularly instructive, but importantly it is strictly decreasing with $q$, and so we can determine the value of $q$ beyond which $g(q) \leq 1/\mu_{\Lambda^*}$. Using the known value $\mu_{\Lambda^*} = \sqrt{2+\sqrt{2}} \approx 1.85$ for the hexagonal lattice \cite{Duminil2012} (which is dual to the triangular lattice), we get the bound $q \geq 88$ quoted in \cref{thm:4local}.

	\subsection{Brickwork circuits}
	
	For brickwork circuits of depth $d_C = 4$, we use the fact that we can compile layers 1 and 2 together, as well as layers 3 and 4, which results in a circuit of the same architecture as the depth-$d_C = 2$ 4-local circuits (see Fig.~\ref{fig:2dArch}). Accordingly, our approach is very similar to before: We use cells of size $8 \times 4$, and absorb the final two layers of gates (which all act within single cells) into the local measurement channels $\mathcal{T}^{Q_i \rightarrow M_i}$. The decomposition Eq.~\eqref{eq:Theta4LocalDecomp} still holds, but this time, the maps $\Theta_{i,a}$ will not correspond to local 2-design measurements on each quartet of qudits, since the underlying random unitaries are 2-local rather than 4-local; thus Eq.~\eqref{eq:ThetaContractiveLambda} no longer holds. For this reason, we split up each quartet further into two pairs of qudits $a = (b_1, b_2)$, such that the qudits in each pair $b_{1,2}$ are acted on by the same unitary in the final layer of gates. We claim that the conjugated map $\tilde{\Theta}_{i,a}$ for each quartet satisfies
	\begin{align}
		\|\tilde{\Theta}_{i,a}[X_a]\|_2 \leq \|X_a\|_2 + \frac{\sqrt{3}}{q} \left(\|X_{b_1}\|_2 + \|X_{b_2}\|_2\right)
		\label{eq:ThetaContractiveBrickwork}
	\end{align}
	where $X_{b_1} = \Tr_{b_2}[X_a]$ is the result of a partial trace on $b_2$, and similar for $X_{b_2}$. Evidently, for large $q$, these latter contributions which were not present in Eq.~\eqref{eq:ThetaContractiveLambda} will be suppressed.
	
	To prove Eq.~\eqref{eq:ThetaContractiveBrickwork}, we first decompose $\Theta_{i,a}$ as
	\begin{align}
		\Theta_{i,a} &= \mathcal{T}_{i,a} \circ \left(\mathcal{P}_{b_1} \otimes \mathcal{P}_{b_2} +\mathcal{P}_{b_1}\otimes\mathcal{D}_{b_2}  + \mathcal{D}_{b_1}\otimes \mathcal{P}_{b_2} \right) \nonumber\\
		&\eqqcolon 
		\Theta_{i,b_1b_2} + \Theta_{b_1} + \Theta_{b_2} 
		\label{eq:ThetaDecompBrickwork}
	\end{align}
	where $\mathcal{P}_Y \coloneqq \text{id}_Y - \mathcal{D}_Y$ is a map that projects onto traceless operators on some region $Y$. Defining $\tilde{\Theta}_{i,b_1}[\cdot] = \sigma_{i,b_1}^{-1/2}\Theta_{i,b_1} \sigma_{i,b_1}^{-1/2}$ by analogy to Eq.~\eqref{eq:ThetaTildeI}, and similarly for the other two maps in \eqref{eq:ThetaContractiveBrickwork}, our aim is to compute norms $\|\tilde{\Theta}_{i,b_1}[X]\|_2$ for an arbitrary operator $X$, so we can employ the triangle inequality to bound this by the sum of three contributions corresponding to the three terms in the above expansion.
	
	Using the same steps that led to Eq.~\eqref{eq:ContractiveHaarIntegral} (specifically for the case $d' = 1$, so we can take $L_0 = \ket{0}\bra{0}$), we can compute
	\begin{align}
		&\|\tilde{\Theta}_{i,b_1b_2}[X]\|_2^2 \leq  q^8 \int \dif \mu_{\rm Br}(U_{b_1b_2}) \nonumber\\
		\times & \braket{0|U_{b_1b_2} \big((\mathcal{P}_{b_1}\otimes \mathcal{P}_{b_2})[X]\big) U_{b_1b_2}^\dagger |0}^2,
	\end{align}
	where the unitary $U_{b_1b_2}$ acting on the quartet $a = (b_1, b_2)$ is no longer sampled from the Haar measure $\dif \mu_H$, but rather the distribution obtained from concatenating four Haar-random two-qudit gates in the brickwork arrangement (these correspond to the gates appearing in the third and fourth layers of the original circuit). The analogous expression for $\tilde{\Theta}_{i,b_1}$ involves the channel $\mathcal{P}_{b_1}$ being replaced by $\mathcal{D}_{b_1}$.
	
	With some work, the necessary Haar integrals can be evaluated and we find
	\begin{align}
		\|\tilde{\Theta}_{i,b_{1}b_2}[X]\|_2^2 &= \frac{q^8\Tr[\tilde{X}^2]}{(q^4-1)^2}\left(1 - \frac{2}{q^2}\left[\frac{2q}{q^2+1}\right]^2 + \frac{1}{q^4}\right) \nonumber\\
		&\leq \Tr[X^2],
	\end{align}
	where $\tilde{X} = X - \mathcal{D}[X]$ is the traceless part of $X$. Similarly,
	\begin{align}
		\|\tilde{\Theta}_{i,b_{1}}[X]\|_2^2 &= \frac{q^4\Tr[\tilde{X}_{b_1}^2]}{q^4 - 1}\left(\left[\frac{2q}{q^2+1}\right]^2 - \frac{1}{q^2}\right) \nonumber\\
		&\leq \frac{3}{q^2}\Tr[X_{b_1}^2],
	\end{align}
	where $X_{b_1} = \Tr_{b_2}[X]$ is the result of a partial trace on $b_2$. Combining these expressions together using the triangle inequality, we arrive at the expression \eqref{eq:ThetaContractiveBrickwork} quoted before.\\
	
	We can use our new bound to obtain a similar expression to Eq.~\eqref{eq:W23Factorize}. To account for the three separate terms appearing in \eqref{eq:ThetaContractiveBrickwork}, we define a new set of variables $\tau_{i, b} = \pm 1$, where now $b \in [16]$ runs over the sixteen pairs of sites within a given cell. We then find
	\begin{align}
		\|T_r\| \leq \sqrt{d'}\prod_{i \in W^-}\left(
		\sum_{\substack{\{\tau_b\}\\ \vec{\tau} \neq (+1, \ldots, +1)}} \|\sigma^{W^{23}_{i, \tau}}\|_2 \theta_{\tau_b} \right)
		\label{eq:TrProductSumBrickwork}
	\end{align}
	where $W^{23}_{i,\tau}$ are the set of qudits in a cell $i$ that are part of $W^2$ or $W^3$, and for which $\tau_{i,b} = -1$. Here,  we define $\theta_{\tau_b} = \prod_{(b_1,b_2)}(\sqrt{3}/q)^{1 - \delta_{\tau_{b_1} \tau_{b_2}}}$, where the product is over all pairs $(b_1, b_2)$ that make up a quartet. We now have $2^{16} - 1$ terms for each factor of $i$. These include the $255$ terms we had before, corresponding the cases where $\tau_{b_1} = \tau_{b_2}$ for all quartets $(b_1, b_2)$, along with additional contributions, each of which are suppressed by at least one factor of $\sqrt{3}/q$. We see that every term in this sum decays at least as fast as $q^{-1}$, and thus there will exist a finite value of $q$ above which each factor becomes less than $1/\mu_{\Lambda^*}$.

	Finally, to upper bound each factor in the product in Eq.~\eqref{eq:TrProductSumBrickwork}, we use the same decomposition into regions $W^{1,2,3,4}$ as in the previous section. Again we have that every cell contains 12 qudits in $W^3_i$, and at least a pair of qudits in $W^2$ that are in a mixed state due to the trace being applied to $W^+$. This mixed state $\sigma_2$ is not given by the reduced density matrix of a fully random 4-qubit state due to the brickwork structure, but we can compute bounds that are analogous to Eq.~\eqref{eq:PurityBounds4local} for the brickwork case
	\begin{align}
		\mathbbm{E}_{\rm gates} \|\sigma_{1,3}\|_2 &\leq c_1', & \mathbbm{E}_{\rm gates} \|\sigma_{2}\|_2 &\leq c_2',
		\label{eq:PurityBoundsBrickwork}
	\end{align}
	where
	\begin{align}
		c_1'^2 &= \frac{q(q^4+6q^2+1)}{(q^2+1)^3} & c_2' &= \frac{2q}{q^2+1}.
	\end{align}
	(For $\sigma_2$, we have considered both possible orientations of the two-qudit state relative to the gates in the brickwork circuit, and taken the maximum here.) For the same reason as before, every term for which $\tau_{b_1} = \tau_{b_2}$ for all quartets $(b_1, b_2)$ is upper bounded by $c_2$, while all other terms are upper bounded by $\sqrt{3}/q$, which is less than $c_2'$ for $q \geq 3$. Thus, we can very quickly get a bound $\|T_r\|_1 \leq \sqrt{d'}(\sqrt{3}(2^{16}-1)c_2')^{|W^-|}$. The corresponding partition function is then ordered if $(q^2+1)/2q > \sqrt{3}\sqrt{2+\sqrt{2}}(2^{16}-1)$, which is achieved for $q \geq \num{419479}$. If we instead consider upper bounds for each term one by one, we can get a tighter bound. We have performed this task numerically, which yields the sufficient condition $q \geq 134$ quoted in \cref{thm:Brickwork}.
	
	\section{Proof of advantage over shallow classical circuits\label{app:Advantage}}
	
	In this appendix, we use our bounds on MIE to prove that random Clifford circuits with constant qudit dimension $q$ and depth $d_C$ cannot be simulated by constant-depth classical circuits. Specifically, we focus on the holographic model (Section \ref{subsec:Holog}) with maximally entangled states on the bonds, random Clifford unitaries on each site, and Pauli measurements. We will refer explicitly to the square lattice, but the results can be immediately generalized to any planar lattice.
	
	The connection between long-ranged MIE and non-simulability by shallow classical circuits was established in Ref.~\cite{Watts2024}; thus our task is to convert our results into a form for which their theorems can be applied. Specifically, we wish to show that the property which they call \textit{long-ranged tripartite MIE} holds. 
	\begin{definition}[Watts \textit{et al.} \cite{Watts2024}]
		A distribution of states of $n$ qubits is said to exhibit the long-ranged tripartite MIE property if there are constants $(c_1, c_2) \in (0,1)$ such that if a uniformly random triplet of qubits $\{h,i,j\} \subset [n]$ is chosen and the remaining qubits $B = [n] \backslash \{h,i,j\}$ are measured, then with probability at least $c_1$ over the choice of triplets, states, and measurement outcomes, the post-measurement reduced states $\rho_{\alpha, s_B} = \Tr_{\{h,i,j\} \backslash \alpha}[\phi_{s_B}]$ on each qudit $\alpha \in \{h,i,j\}$ simultaneously satisfy
		\begin{align}
			\textup{Tr}[(\rho_{\alpha, s_B})^2] &< c_2 & \forall \alpha \in \{h,i,j\}.
			\label{eq:TripartiteMIECond}
		\end{align}
	\end{definition}
	\noindent The authors of Ref.~\cite{Watts2024} showed that any quantum circuit satisfying the above condition cannot be simulated by any classical circuit whose depth scales less than logarithmically with system size. Thus, our task is to show that the above indeed holds. We will repeatedly make use of the fact that we need only show that the probability of \eqref{eq:TripartiteMIECond} holding is lower bounded by some arbitrarily small constant (not diminishing with system size). Thus, we can condition on certain favourable events happening during the various random processes involved, post-selecting out all other instances. This is valid provided the full collection of events we condition on has a constant probability of occurring.
	
	To convert our results for quidts of dimension $q $ into a qubit representation, we will assume that $q = 2^m$ for some integer $m$, and accordingly think of each qudit as being composed of $m = O(1)$ qubits. In the holographic model on the square lattice, each site hosts four halves of maximally entangled pairs, so there are $4m$ qubits per site. Note that according to the above definition, we need to leave a uniformly random set of three \textit{qubits} unmeasured and consider their entanglement structure. Our first step is to argue that the condition \eqref{eq:TripartiteMIECond} for three unmeasured qubits is implied by an equivalent condition where we leave three full sites completely unmeasured.
	
	First, the probability of qubits $\{h,i,j\}$ all belonging to different sites approaches 1 in the thermodynamic limit. Conditioned on this being true, we can use $\{H,I,J\}$ to denote the corresponding sites to which these qubits belong. Clearly, after conditioning, this subset is itself also uniformly random over all subsets of three sites.  Let us suppose that we have chosen the Clifford unitaries and performed measurements on all sites except for $\{H,I,J\}$. This will yield some state $\phi^{HIJ}$ that depends on the unitaries applied and outcomes obtained. Our assertion is that there is a constant probability $c_1' > 0$ that these states simultaneously satisfy
	\begin{align}
		S(\rho^X) &\geq c_2' & \forall X \in \{H, I, J\}.
		\label{eq:TripartiteMIESites}
	\end{align}
	for some $c_2' > 0$.

	By Result 1 of Ref.~\cite{Bravyi2006}, any tripartite stabilizer state is equivalent up to local Clifford unitaries on $H$, $I$, and $J$ to a tensor product of some collection of $\ket{0}$ states, Bell pairs $\ket{\Phi^+} = \frac{1}{\sqrt{2}}(\ket{00} + \ket {11})$, and Greenberger-Horne-Zeilinger (GHZ) states $\ket{\text{GHZ}} = \frac{1}{\sqrt{2}}(\ket{000} + \ket {111})$. If Eq.~\eqref{eq:TripartiteMIESites} holds, there are two (non-mutually exclusive) possibilities: 1) $H$, $I$, and $J$ share at least one GHZ state, or 2) There is at least one Bell pair between every pair $(H,I)$, $(I,J)$, $(J,H)$. Now, once this state is generated, we pick a qubit uniformly from each site $h \in H$, $i \in I$, $j \in J$ and sample the final three Clifford unitaries. These random processes are statistically independent from any processes that occurred before, and so we can also condition favourable choices of these variables, while only reducing the probability $c_1$ by a constant factor. Finally, we measure all qubits except for $\{h,i,j\}$. The net effect of this process is equivalent to projecting each site onto some rank-2 subspace spanned by a projector $\Pi_{X}$ ($X = H, I, J$) which itself is a stabilizer state. States corresponding to different measurement outcomes are related to one another by Pauli rotations, which do not change the entanglement structure, and so for our purposes we can work on the basis that $\Pi_X$ only depends on the choice of the Clifford unitaries and sites $h,i,j$. 
	
	For possibility 1) described above, there is some probability that the Clifford unitary and choice of unmeasured qubit at each site are such that the effective projector $\Pi_X$ on each site has no effect on the qubits that host the GHZ state. The probability of this occurring is at least $(|\mathcal{C}|_{4m})^{-3}$, where $|\mathcal{C}|_{4m}$ is the cardinality of the Clifford group on $4m$ qubits (ignoring signs of stabilizers). Since $m$ is system size-independent, this gives us a constant probability. For possibility 2), there is a Bell pair between each pair of sites, so each site hosts two halves of Bell pairs. Over the choices of Clifford unitaries and unmeasured qubits, there is some probability that their net effect is to project the two halves on each site onto the two-dimensional subspace spanned by $\ket{\Phi^+}$ and $\ket{\Phi^-} = \frac{1}{\sqrt{2}}(\ket{00} - \ket{11})$ (this is equivalent to measuring the operator $ZZ$ on the two halves of each Bell pair, and projecting out all other degrees of freedom). The net result is a GHZ state up to some local Clifford unitaries. Again, because the number of Cliffords and outcomes on each site is bounded when $m$ is fixed, this probability is constant. Thus, if Eq.~\eqref{eq:TripartiteMIESites} is satisfied with some constant probability $c_1' > 0$, then Eq.~\eqref{eq:TripartiteMIECond} is satisfied for some (possibly smaller) constant probability $c_1 > 0$.
	
	Having established this fact, we now need only prove our assertion, namely that there is some constant probability that Eq.~\eqref{eq:TripartiteMIESites} is indeed satisfied. First note that because of the stabilizer structure of the problem, $S(\rho^X)$ is quantized to an integer multiple of $\log 2$. Thus, taking $c_2' = 1$, Eq.~\eqref{eq:TripartiteMIESites} becomes the condition that $\rho^X$ is not pure for each $X$. Now we use the union bound to reduce to a single random variable: If we can find a constant $c_3' < 1/3$ such that for any fixed $X \in \{H,I,J\}$, the probability $\text{Pr}(\rho^X \text{ is pure}) \leq c_3'$, then we have
	\begin{align}
		\text{Pr}\left(\textstyle \bigcup_{X \in \{H,I,J\}} \rho^X \text{ is pure}\right) \leq 3c_3' < 1,
	\end{align}
	and thus we will have succeeded. For this purpose, we will need a concentration inequality of the kind given in Eq.~\eqref{eq:EntropyConcDelta}, which is expressed in terms of the distillation error $\overline{\epsilon}$. In fact, since we are only looking to determine whether the relevant state is pure or not, we can consider the problem of attempting to distill a Bell pair of dimension $d' = 2$ between $X$ and $\{H,I,J\}\backslash X$, and because any pure state will yield an error of $\epsilon = 1$, we can simply use $\text{Pr}(\rho^X\text{ is pure}) \leq \text{Pr}(\epsilon \geq 1) \leq \mathbbm{E}[\epsilon]$.

	Finally we use Lemma \ref{lem:SAWQeBound} to upper bound the mean error $\overline{\epsilon}$ in the case where $I$ and $J$ are traced out, and $H$ is sent through the distillation channel. Again we can look at a sum over separating domain walls with Boltzmann weights given by $2^{-m|W|/2}$, i.e.~the effective inverse temperature is $\beta = \frac{1}{2}m \log 2$. In this case, since $H$, $I$, and $J$ are in the bulk of the system, the relevant domain walls now include loop configurations of the following kinds: 1) a single loop encircling $H$, 2) a single loop encircling $I$ and $J$ together, and 3) two separate loops encircling $I$ and $J$ separately. As we will see, configurations with long loops are strongly suppressed. For a $\sqrt{n} \times \sqrt{n}$ square lattice, we will find it useful to condition on the event that $H$, $I$, and $J$ are all a $\ell_1$ distance of at least $\sqrt{n}/4$ from the edges of the system, and also all pairwise separated by at least $\sqrt{n}/4$ (this occurs with constant probability). This effectively allows us to ignore configurations of type 2) above, and also any configurations with domain walls that terminate on the edges of the system (and thus do not form closed loops).
	
	Loops that do not intersect themselves are the closed-path analogues of self-avoiding walks, and are sometimes referred to as lattice polygons. Their statistical mechanics has been studied for a long time, and it is known that there is an ordered phase for exactly the same values of $\beta$ as for self-avoiding walks \cite{Hammersley1961}, namely when $\beta > \beta_{\rm crit} = \log \mu$, where here $\mu$ is the connective constant of the square lattice. This fact can be proven starting from upper bounds on the total number  of such lattice polygons $c_l$ of a fixed length $l$ that intersect the origin. We will use the bound $c_l \leq \mu^l$ given in Ref.~\cite{Whittington2009}, which implies that the total weight coming from domain walls of length $l \geq l_0$ (irrespective of boundary conditions) is at most
	\begin{align}
		w(l_0) \coloneqq \sum_{l=l_0}^\infty c_l e^{-\beta l} &\leq \frac{\nu^{l_0}}{1 - \nu} & \nu < 1
	\end{align}
	where $\nu = e^{\log \mu - \beta}$. This bound justifies our neglect of configurations of type 2).
	
	Now, we will bound the contribution to the partition function from configurations of type 1). For a given lattice polygon intersecting the origin of length $l$ and area $A$, the number of ways we can translate this polygon such that it encircles the site $H$ will be given by $A$, which is at most $l^2/16$. Therefore, we can bound the total contributions of this type as
	\begin{align}
		\sum_{l=4}^{\infty} \frac{l^2}{16} c_l e^{-\beta l} &\leq \frac{2\nu^4}{(1-\nu)^3} & \nu < 1
	\end{align}
	Finally, we can upper bound configurations of type 3) by the square of the above, since these consist of separate loops encircling $I$ and $J$. This gives us a bound on the partition function, which we can substitute into Eq.~\eqref{eq:QeBoundFinal}, and we need only show that this yields a value of $\overline{\epsilon}$ that is less than 1/3 by some constant amount. Since our bound on $\overline{\epsilon}$ is monotonically decreasing in $\nu$, we can numerically verify that $\nu \leq 1/3$ is more than sufficient, and thus for $\frac{1}{2} m \log 2 \geq \log(3\mu)$ we will have proved that the condition \eqref{eq:TripartiteMIECond} holds. This is satisfied for $m = 6$, namely $q = 64$ as quoted in \cref{thm:AdvantageFormal}.

	\bibliography{shallow_2d}
	
\end{document}